\documentclass[printer]{aa} 
\usepackage{threeparttable}
\usepackage{lscape}
\usepackage{multirow}
\usepackage[normalem]{ ulem }
\usepackage{soul}
\usepackage{colortbl}
\usepackage{grffile, graphicx}
\usepackage[T1]{fontenc}
\usepackage{lmodern}
\usepackage{txfonts}

\newcommand{\Teff}{\ensuremath{\mathrm{T_{eff}}}}
\newcommand{\MJup}{\ensuremath{\mathrm{M_{Jup}}}}

\newcommand{\RJup}{\ensuremath{\mathrm{R_{Jup}}}}

\newcommand{\Msun}{\ensuremath{\mathrm{M_{\odot}}}}
\newcommand{\mic}{\ensuremath{\mathrm{\mu m}}}

\newcommand{\Lum}{\ensuremath{\mathrm{log(L/L_{\odot})}}}

\newcommand{\Msolid}{\ensuremath{\mathrm{M_{solid}}}}
\newcommand{\Mplanet}{\ensuremath{\mathrm{M_{planet}}}}
\newcommand{\Msolidini}{\ensuremath{\mathrm{M_{solid\,ini}}}}
\newcommand{\Mgas}{\ensuremath{\mathrm{M_{gas}}}}
\newcommand{\Maccreted}{\ensuremath{\mathrm{M_{accreted}}}}
\newcommand{\Mini}{\ensuremath{\mathrm{M_{ini}}}}
\newcommand{\fsg}{\ensuremath{\mathrm{f_{s/g}}}}
\newcommand{\kms}{\ensuremath{~\mbox{km\,s$^{-1}$}}}
\newcommand{\ncs}{\ensuremath{n_{\mathrm{C},\,\mathrm{s}}}}
\newcommand{\ncg}{\ensuremath{n_{\mathrm{C},\,\mathrm{g}}}}
\newcommand{\nos}{\ensuremath{n_{\mathrm{O},\,\mathrm{s}}}}
\newcommand{\nog}{\ensuremath{n_{\mathrm{O},\,\mathrm{g}}}}

\DeclareUnicodeCharacter{2212}{-}

\begin{document} 
   \title{A medium-resolution spectrum of the exoplanet HIP~65426\,b\thanks{Based on observations collected at the European Southern Observatory under ESO programs 0101.C-0840(A).}}
\titlerunning{Medium-resolution spectra of the exoplanet HIP~65426\,b}

   \author{S. Petrus\inst{1}, M. Bonnefoy\inst{1}, G. Chauvin\inst{2,1}, B. Charnay\inst{3}, G-D. Marleau\inst{4,5,6}, R. Gratton\inst{5},  A.-M. Lagrange\inst{1}, J. Rameau\inst{1}, C. Mordasini\inst{5}, M. Nowak\inst{7,8}, P. Delorme\inst{1}, A. Boccaletti\inst{3}, A. Carlotti\inst{1}, M. Houll\'{e}\inst{9}, A. Vigan\inst{9}, F. Allard\inst{10}\thanks{Deceased}, S. Desidera\inst{8}, V. D'Orazi\inst{8}, H. J. Hoeijmakers\inst{11}, A. Wyttenbach\inst{1}, B. Lavie\inst{11}} 
   
   \authorrunning{Petrus et al.}

   \institute{\inst{1} Univ. Grenoble Alpes, CNRS, IPAG, 38000 Grenoble, France \\
   \inst{2} Unidad Mixta Internacional Franco-Chilena de Astronom\'{i}a, CNRS/INSU UMI 3386 and Departamento de Astronom\'{i}a, Universidad de Chile, Casilla 36-D, Santiago, Chile\\
   \inst{3} LESIA, Observatoire de Paris, Universit\'{e} PSL, CNRS, Sorbonne Universit\'{e}, Univ. Paris Diderot, Sorbonne Paris Cit, 5 place Jules Janssen, 92195 Meudon, France.\\
   \inst{4} Institut fur Astronomie und Astrophysik, Universit\"{a}t T\"{u}bingen, Auf der Morgenstelle 10, 72076 T\"{u}bingen, Germany\\
 \inst{5} Physikalisches Institut, University of Bern, Gesellschaftsstrasse 6, 3012 Bern, Switzerland\\
   \inst{6} Max-Planck-Institut f\"{u}r Astronomie, K\"{o}nigstuhl 17, 69117 Heidelberg, Germany\\
    \inst{7} Institute of Astronomy, University of Cambridge, Madingley Road, Cambridge CB3 0HA, United Kingdom\\
   \inst{8} Kavli Institute for Cosmology, University of Cambridge, Madingley Road, Cambridge CB3 0HA, United Kingdom\\
   \inst{9} Aix Marseille Univ., CNRS, CNES, LAM, Marseille, France\\
   \inst{10} Univ Lyon, Ens de Lyon, Univ Lyon 1, CNRS, Centre de Recherche Astrophysique de Lyon UMR5574, F-69007, Lyon, France\\
   \inst{11} Observatoire de Gen\`{e}ve, University of Geneva, Chemin des Maillettes, 1290, Sauverny, Switzerland ; Center for Space and Habitability, Universit\"{a}t Bern, Gesellschaftsstrasse 6, 3012, Bern, Switzerland}
   \date{Received 15/07/2020; accepted 02/12/2020}

\abstract{Medium-resolution integral-field spectrographs (IFS) coupled with adaptive-optics such as Keck/OSIRIS, VLT/MUSE, or SINFONI are appearing as a new avenue for enhancing the detection and characterization capabilities of young, gas giant exoplanets at large heliocentric distances ($>$5 au). We analyzed K-band VLT/SINFONI medium-resolution ($R_\lambda\sim$ 5577) observations of the young giant exoplanet HIP~65426\,b. Our dedicated IFS data analysis toolkit (\texttt{TExTRIS}) optimied the cube building, star registration, and  allowed for the extraction of the planet spectrum. A Bayesian inference with the nested sampling algorithm coupled with the self-consistent forward atmospheric models \texttt{BT-SETTL15} and \texttt{Exo-REM} using the \texttt{ForMoSA} tool yields \Teff=1560$\pm100$\,K, log(g)$\leq$4.40 \,dex, [M/H]=$0.05
^{+0.24}_{-0.22}$\,dex, and an upper limit on the C/O ratio ($\leq$0.55). The object is also re-detected with the so-called \textit{``molecular mapping''} technique. The technique yields consistent atmospheric parameters, but the loss of the planet pseudo-continuum  in the process degrades or modifies the constraints on these parameters. The solar to sub-solar C/O ratio suggests an enrichment by solids at formation if the planet was formed beyond the water snowline ($\geq$20\,au)  by core-accretion. However, a formation by gravitational instability can not be ruled out. The metallicity is compatible with the bulk enrichment of massive Jovian planets from the Bern planet population models. Finally, we measure a radial velocity of 26$\pm$15 \kms\, compatible with our revised measurement on the star. This is the fourth imaged exoplanet for which a radial velocity can be evaluated, illustrating the potential of such observations for assessing the coevolution of imaged systems belonging to star forming regions, such as HIP~65426.}


   \maketitle

\section{Introduction}

\begin{table*}[t]
\centering
\caption{Observing log}
\label{Tab:Obs_Log}
\small
\begin{tabular}{ccccccccccc}
\hline
\hline
Date & UTC Time &  Instrument	&	Setup  &  DIT & N$_{exp}$   &  $\mu$ & seeing & $<$Strehl$>$ & $\Delta \pi$ &    Notes 	\\
DD/MM/YYYY  & hh:mm &      &       &   (s) &   &   & (")   & ($\%$) &   ($^{\circ}$) & \\    
\hline
25/05/2018  & 00:47-00:48 & SINFONI &   K-0.025"/pix    &   10  & 2 &  1.16 &   0.7 &   28.0    &   0.28    &   On-axis \\
25/05/2018  & 00:51-02:29 & SINFONI &   K-0.025"/pix    &   100 & 53 & 1.13 & 0.6 & 33.0 & 47.3 & Star offcentered \\  
26/05/2018  & 00:47-00:48 & SINFONI &   K-0.025"/pix    &   10  & 2 &  1.15 &  1.0  &   14.8    &   0.30    &   On-axis \\
26/05/2018  & 00:51-02:29 & SINFONI &   K-0.025"/pix    &   100 & 41 & 1.13 & 1.6 & 9.9 & 36.5 & Star offcentered \\  
\hline
\end{tabular}
\tablefoot{Average Strehl ratio as measured by the real-time computer. $\Delta \pi$ corresponds to the amount of field rotation.}
\end{table*}

Direct imaging can provide high-fidelity spectra of young (age $<$150\,Myr) self-luminous giant exoplanets. These spectra are typically made of tens to thousands of data-points over a broad wavelength range ($0.5-5\,$\mic) and can be acquired in a few hours of telescope time. The set of spectro-photometric data collected thus far on the few dozen of imaged planetary-mass companions \citep[e.g.,][]{2005A&A...438L..29C, 2007ApJ...657.1064M, 2008ApJ...689L.153L, 2010A&A...517A..76P, 2010ApJ...710L..35J, 2013Sci...339.1398K, 2014A&A...562A.127B, 2014ApJ...780L...4B, 2017AJ....153..182C, 2017A&A...603A..57S} exhibits numerous molecular and atomic lines (e.g., H$_{2}$O, $^{12}$CO, K I). They enable to probe the chemical and physical phenomena at play in exoplanetary atmospheres with effective temperature (\Teff) in the range $\sim600-2300$\,K, similar to those of mature brown-dwarfs \citep[hereafter BDs, ][]{1988Natur.336..656B, 1995Natur.378..463N}, but with surface gravities 100 to 1000 times smaller given the larger radii at the very early evolution stages. Planetary formation models indicate the gaseous envelope of these young Jupiters formed in circumstellar disks should undergo a differential chemical enrichment depending on the detailed formation history  \citep[e.g.,][]{2011ApJ...735...30B, 2011ApJ...743L..16O, 2011ApJ...733...65B, 2014Life....4..142H, 2016ApJ...831L..19O, 2016ApJ...832...41M, 2018arXiv181110904V}.

Exoplanet spectra obtained in direct imaging therefore offer the opportunity to determine  the physical characteristics of the studied objects (age, mass, radius, \Teff, log(g)),  exploring the impact of surface gravity on the atmospheres \citep[e.g., ][]{2010A&A...517A..76P, 2014A&A...562A.127B}, revealing any deviation from the host-star chemical composition (e.g. C/O ratio, [M/H]; \citealt{2013Sci...339.1398K, 2017AJ....154...91L, 2019ESS.....440407N}), and ultimately tracing their mechanism of formation. Atmospheric models, including the formation of clouds of different compositions (silicates, iron, sulfites) have been considered for more than a decade to reproduce the spectral features, colors, and luminosity of directly imaged planets \citep{2001ApJ...556..357A, 2001ApJ...556..872A, 2008MNRAS.391.1854H, 2011ApJ...733...65B, 2011ApJ...737...34M, 2012RSPTA.370.2765A,  2012ApJ...756..172M,2018ApJ...854..172C,  2019A&A...627A..67M}, and are now massively used to derive the physical properties of the atmosphere of imaged exoplanets. 

With the direct imaging technique, the challenge of accurately removing the dominant stellar flux on broad wavelength intervals represents the main hurdle for extracting unbiased exoplanet spectra. Since 2013, the  high-contrast imagers and spectrographs such as GPI \citep{2006SPIE.6272E..0LM}, SPHERE \citep{2019A&A...631A.155B} or SCExAO/CHARIS \citep{2015SPIE.9605E..1CG, 2015PASP..127..890J} have gathered  low-resolution ($R_{\lambda} = 30-350$), near-infrared ($0.95-2.45\,$\mic) spectra of exoplanets down to 100\,mas from their host star. The spectra confirmed some of the results inferred earlier on from photometric spectral energy-distributions \citep[e.g.,][]{2008Sci...322.1348M, 2013A&A...555A.107B}: the reduced surface gravity affects the vertical mixing and the gravitational settling of condensates leading to thicker cloud layers, upper atmosphere sub-micron hazes, and cloud opacities remaining down to $\Teff=600$\,K at early ages  \citep{2016A&A...587A..58B, 2017AJ....154...10R, 2017A&A...605L...9C, 2017A&A...603A..57S, 2018AJ....155..226G, 2018A&A...617A..76C, 2020AJ....159...40U, 2017A&A...608A..79D, 2020MNRAS.495.4279M}. These spectra have however led to conflicting conclusions on the derivation of the atmospheric composition \citep{2017A&A...603A..57S, 2017AJ....154...10R}, possibly owing to the limited resolution of the observations and uncertainties in the atmospheric models. Spectroscopy at medium-resolving powers (R of a few 1000s) is warranted for accessing that information \citep[see][]{2017AJ....154...91L} and testing the models further. 

\begin{figure*}[t]
  \centering
  \includegraphics[width=19cm]{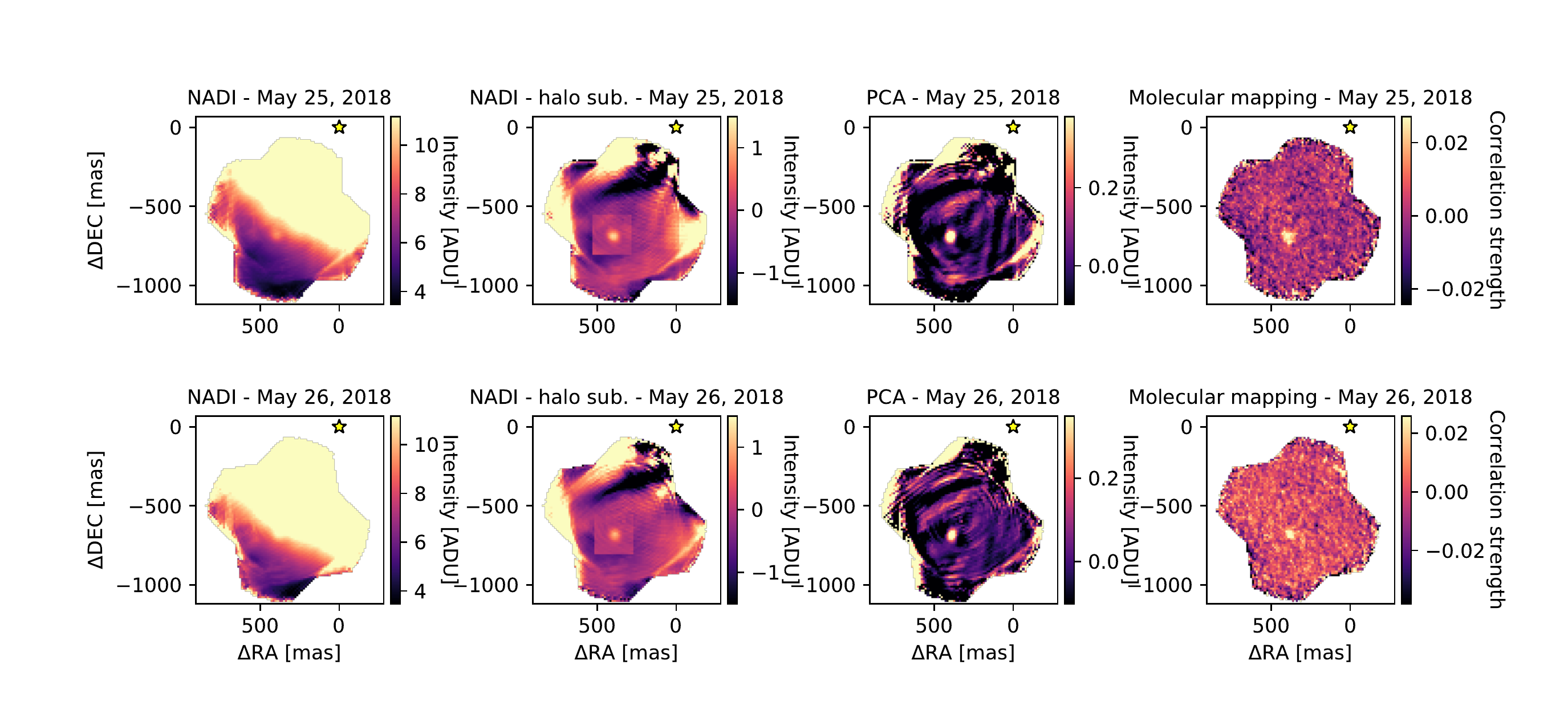} 
  \caption{Detection of HIP~65426\,b with SINFONI using (from \textit{left} to \textit{right}) a simple derotation and stack of the individual cubes (nADI), a removal of the halo at each wavelength using a circular profile and a local fit with a 2D polynomials, the PCA algorithm (3 modes), and the molecular mapping with an \texttt{Exo-REM} model at \Teff=1600\,K, log(g)=4.0\,dex, [M/H]=0\,dex, C/O=0.35 (at 20 \kms).}
  \label{Fig:detecSINF}
\end{figure*}

\label{Sec: Data_collection}

Adaptive-optics-fed integral-field spectrographs (IFS) operating at medium spectral resolving power ($R_{\lambda} =2000-5000$) in the near-infrared such as SINFONI at VLT, NIFS at Gemini-North, or OSIRIS at Keck offer to partly deblend a rich set of molecular absorption contained in the exoplanet spectra.  They have been used for gathering the spectra of young low-mass companions at large distance from their host stars ($>$1") or in systems with moderate contrasts in the near infrared \citep{2007ApJ...656..505M, 2007A&A...463..309S, 2008A&A...491..311S, 2009ApJ...704.1098L, 2011ApJ...743..148B, 2014A&A...562A.127B, 2014A&A...566A..85S, 2015ApJ...811L..30B, 2017AJ....154..165B, 2017A&A...608A..71D}. 
More recently, high-contrast imaging techniques such as the Angular-Differential-Imaging \citep[e.g. ADI,][]{2006ApJ...641..556M} have been implemented on these instruments and allowed for characterizing companions at shorter separations \citep[$\rho<$1";][]{2010ApJ...723..850B, 2015MNRAS.453.2378M, 2018A&A...617A.144H, 2019A&A...624A...4M, 2019ApJ...877L..33C}. 
Using OSIRIS, \cite{2013Sci...339.1398K} and \cite{2015ApJ...804...61B} detected carbon monoxide and water in the spectrum of HR\,8799\,b and c which they confirmed  using the cross-correlation of the spectra with templates of pure H$_{2}$O, CO, and CH$_{4}$ absorption. The cross-correlation approach has been generalized to all spaxel contained in IFS datacubes by \cite{2018A&A...617A.144H} following a modified prescription of the algorithm described in \cite{2002ApJ...578..543S}. This ``molecular mapping'' method allows for getting rid of the remaining stellar flux residuals including the quasi-static noise created by the IFS optical elements \citep{2008A&A...478..597J} while providing an interesting mean to identify unambiguously cool companions and characterize their properties (\Teff, log(g), [M/H], composition, and radial velocity). The approach has thus far been used on a limited set of data  \citep[$\beta$ Pictoris b, HR8799\,b;][]{2018A&A...617A.144H, 2018A&A...616A.146P}, and its performances and limitations are yet to be investigated on additional data and against alternative  methods exploiting the IFS data diversity \citep[e.g.,][]{2006ApJ...641..556M, 2007MNRAS.378.1229T,  2019AJ....158..200R}.

HIP~65426\,b \citep{2017A&A...605L...9C} is the first exoplanet discovered with the VLT/SPHERE instrument, at a projected separation of 92\,au from the young ($14\pm4$\,Myr) intermediate-mass star HIP~65426 (A2V, M=1.96$\pm$0.04\,M$_\odot$). The star is located at a distance of $111.4\pm3.8$\,pc \citep{2016A&A...595A...2G}, and belongs to the Lower Centautus-Crux (hereafter LCC) association \citep{1999AJ....117..354D,2011MNRAS.416.3108R}. The low-resolution Y to H-band photometry and spectra of HIP~65426\,b   indicate the object has a L$6\pm1$ spectral type with clear signatures of reduced surface gravity \citep{2017A&A...605L...9C}.  The hot-start evolutionary models \citep{2003A&A...402..701B, 2000ApJ...542..464C, 2013A&A...558A.113M} predict $M_{b}\,=\,6\,-\,12$\,\MJup\, in the planetary-mass range. Atmospheric models estimate \Teff~in the range $1100-1700$\,K, a surface gravity log(g) lower than 5 dex, and a radius from 1 to 1.8\,\RJup. These physical properties were  confirmed by \cite{2019A&A...622A..80C} who analysed the SPHERE data together with new VLT/NaCo L and M bands (3.49-4.11 \mic\, and 4.48-5.07\,\mic) observations. They found $\Teff\,=\,1618\pm7$\,K, $log(g)\,=\,3.78_{-0.03}^{+0.04}$\,dex and $R\,=\,1.17\pm0.04$\ \RJup~ and estimate a new mass $M\,=\,8\pm1$\,\MJup\,(statistical errors). The relatively low mass ratio of HIP~65426\,b with A ($M_{b}/M_{A}\sim$0.004), its intermediate semi-major axis \citep[$110^{+90}_{-30}$ au,][]{2019A&A...622A..80C}, and the non-detection of a debris disk in the system \citep{2017A&A...605L...9C} question the origin of HIP~65426\,b and makes the system an interesting testbed for planet formation theories. \cite{marleau2019a} explored formation scenarios for HIP~65426\,b, including the evolution of the protoplanetary disc and the gravitational interaction with possible further companions in the system.

In this paper, we analyse medium-resolution ($R_\lambda\sim$5500) K-band data of HIP~65426\,b collected with the VLT/SINFONI IFS to characterize the exoplanet and evaluate the potential of the molecular mapping technique. We describe in Section \ref{Sec: Data_collection} the observations and data reduction based on the molecular mapping and classical ADI extraction techniques. We characterize the planet in Section \ref{Sec:Physical_properties} using the new spectroscopic data, and discuss our results in the Section  \ref{Sec:Discussion}.

\section{Observations and data reduction}
\label{sec:obs}

\subsection{Observations}
HIP~65426 was observed on May 25 and May 26, 2018 with the VLT/SINFONI instrument (program 0101.C-0840; PI Hoeijmakers) mounted at the Cassegrain focus of VLT/UT4\footnote{A third observation was attempted on May 27, 2018 but the conditions were poor and we did not considered it here.} (see Table \ref{Tab:Obs_Log}). SINFONI is made of a custom adaptive optics module (MACAO) and of an IFS (SPIFFI). SPIFFI cuts the field-of-view into 32 horizontal slices (slitlets), which are re-aligned to form a pseudo-slit, and  dispersed by a grating on a Hawaii 2RG (2k$\times$2k) detector \citep{2003SPIE.4841.1548E, 2004Msngr.117...17B}. 

The instrument was operated with pre-optics and a grating sampling a 0.8"$\times$0.8" field of view with rectangular spaxels of 12.5$\times$25 mas size, from 1.928-2.471 \mic, at a spectral resolution $R_\lambda=\lambda / \Delta \lambda \sim$ 5220 (from the width of the line-spread function). MACAO was used at all time during the observations with the primary star as a reference for the wave-front sensing.  The de-rotator at the telescope focus was in addition turned off to allow for pupil-tracking observations.

At each night, the star was first placed inside the field-of-view and two 10\,s integrations were obtained. The core of the star point-spread-function (PSF) was then moved outside of the field of view with the instrument field selector and a sequence of 100\,s integrations centered on the expected position of the planet were performed. This strategy is similar to the one adopted on $\beta$ Pictoris \citep[][Bonnefoy et al. in prep]{2018A&A...617A.144H}. It reduces the impact of the read-out in the noise budget while avoiding any persistence on the detector of SPIFFI. \\

\subsection{Cube building and registration}
\label{Sec: Cube_building}
The data were initially reduced with the SINFONI data handling pipeline v3.2.3 \citep{2006NewAR..50..398A} through the EsoReflex environment \citep{2013A&A...559A..96F}. The pipeline used a set of calibration frames obtained at day time to perform basic cosmetic steps on the raw bi-dimensional science frames and correct them from the distortion. The pipeline also identified the position of slitlets on the frames and the wavelength associated to each pixel before building a serie of datacubes corresponding to each exposure. The sky emission was evaluated and removed through the field of view using the methods from \cite{2007MNRAS.375.1099D}.

We used on top of the ESO recipes the \textit{Toolkit for Exoplanet deTection and chaRacterization with IfS} (hereafter \texttt{TExTRIS}) to optimize our  cubes for the high-contrast science (The main steps are described below, see Bonnefoy et al. in prep for another application of this tool). The toolkit first corrected the raw SINFONI data from various noises occurring on the detector described in \cite{2014A&A...562A.127B}. 

\texttt{TExTRIS} also mitigated two different effects affecting the cubes:  
\begin{itemize}
    \item we corrected an incorrect estimate of the slitlet positions on the raw science frame and masked the first two and last two columns of spaxels in the cubes which are  affected by cross-talk at the slitlet edges. This step is critical for removing important stellar flux residual at the edges of the field of view and getting an accurate knowledge of the star position outside the field of view. 
    \item we re-examined the wavelength calibration performed by the ESO pipeline by comparing the many telluric absorption lines contained in each spaxel to a model generated for the observing conditions with the python module \texttt{skycalc\_ipy}\footnote{https://github.com/astronomyk/skycalc\_ipy}  and based on the ESO Skycalc tool \citep{refId0, refId1}.  The method was shown to be critical for yielding a robust radial velocity measurement of $\beta$ Pictoris b (Bonnefoy et al, in prep) and improving the capabilities of the molecular mapping technique (see Section \ref{subsubsec:molmap}). It is applicable whenever the spaxels contain enough stellar light to evidence the telluric absorptions. The shifts were evaluated on each input  cube across the field of view. A master median-combined map of velocity shift was then created and  each individual cube was corrected from the effect. We find spaxels are red-shifted by 21.6 and 16.4 \kms\, on average (0.6 and 4.2 \kms\, dispersion along the sequences) on the first and second night of observations, respectively. We found and corrected additional relative $\sim$\,\kms\, shifts from spaxel to spaxel across the field. These smaller shifts are visible at the same position for all cubes along the sequence and point toward differential errors on the wavelength solutions between the slitlets. We validated this re-calibration on the final products in the Appendix \ref{Appendix:A} using an independent method relying on OH sky emission lines. 
\end{itemize}

We used \texttt{TExTRIS} to re-compute the parallactic angles and rotator angular offsets following the method we described in \cite{2015MNRAS.453.2378M} to allow for realigning the cubes to the North. 

To conclude, the software retrieved the position of the star outside the field of view. 
SINFONI lacks an atmospheric dispersion corrector, which makes the registration more complex. We followed a three-step strategy:\begin{itemize}
        \item the position of the star was first evaluated at each wavelength in the last cube corresponding to the 10\,s exposure using a Moffat function. The drift was fitted with a low-order polynomial;
        \item we accounted for the field selector offsets, the field rotation, and the change in atmospheric refraction to measure the star position outside the field of view in the first 100\,s exposure. The theoretical change of the star position due to atmospheric refraction was computed using empirical formulas reported into the source code of the SINFONI data handling pipeline recipes and adapted to the case of the pupil-tracking mode (see the pipeline user manual\footnote{http://www.eso.org/sci/software/pipelines/index.html\#pipelines\_table});  
        \item we iterated over the next 100\,s exposures using our computed change of theoretical atmospheric refraction and the evolution of the parallactic angle.
\end{itemize}
We assumed for that purpose a pivot point located at the center of the field of view (X=32, Y=32; see the ESO manual\footnote{https://www.eso.org/sci/facilities/paranal/decommissioned/sinfoni/ doc/VLT-MAN-ESO-14700-3517\_v101.0.pdf}). The accuracy on the position of that pivot point is however tied to the proper correction of the slitlet edges which we evaluate to be accurate to $\sim$1 pixel. We therefore explored different pivot point locations within a 3 pixel-wide box centered on the theoretical ones, with 0.5 pixel increments and concluded that the theoretical value was offering the strongest cross-correlation signal (see section \ref{Sec:Physical_properties}) on the planet on both night in our data. The strategy was vetted on similar data obtained on $\beta$ Pictoris b (Bonnefoy et al., in prep), and provides a centering with an accuracy better than a spaxel ($\sim$12.5\,mas) along the sequence.

\begin{figure*}
  \centering
  \includegraphics[width=12cm]{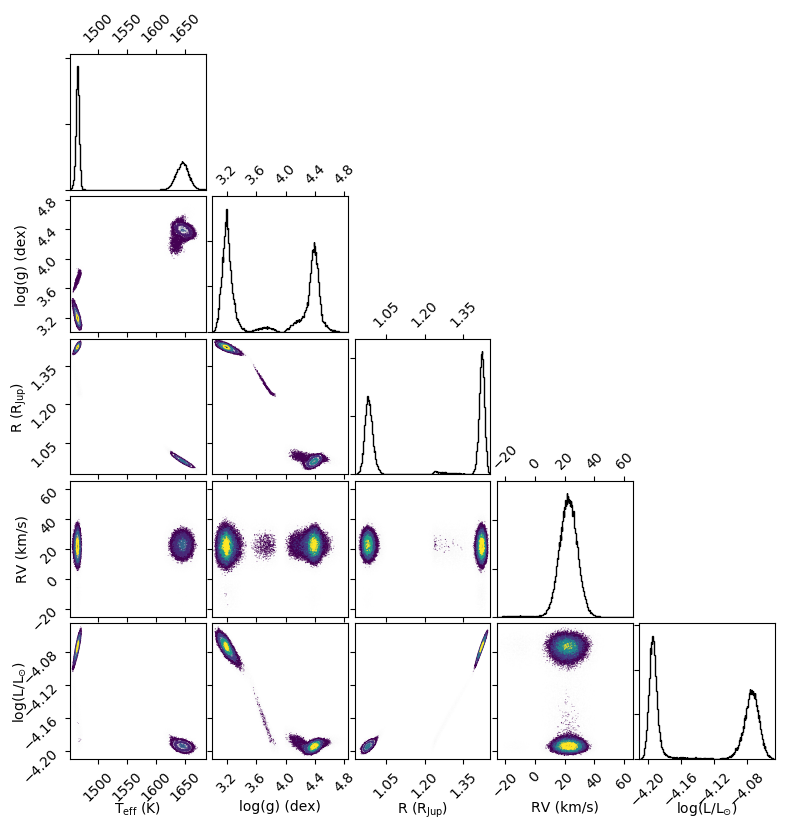}\\
  \includegraphics[width=8cm]{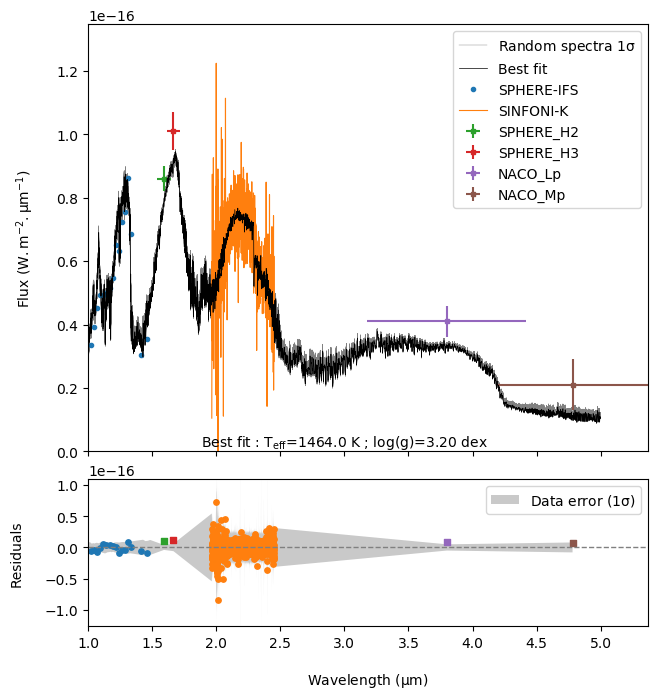}\hspace{2cm}
  \includegraphics[width=8cm]{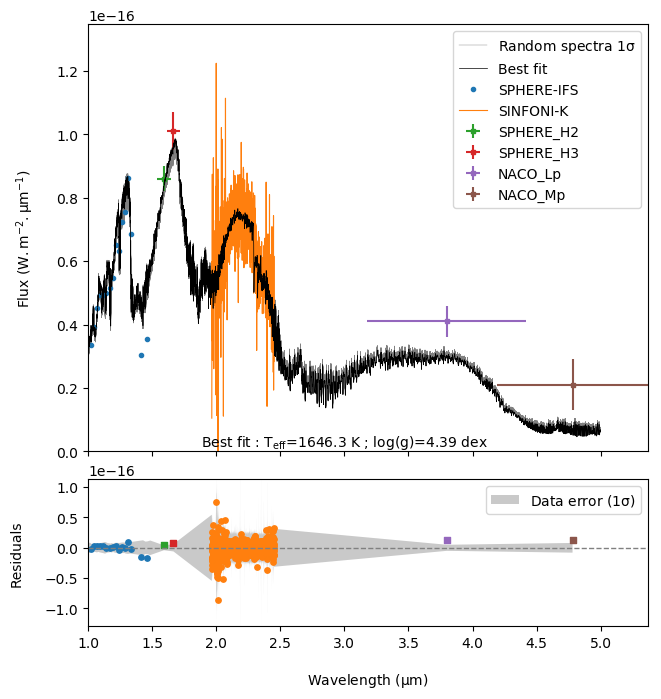}\hspace{2cm}
  \caption{\textit{Top} : Bi-modal posterior distribution inferred from \texttt{ForMoSA} using the \texttt{BT-SETTL15} models. \textit{Bottom left} and \textit{right} : comparison between the best fit and the observed data for the low \Teff\,  and the high \Teff\, modes, respectively. The data points were obtained using the low-resolution SPHERE-IFS spectrum \citep[R$\simeq$30-50; \textit{blue} points,][]{2017A&A...605L...9C}, the SPHERE-IRDIS H-band photometry \citep[\textit{green} and \textit{red} points; ][]{2019A&A...622A..80C}, our medium-resolution SINFONI spectrum (R$\simeq$5500); \textit{orange} points, this work), and the NaCo L' and M'-band photometry  \citep[\textit{purple} and \textit{brown} points, respectively;][]{2019A&A...622A..80C}. We overlaid 50 synthetic spectra picked randomly within the 1-$\sigma$ envelop of each posterior in \textit{grey}.}
  \label{Fig:ForMoSA_btsettl}
\end{figure*}

\subsection{Extraction of emission spectra using the ADI}
We applied the angular differential imaging technique \citep{2006ApJ...641..556M} on each cube slice to evaluate and remove the halo. Both the classic-ADI approach and the PCA\footnote{We used for that purpose the \texttt{numpy-LAPACK} implementation of the PCA implemented as part of the \texttt{VIP\_HCI} package \citep{Gomez_Gonzalez_2017}.} (5 modes) allow to retrieve the planet at each epoch with a shape and size compatible with the PSF (Figure \ref{Fig:detecSINF}). This validates further the star registration process. The post-ADI cubes are affected by residuals from the diffraction pattern created by the telescope spiders (the negative and positive lines near the edges of the field of view). The PCA appears to mitigate some of those residuals. We also used the nADI approach considering a simple re-alignment of the individual cubes to the North and a stack. The nADI reveals the planet on top of the stellar halo. We found the halo could be subtracted on data from May 25, 2018 with the nADI approach removing a circular profile at first and fitting the residuals in a 20 pixel square box around the planet with a 2D polynomial. The box size was chosen to avoid introducing artefacts at the edges of the field of view that could bias the fit while allowing for having a stable (converged) fit of the halo at each wavelength (Appendix \ref{Appendix:extraction}). This method fails to provide an accurate removal of the halo in the data from May 26, 2018 obtained in poorer conditions.
The companion spectrum was extracted within a 75\,mas circular aperture in the PCA and the halo-subtracted nADI cubes. We used the average flux of the residuals in the halo-substracted nADI cubes in an annuli of radius 50-113mas to estimate the error bars on the planet flux.
A spectrum of the star was obtained from the short-exposures datacubes (see Table \ref{Tab:Obs_Log}) within the same aperture and allowed to compute the flux ratio with the planet at each wavelengths. This ratio was multiplied by a BT-NEXTGEN  synthetic spectrum \citep{2012RSPTA.370.2765A} at \Teff=8400K,   log g = 3.5 dex, and M/H=0 dex degraded to the resolution of SINFONI and flux-calibrated on the TYCHO (B, V), 2MASS (J, H, K), DENIS (K), WISE (W1 to W4) and AKARI (S9W) photometry \citep{2000A&A...355L..27H, 2003yCat.2246....0C, 2005yCat.2263....0D, 2012yCat.2311....0C} of the star extracted from the VOSA \citep{2008A&A...492..277B} SED analyser.\footnote{We explored \Teff~ from 8000 to 9000K, and log g from 3.5 to 4.5 dex and determine the best-matching model using a $\chi^{2}$ minimization.} This allowed us to obtain final telluric-free spectra of the planet. The spectra of HIP~65426\,b extracted from the PCA on both nights have an identical shape but are noisier than the spectrum extracted from the nADI-processed datacubes obtained at one epoch (see Appendix \ref{Appendix:extraction}). The later spectrum absolute flux is in excellent agreement with the K1 and K2 photometry reported in \cite{2019A&A...622A..80C}.  We chose to use it for the analysis presented in Section\,\ref{Sec:Physical_properties}.

\begin{figure*}
  \centering
  \includegraphics[width=12cm]{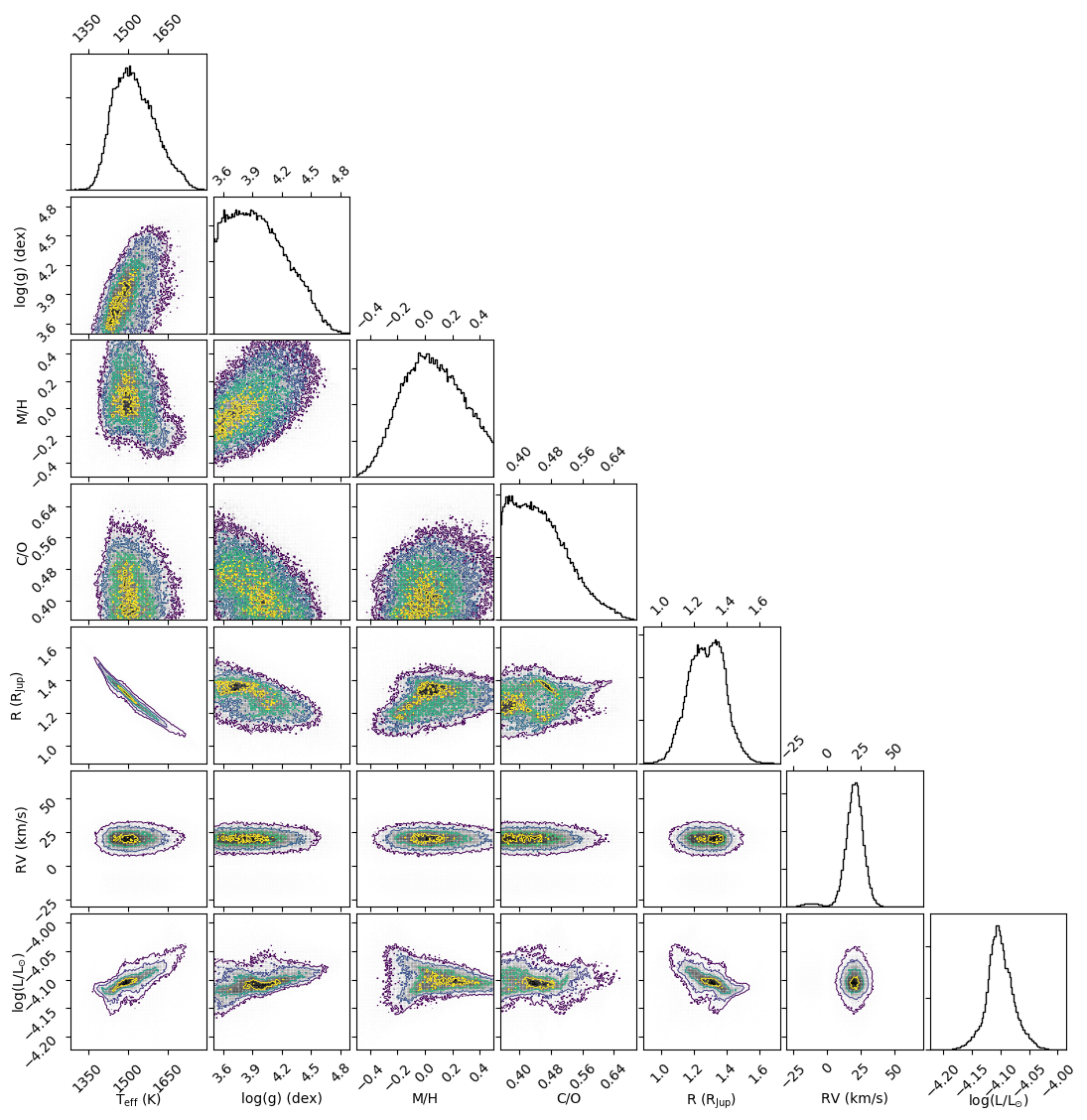} \\
  \includegraphics[width=10cm]{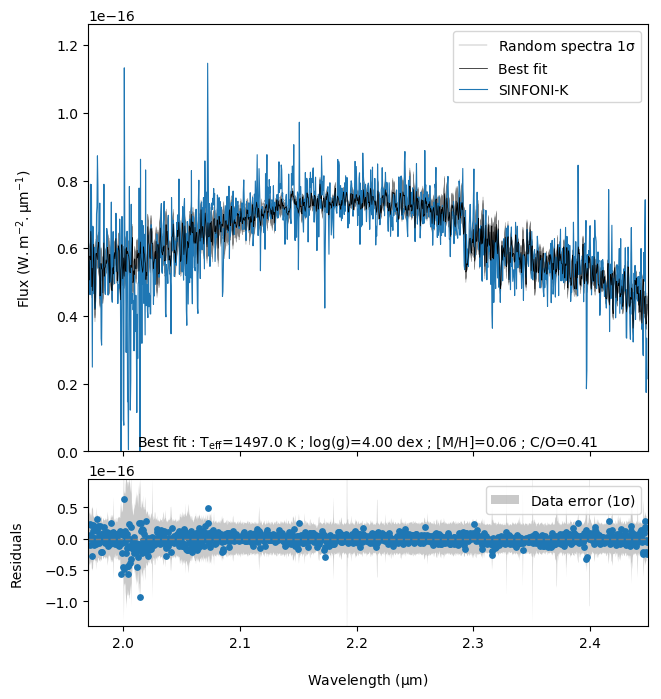}\hspace{2cm}
  \caption{Same as Figure \ref{Fig:ForMoSA_btsettl} but with the grid of \texttt{Exo-REM} spectra and applicable to the K-band SINFONI data.}
  \label{Fig:ForMoSA_exorem}
\end{figure*}

\subsection{Molecular mapping}
\label{subsubsec:molmap}
We implemented in \texttt{TExTRIS} the ''molecular mapping'' technique following the main steps described in \cite{2018A&A...617A.144H}. A reference spectrum of the star was first extracted in each cube from the median-combination of the 1\% most illuminated spaxels in the field-of-view. This reference spectrum contains a representative set of telluric and stellar features (here Brackett $\gamma$), but does not model the changing slope of the spectra of each individual spaxel across the field-of-view because of the evolution of the SINFONI PSF with wavelength. A model of the stellar emission was then created for each spaxel (i,j) computing the ratio between the spaxel (i,j) and the referenced spectrum. The ratio was smoothed with a Gaussian boxcar of width 22\,\AA\,(10 spectral elements) and used for computing a slope-corrected  model of each spaxel (i,j). The model was then subtracted to the spaxels to create a cube of residuals where both the continuum emission of the star and of any companion is removed. This step was repeated onto each datacube for the sequence once.  

We then cross-correlated each spaxel with a template spectrum (see Section \ref{Sec:atmomod}) smoothed to the spectral resolution of SINFONI and for which the continuum had been removed beforehand using a gaussian boxcare of 20\AA\, too. The cross-correlation function (CCF) was performed with the Python module \texttt{crosscorrRV} from the \texttt{PyAstronomy} package\footnote{https://www.hs.uni-hamburg.de/DE/Ins/Per/Czesla/PyA/PyA/index.html} customized to include a normalization of the CCF following \cite{10.1117/12.421129}. The CCF explores velocities from -100 to +100 \kms\, in steps of 8\kms\,($\sim$4 times the original sampling of the data in velocity). The individual cubes of cross-correlation function were then corrected from the barycentric velocity using the Python module \texttt{barycorrpy}\footnote{https://github.com/shbhuk/barycorrpy} \citep{2018ascl.soft08001K, 2018RNAAS...2....4K} and combined once phased in velocity. The planet is re-detected at a separation  $\rho=812\pm18$ mas and PA$=150.2 \pm 1.2^{\circ}$, and  $\rho=796\pm18$ mas and PA$=150.6\pm1.2^{\circ}$ on May 25$\rm^{th}$ and 26$\rm^{th}$, 2018, respectively (the error bars correspond to 1 pixel in the SINFONI field).  The SINFONI astrometry is compatible within $2\sigma$ with the one measured on VLT/SPHERE data from May 12$\rm^{th}$, 2018 ($\rho=822.9\pm2.0$mas, PA$=149.85\pm0.15^{\circ}$) by \cite{2019A&A...622A..80C} while it does not rely on a proper calibration of the spaxel size and absolute orientation of the instrument.  We present an analysis of the correlation signals in Section \ref{subsec:molmapcarac}.





\section{Physical properties}
\label{Sec:Physical_properties}
We adopted two independent methods for deriving the physical properties of HIP~65426\,b's atmosphere. We used a classical forward modelling approach that compares observed spectra with grids of pre-computed synthetic spectra using Bayesian inference methods to estimate posteriors on a set of parameters. We also applied the molecular mapping technique studying the evolution of the cross-correlation signal at the planet location for different molecular templates or grids of atmospheric models used as input. For both methods, we considered grids of  \texttt{BT-SETTL15}\ and  \texttt{Exo-REM}\  \citep{2012RSPTA.370.2765A, 2018ApJ...854..172C} synthetic spectra. The corresponding models are described below.

\subsection{Atmospheric models}
\label{Sec:atmomod}
\subsubsection{\texttt{BT-SETTL15}}
\texttt{BT-SETTL15} includes a 1D cloud model where the abundance and size distribution of  55 types of grain are determined comparing the timescale of the condensation, the coalescence, the gravitational settling and the mixing of these solids supposed spherical. The details of each solids and chemical elements included in the model are described in \cite{2018AeA...620A.180R}. The opacities in the spectral energy distribution (hereafter SED) are  calculated line by line and the overall radiative transfer is carried on by the \texttt{PHOENIX} code  \citep{1997ApJ...483..390H, 2001ApJ...556..357A}.  The convection is handled following the mixing-length theory, and works at hydrostatic and chemical equilibrium. The model also accounts for non-equilibrium chemistry  between CO, CH$_{4}$, CO$_{2}$, N$_{2}$, and NH$_{3}$. The grid we used (CIFIST2011c\footnote{https://phoenix.ens-lyon.fr/Grids/BT-Settl/CIFIST2011c/}) considers a \Teff\,from 300 to 7000\,K in steps of 100\,K, a surface gravity (hereafter log(g)) from 2.5 to 5.5\,dex in steps of 0.5 dex. For computing efficiency, we used models  with \Teff\ in the range of 1200-2000\,K which is a conservative prior considering the previous spectral characterization of HIP~65426\,b. The grid does not allow for an exploration of non-solar metallicities in that temperature range.

\subsubsection{\texttt{Exo-REM}}

\texttt{Exo-REM} is an atmospheric radiative-convective equilibrium model including in addition a cloud description well suited for reproducing the spectra of brown dwarfs and exoplanets where dust dominates, especially at the L-T transition. Much like the \texttt{BT-SETTL15}\, models, the atmosphere is cut into 64 pressure levels. The flux over these layers is calculated iteratively using a constrained linear inversion method, for a radiative-convective equilibrium. 
The initial abundances of each chemical element are first established using the values tabulated in \cite{2010ASSP...16..379L}. The model includes the collision-induced absorptions of H$_{2}$–H$_{2}$ and H$_{2}$–He, ro-vibrational bands from nine molecules (H$_{2}$O, CH$_{4}$, CO, CO$_{2}$, NH$_{3}$, PH$_{3}$, TiO, VO, and FeH), and resonant lines from Na and K. As \texttt{BT-SETTL15}, \texttt{Exo-REM} accounts for non-equilibrium chemistry between CO, CH$_{4}$, CO$_{2}$, and NH$_{3}$ due to vertical mixing. The abundances of the other species are computed at thermochemical equilibrium. The vertical mixing is parametrized by an eddy mixing coefficient from cloud-free simulations. The cloud model includes the formation of iron, silicate, Na$_{2}$S, KCl, and water clouds. This grid was generated exclusively for this study and provides spectra from 1.887 \mic\, to 2.500 \mic\, and includes four free parameters: The \Teff\, from 300 to 1800\,K in steps of 50\,K, the log(g) from 3.5 to 5.0\,dex in steps of 0.5 dex, the [M/H] for -0.5 to 0.5 in steps of 0.5 dex and the [C/O] from 0.35 to 0.8, in steps of 0.05. We notice that synthetic spectra are available at higher \Teff~but we have chosen to not include them due to the non convergence of these models beyond \Teff=1800\,K.

\subsection{Bayesian inference of the spectro-photometry}
\label{Sec:Bayse}

\begin{table*}[!h]
 \centering
 \begin{threeparttable}
\renewcommand{\arraystretch}{1.3}
  \caption{Estimation of the atmospheric parameters of HIP~65426\,b for each grid of synthetic pre-computed spectra and for each method. The errors are statistical and result from the Bayesian inversion. In the case of posteriors located at the edge of the grid, an upper/lower limit is defined for the corresponding parameter.}
\tiny   
  \begin{tabular}{c|cc|ccc}
  \hline   
  \hline
\multicolumn{6}{c}{\textbf{\texttt{BT-SETTL15}}} \\
\hline
    & \Teff         & log(g)    &       R       &    RV     &   \Lum   \\
    & (K)           & (dex)     &      (\RJup)  &   (\kms)  &     \\
\hline
\hline
\multicolumn{6}{c}{\texttt{ForMoSA}} \\
\hline
1.0 - 4.7 \mic\,Low \Teff\,with continuum & 1464$_{-3}^{+3}$  &  3.20$_{-0.06}^{+0.09}$ &  1.42$_{-0.01}^{+0.01}$ & 21.8$_{-6.2}^{+6.3}$  &   -4.07$_{-0.01}^{+0.01}$ \\
1.0 - 4.7 \mic\,High \Teff\,with continuum & 1645$_{-11}^{+10}$  &  4.38$_{-0.11}^{+0.07}$ &  0.98$_{-0.01}^{+0.02}$ & 22.8$_{-5.6}^{+5.5}$  &   -4.19$_{-0.01}^{+0.01}$ \\
\hline
K band Low \Teff\,with continuum & 1461$_{-6}^{+6}$  & $\leq$ 3.25 & 1.43$_{-0.02}^{+0.02}$  & 21.7$_{-6.2}^{+6.1}$  & -4.09$_{-0.02}^{+0.02}$ \\
K band High \Teff\, with continuum & 1610$_{-37}^{+20}$  & 4.19$_{-0.13}^{+0.14}$   & 1.03$_{-0.04}^{+0.05}$  & 22.3$_{-6.1}^{+6.0}$  & -4.19$_{-0.01}^{+0.01}$ \\
\hline
K band without the continuum & 1633$_{-260}^{+234}$  & $\leq$ 4.00   & -  & 18.7$_{-16.8}^{+14.0}$  & - \\
\hline
\hline
\multicolumn{6}{c}{\texttt{TExTRIS}} \\
\hline
K band Low \Teff\,without the continuum & 1400  & 3.0    & -  & 20$_{-8}^{+8}$  & - \\
K band High \Teff\,without the continuum & 1700  & 3.5    & -  & 20$_{-8}^{+8}$  & - \\
\hline
\hline
\multicolumn{6}{c}{$\chi^{2}$ minimization} \\
\hline
K band Low \Teff\,without the continuum & 1400  & 3.5    & -  &  - & - \\
K band High \Teff\,without the continuum & 1800  & 3.5   & -  &  - & - \\
\hline
\hline
 \end{tabular}

\label{Tab:ParamSETTL}
 
~\\

  \begin{tabular}{c|cccc|ccc}
  \hline   
  \hline
\multicolumn{8}{c}{\textbf{\texttt{Exo-REM}}} \\
  \hline
    & \Teff         & log(g)   & [M/H] & C/O     & R         &    RV     &   \Lum  \\
    & (K)           & (dex)  &     &       & (\RJup)   &   (\kms)  &    \\
\hline
\multicolumn{8}{c}{\texttt{ForMoSA}} \\
\hline
K band with continuum & 1518$_{-71}^{+88}$  &  $\leq$ 4.20 & 0.05$_{-0.22}^{+0.24}$ & $\leq$ 0.50  &  1.28$_{-0.11}^{+0.10}$ & 20.2$_{-6.4}^{+6.0}$  &  -4.10$_{-0.2}^{+0.2}$ \\
\hline
K band without the continuum  & 1639$_{-187}^{+141}$  & $\leq$ 4.5  & 0.02$_{-0.35}^{+0.35}$ & $\leq$ 0.55  & -  & 17.0$_{-15.0}^{+10.8}$  & - \\
\hline
\multicolumn{8}{c}{\texttt{TExTRIS}} \\
\hline
K band without the continuum  & 1700  & 3.5  & 0.0  & 0.40   & -  & 20$_{-8}^{+8}$  &  - \\
\hline
\multicolumn{8}{c}{$\chi^{2}$ minimization} \\
\hline
K band without the continuum  & 1500  & 3.5  & 0.0  &  0.40  & -  & -  & - \\
\hline
\hline
 \end{tabular}
\label{Tab:param_ForMoSA}
\end{threeparttable}

\end{table*}

We used the  \texttt{ForMoSA} code presented in \cite{2020A&A...633A.124P} to compare the synthetic spectra to the data following the forward-modelling approach. \texttt{ForMoSA} relies on the Nested Sampling algorithm \citep{skilling2006} to determine the posterior distribution function of a set of free parameters in the models. This  method performs a global exploration of the model parameter space to look for local maxima of likelihoods following an iterative method which isolates progressively restrained area of iso-likelihood while converging toward the maximum values. That avoids missing local maxima of likelihood and allows to evaluate the Bayesian evidence that can be used for performing model selection \citep{2008ConPh..49...71T}.

Because the Bayesian inference implies to generate random points with continuous distribution inside the parameter space, we chose to reduce the step of each model grids by interpolating them with the N-dimensional linear interpolation python package \texttt{griddata} which is triangulating the input data with \texttt{Qhull}\footnote{http://www.qhull.org/} before applying barycentric interpolation. The interpolated steps of the \texttt{BT-SETTL15} grids are 10\,K for the \Teff\, and 0.1 dex for the log(g) while  the interpolated steps of the \texttt{Exo-REM} grids are 10\,K for the \Teff, 0.1 dex for the log(g), 0.1 for the metallicity and 0.05 for the C/O ratio. It increases the accuracy of the second interpolation step which occurs over the course of the nested sampling process when a new random  point is generated in the parameter space. That interpolation is based on N-dimensional weighted mean  of the closest neighboring spectra inside the interpolated grid. The physical parameters derived by \texttt{ForMoSA} are summarized in Table \ref{Tab:param_ForMoSA}. We stress that the errors given by \texttt{ForMoSA} are statistical and have been determined for each parameter as the range which encompasses 68\% of the solutions in each posterior, they do not include possible systematical errors in models that are difficult to quantify (see \citealt{2020A&A...633A.124P}). Whenever the posterior distribution do not show a maximum (e.g. C/O ratio), we defined an upper limit on the given parameter encompassing 68\% of the  posterior distribution starting from the posterior maximum. Extended grids would be needed to constrain these parameters, but such analysis is left for future work. The best fits to the data, obtained by using 20 000 living points in the nested sampling algorithm, are presented in Figures \ref{Fig:ForMoSA_btsettl} and \ref{Fig:ForMoSA_exorem} for both families of atmospheric models.

When the 1.0-4.7\,$\mic$ spectral energy distribution is compared to the \texttt{BT-SETTL15} models (Figure \ref{Fig:ForMoSA_btsettl}), two distinct groups of solutions emerge: at ``low-\Teff''  (\Teff=1464$\pm$3\,K) and at ``high-\Teff''  (\Teff=1645$\pm$10\,K). Both sets of solutions were explored within separate runs using flat priors between 1200-1500\,K and 1500-2000\,K, respectively. The results are shown in Table \ref{Fig:ForMoSA_btsettl} together with the data and the distribution of 50 synthetic spectra corresponding to draws within the 1-$\sigma$ interval of each posterior. The residuals between the observations and the best solutions are also shown, and compared with the data error bars (\textit{grey area}). 
 
 The L' and M' photometry is up to 1.5$\sigma$ above the predictions. This may be due to the lack of exploration of non-solar atmospheric  composition by the \texttt{BT-SETTL} models, to an imperfect handling of the non-equilibrium chemistry which tend to enhance the L' and M' band fluxes in young  L-T transition objects \citep{2007ApJ...669.1248H, 2014ApJ...792...17S, 2016ApJ...829...66M, 2016ApJ...829...39S} or cloud properties \citep[e.g., ][]{2014ApJ...792...17S},  or even to excess emission from a putative circumstellar disk \citep{2019ApJ...877L..33C}. Similar deviations beyond 2.5 $\mu$m are also seen in the spectral-energy-distribution fit of mid and late-L type brown dwarfs from Upper-Scorpius \citep{2018MNRAS.473.2020L} using the same models. We also note that the H-band fluxes (SPHERE) are not reproduced well in the case of the "low-\Teff" solution while the L' and M' band fluxes and the water-band absorption from 1.35 to 1.45 $\mu$m are better represented. Previous version of the \texttt{BT-SETTL} models already failed representing the shape of the H band  of young L-dwarfs sampled by the SPHERE H-band filters  \citep{2014A&A...564A..55M}. Removing these points do not remove the bi-modal posterior distribution. 

Conversely, one unique set of solutions at intermediate \Teff\,is found with the \texttt{Exo-REM} models (Figure \ref{Fig:ForMoSA_exorem}) which allow for an exploration of non-solar abundances. Conversely, both models tend to predict close mono-modal solutions when the continuum is removed that are in better agreement with the ones found by \cite{2019A&A...622A..80C} made without the SINFONI data albeit with much larger uncertainties.

A consistent radial velocity of 22$\pm$6 \kms\~and 20$\pm$6 \kms\ is found for HIP~65426\,b using the \texttt{BT-SETTL15} and \texttt{Exo-REM} models at K-band, respectively.

We used the \Teff\ infered from \texttt{ForMoSA} and the DUSTY evolutionary model  \citep{2000ApJ...542..464C}  to derive a planet mass between 8.1 and 10.0\,\MJup\, consistent with the values from \cite{2019A&A...624A..20M}, and a radius between 1.48 and 1.53\,RJup\ ~. The radii obtained with \texttt{ForMoSA} and the \texttt{BT-SETTL} models for the "cold-\Teff"\  ~solution are in better agreement with the tracks while those obtained for the "high-\Teff\"~ prediction are clearly at odd with the expected evolution of a young Jupiter such as HIP~65426b. The \texttt{Exo-REM} models also tend to under-predict the radii, but at reduced scale. These differences are well documented already for young L-type planets such as HIP~65426b \cite[e.g.,][]{2012ApJ...754..135M, 2016ApJ...824..121D} and could be explained by the imperfect description of the cloud physics and opacity.





\begin{figure*}[h]
  \centering
  \includegraphics[width=5.3cm]{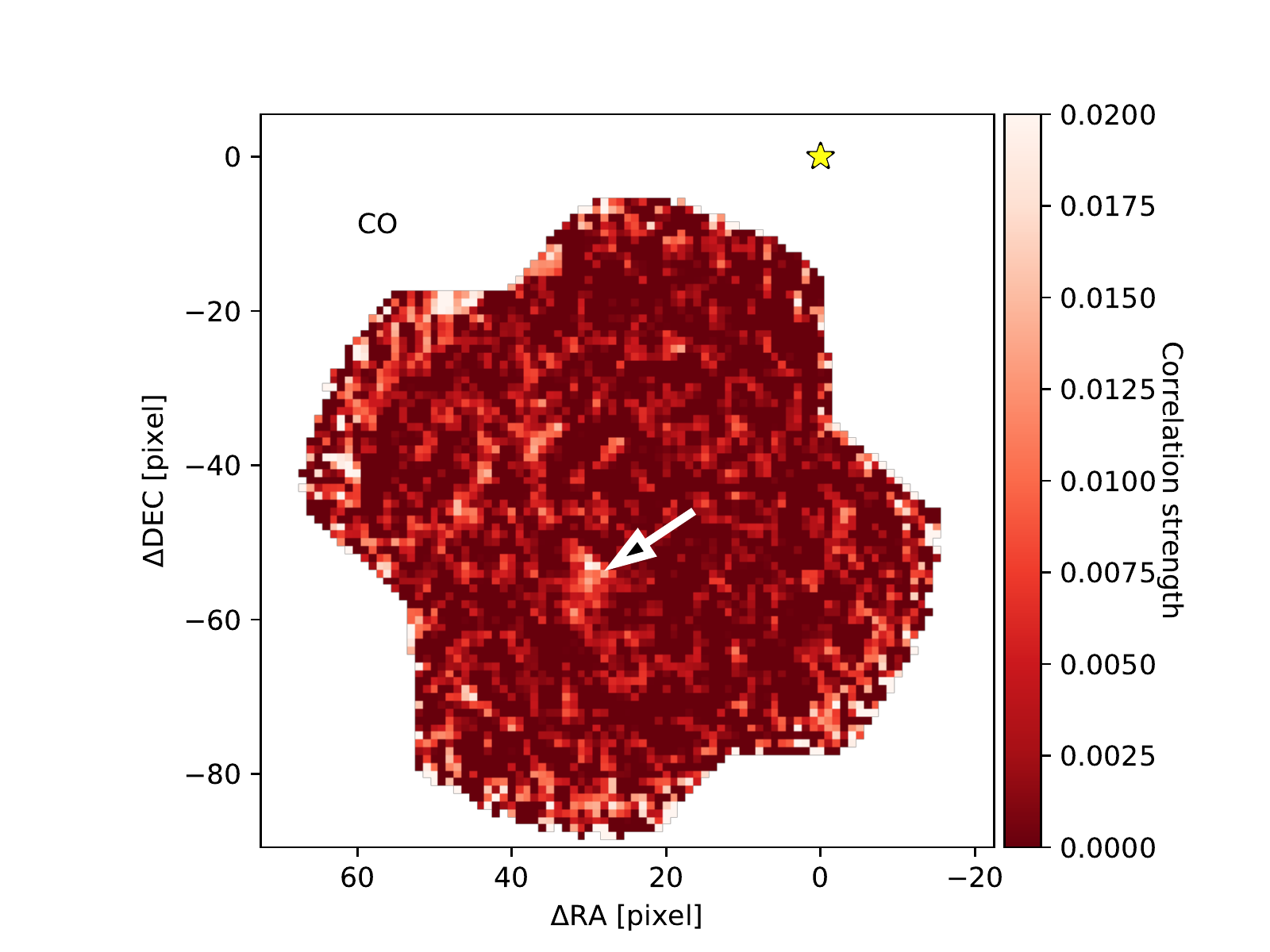}  
  \includegraphics[width=5.3cm]{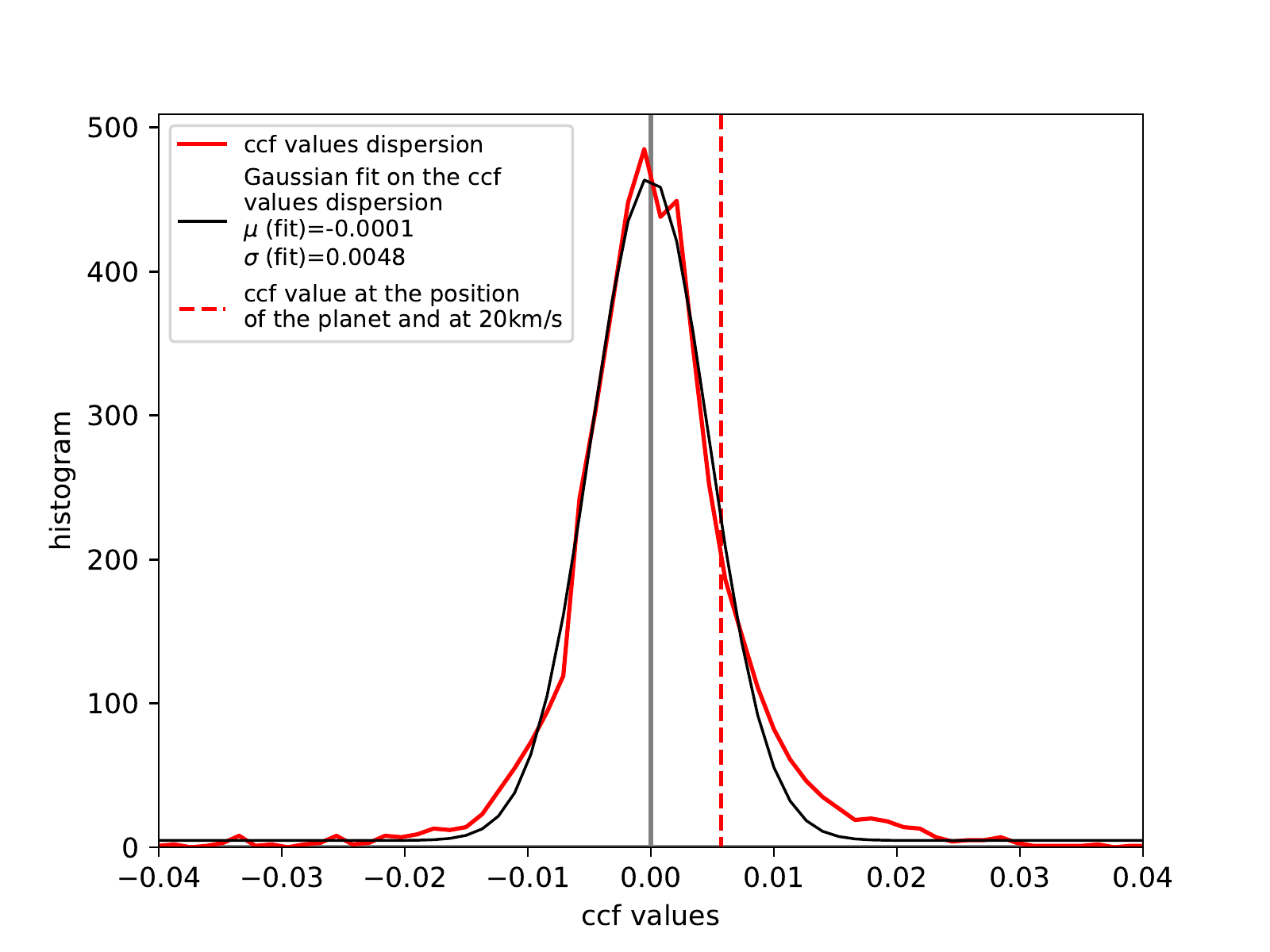}  
  \includegraphics[width=5.3cm]{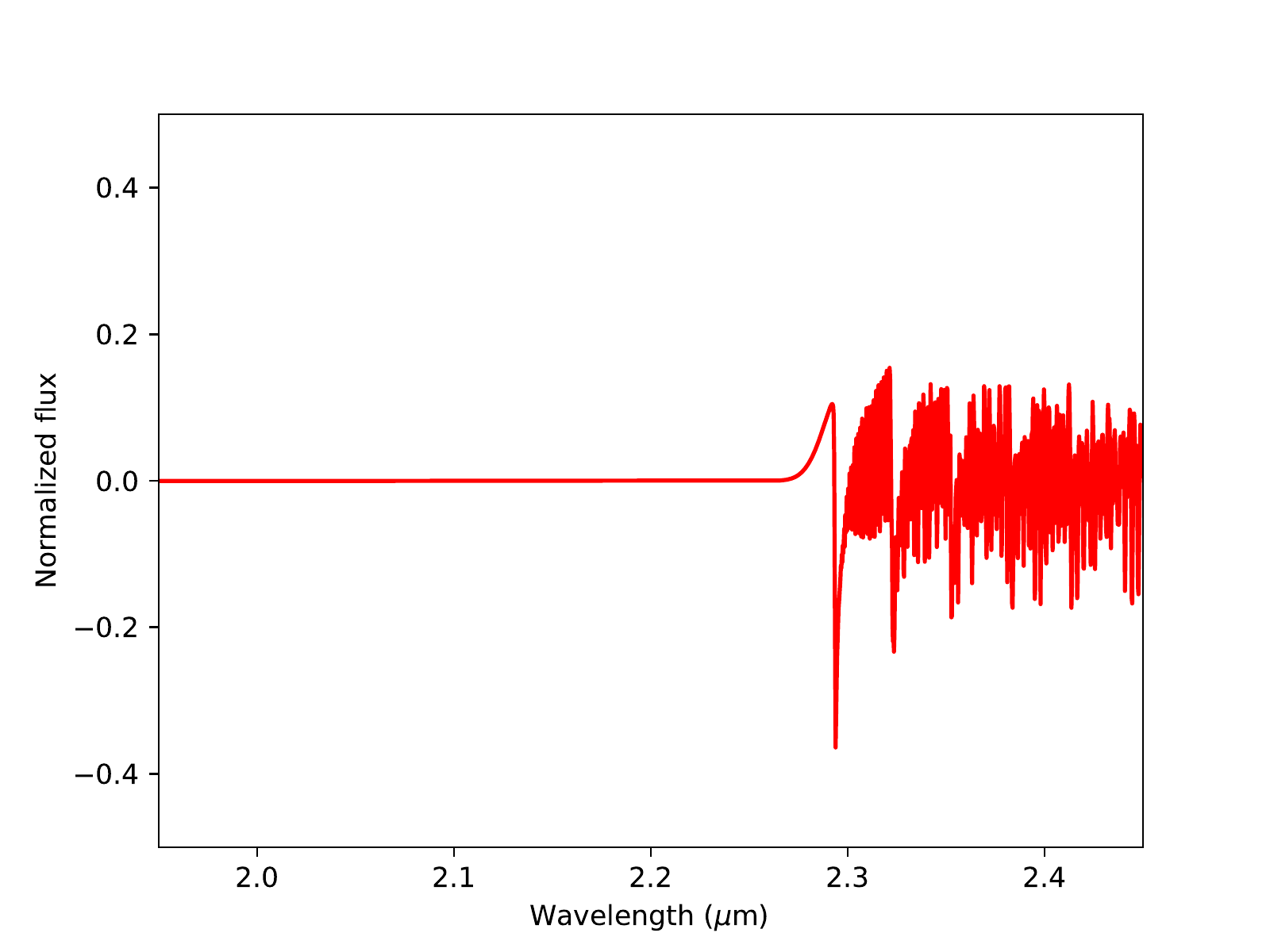}  \\
  \includegraphics[width=5.3cm]{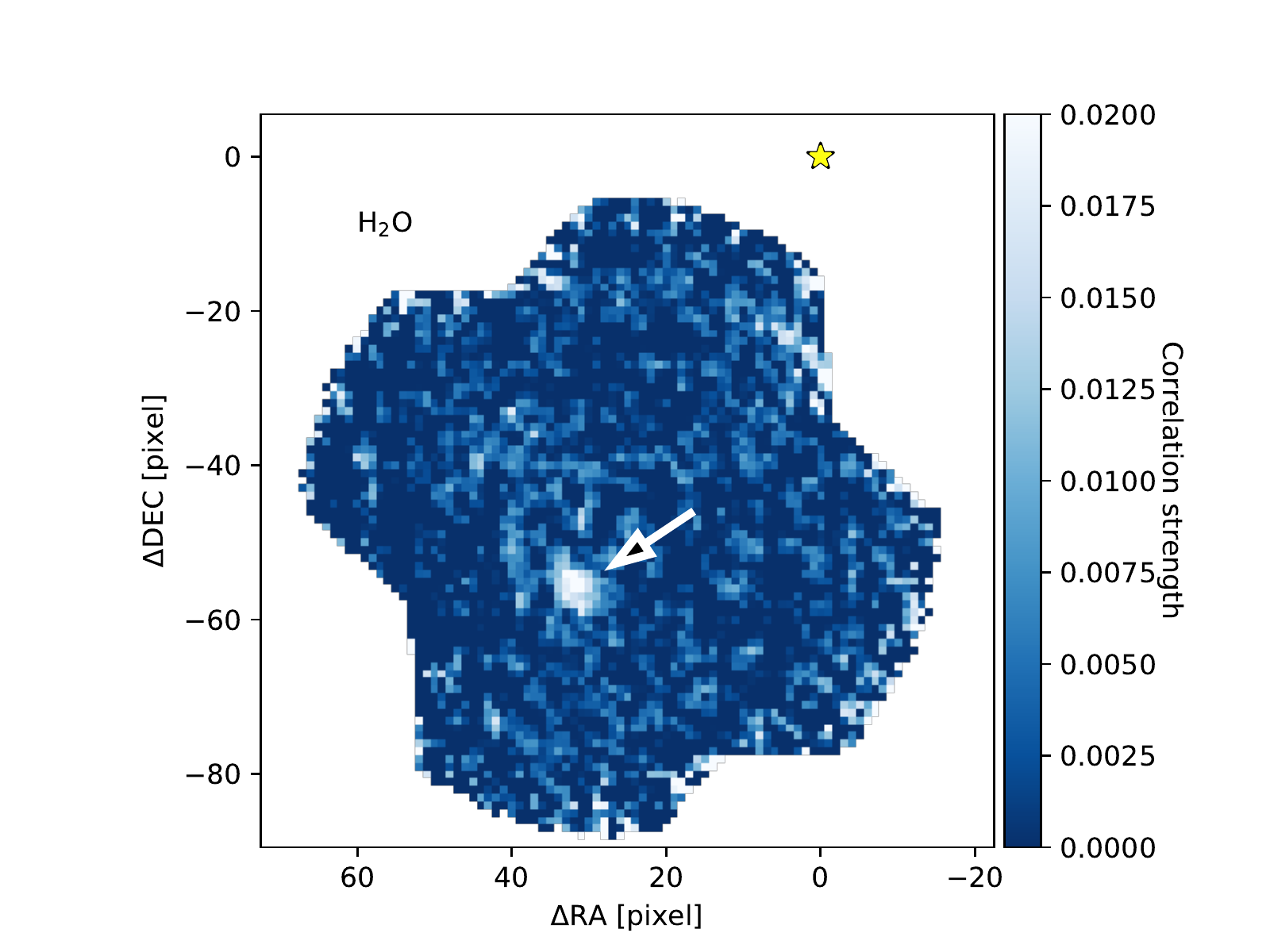}  
  \includegraphics[width=5.3cm]{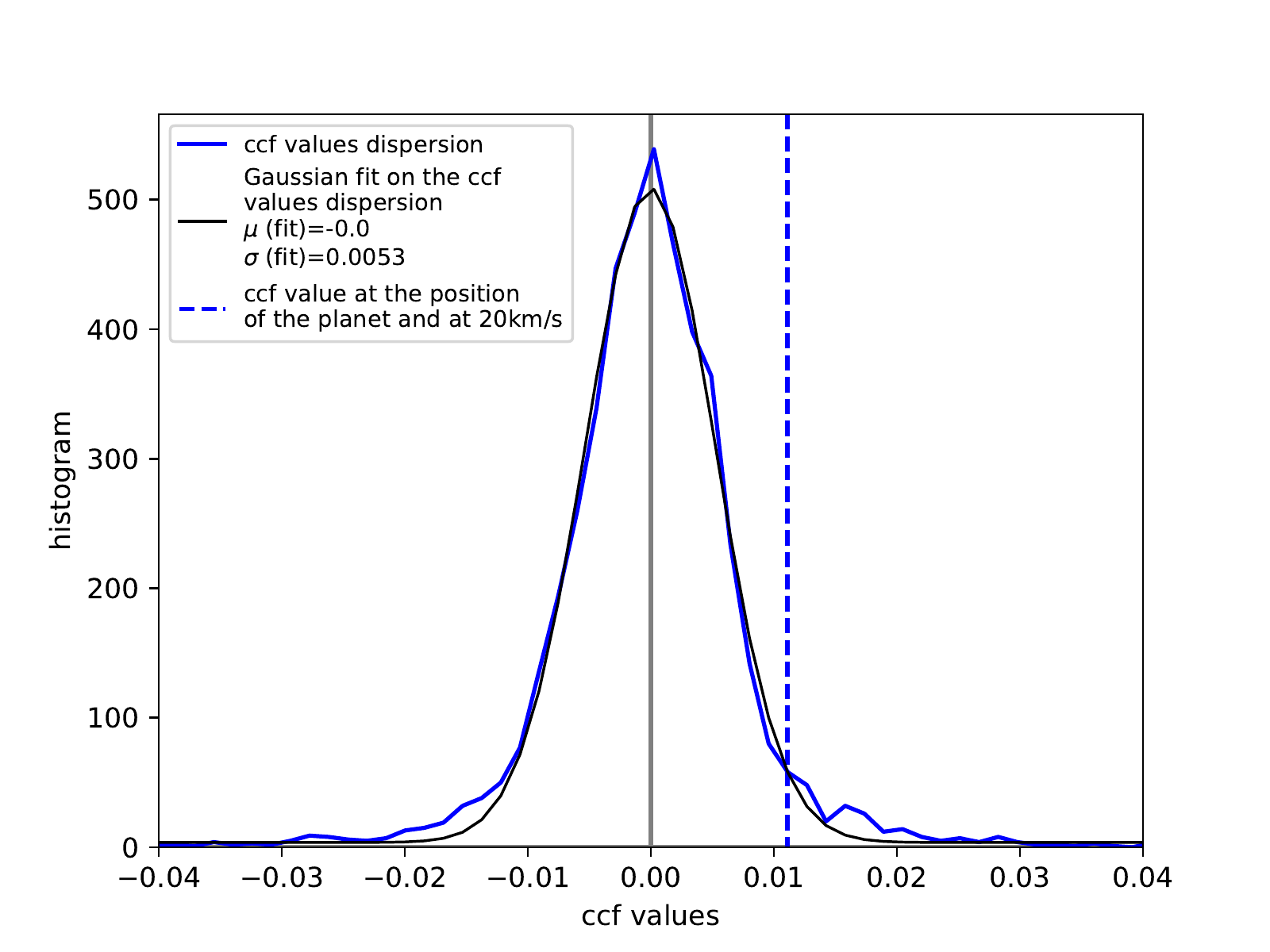}  
  \includegraphics[width=5.3cm]{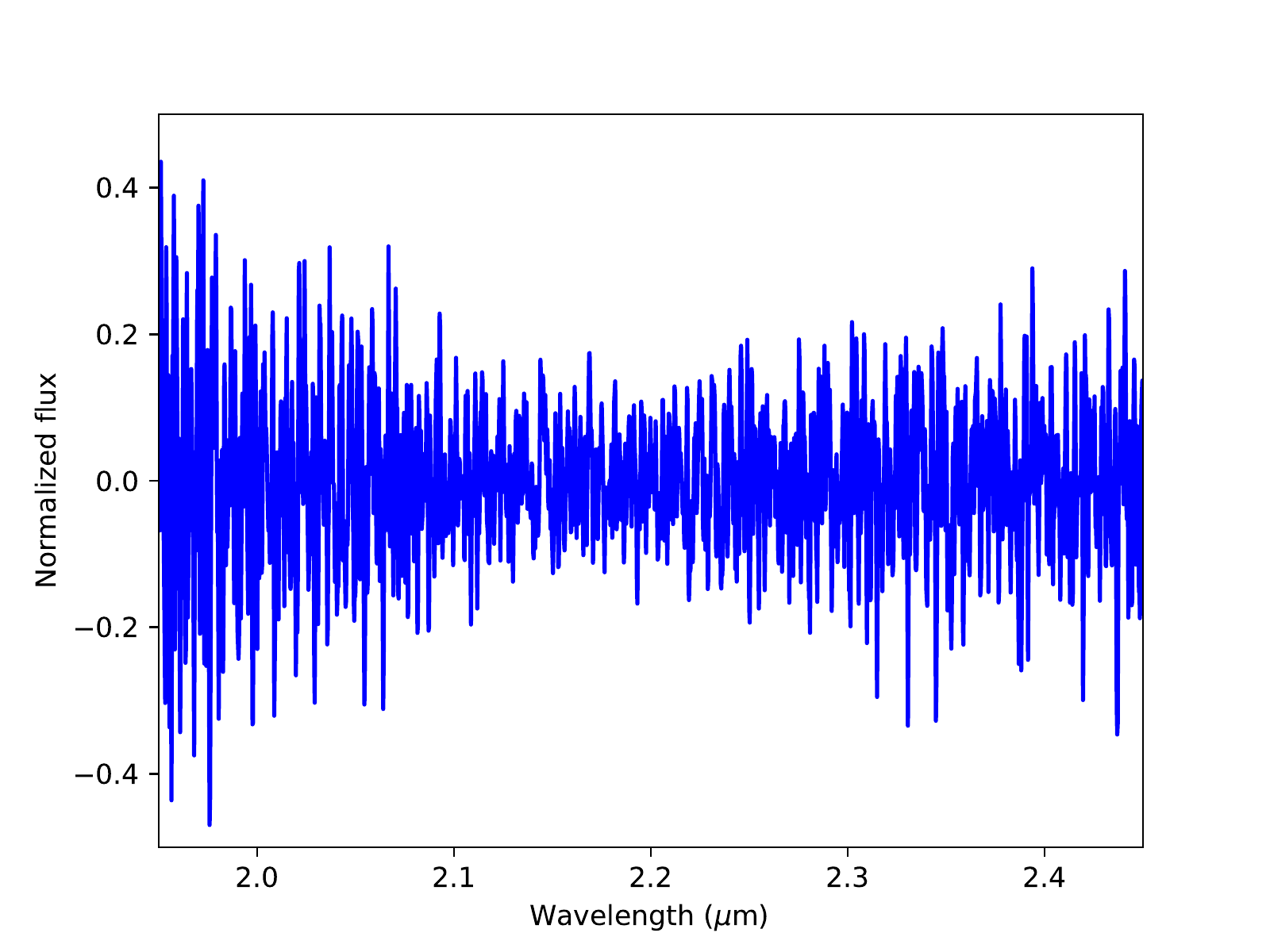} \\
  \includegraphics[width=5.3cm]{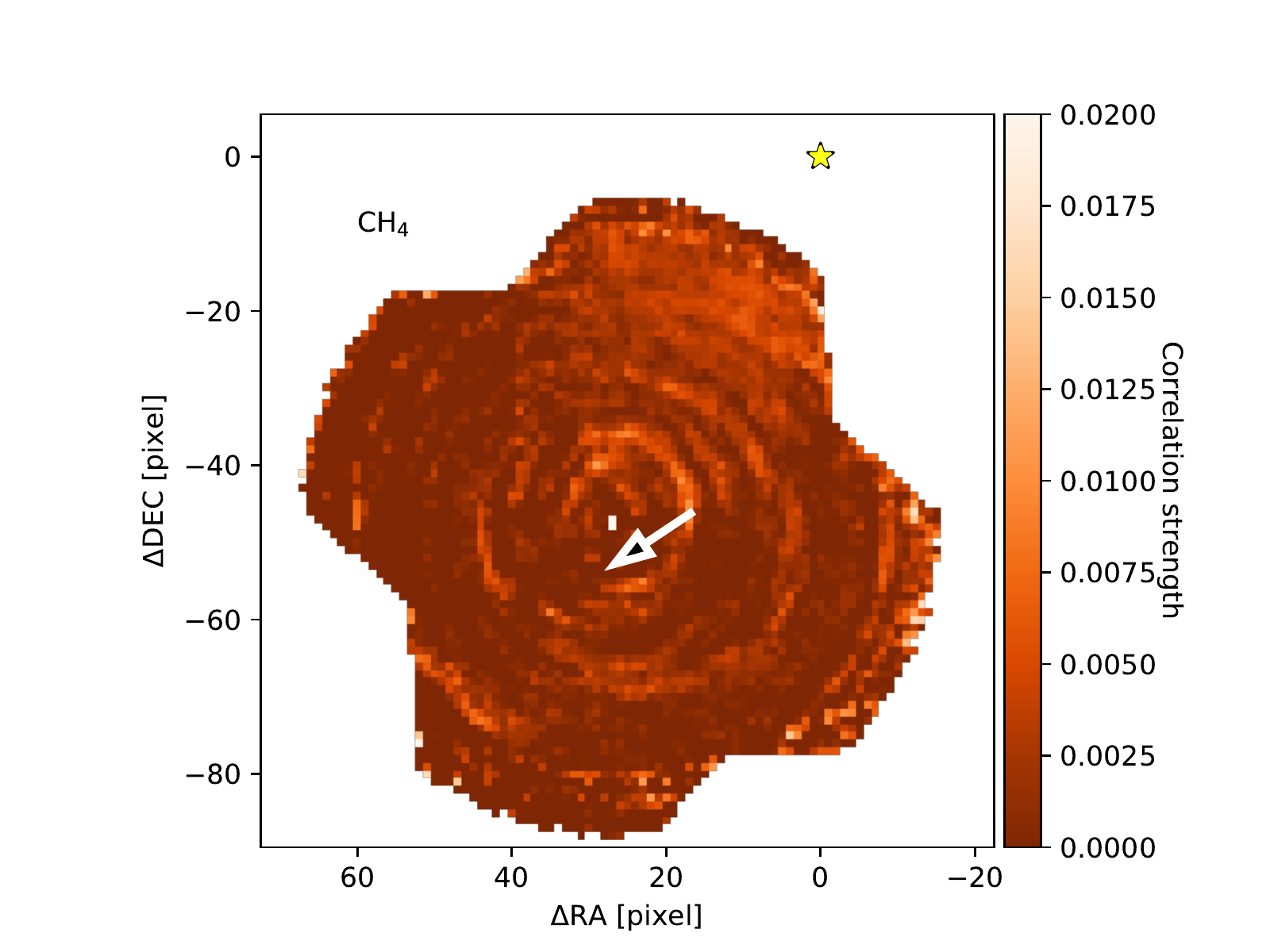}  
  \includegraphics[width=5.3cm]{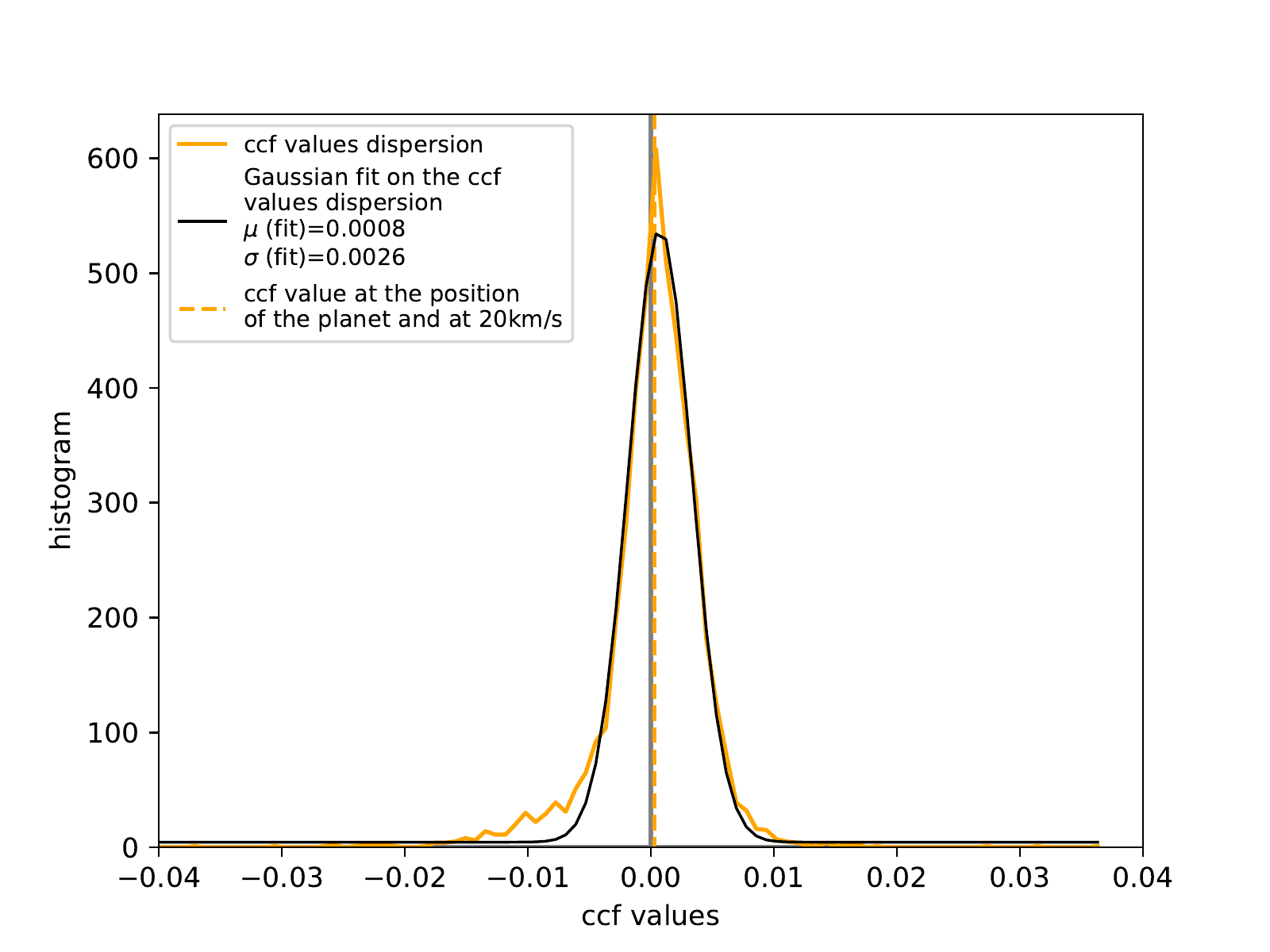} 
  \includegraphics[width=5.3cm]{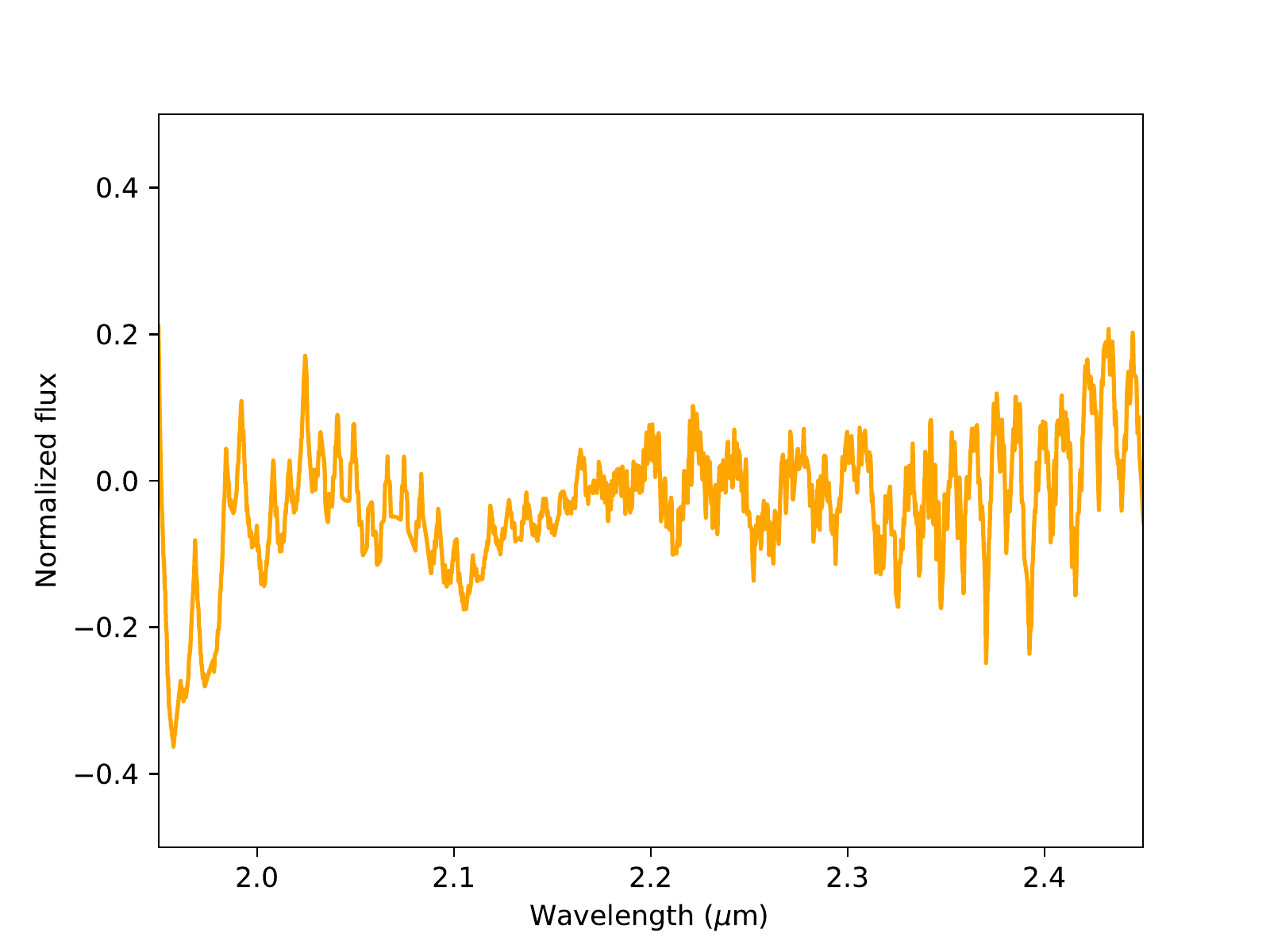}  \\
  \includegraphics[width=5.3cm]{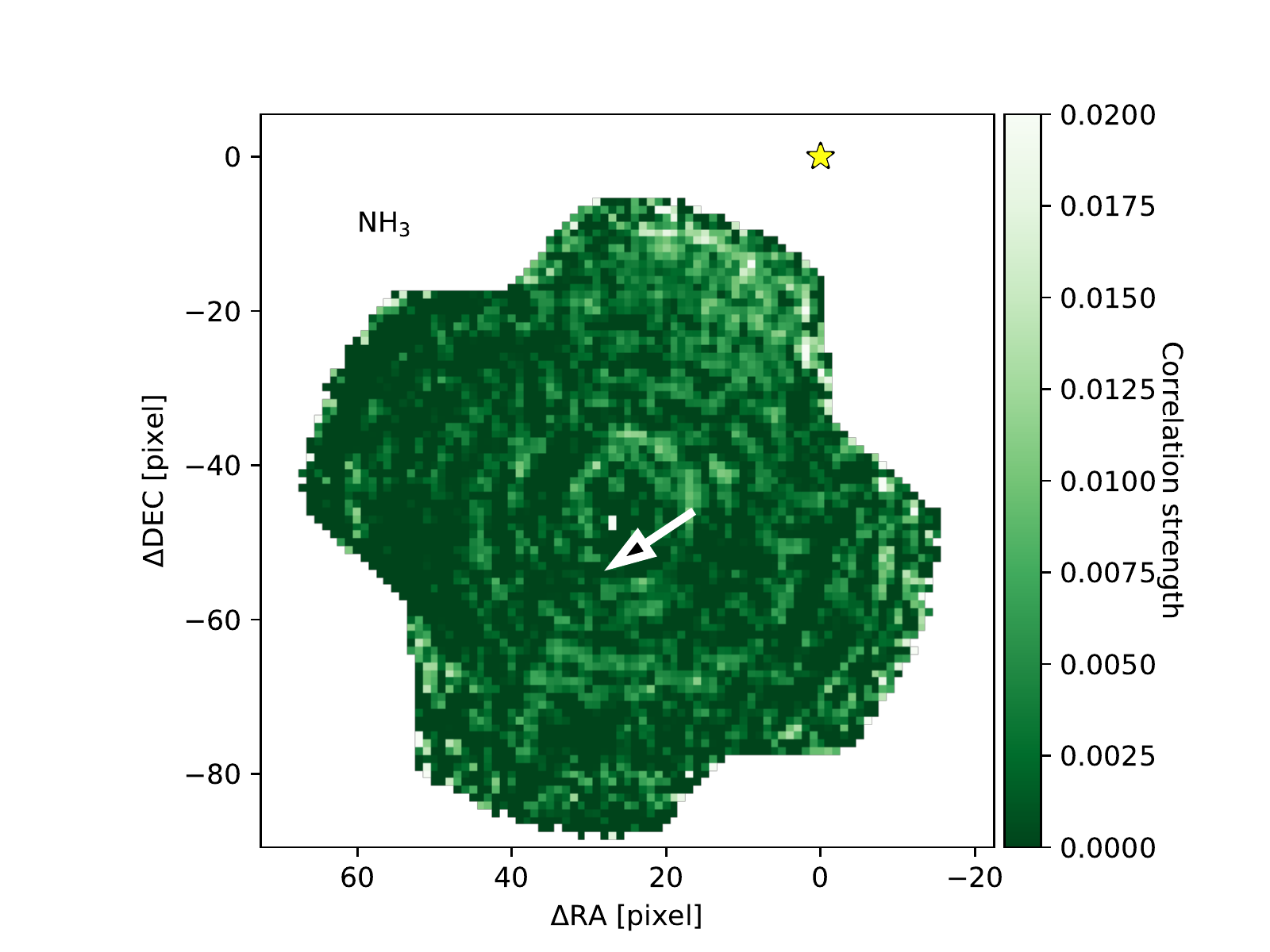}  
  \includegraphics[width=5.3cm]{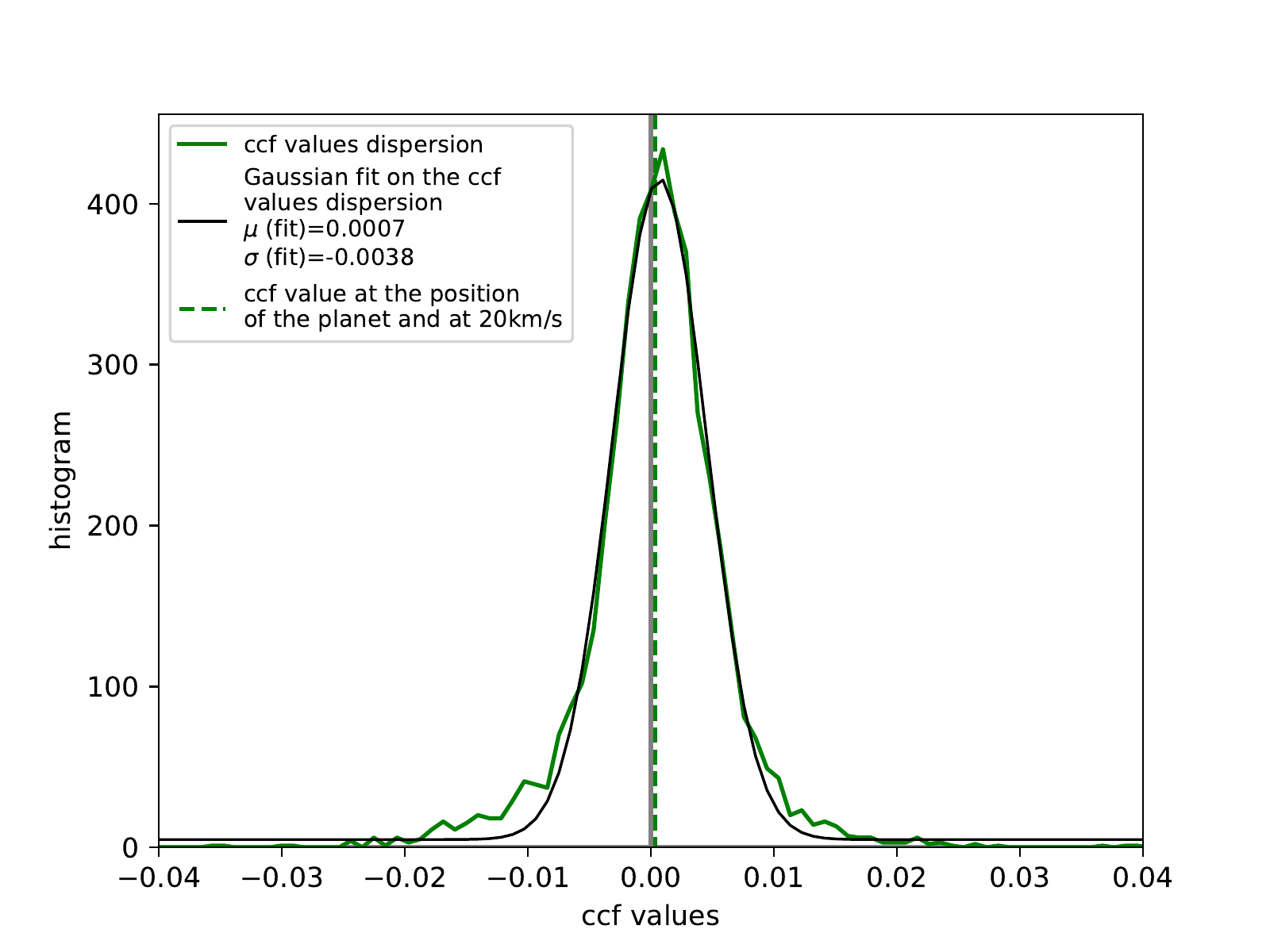}  
  \includegraphics[width=5.3cm]{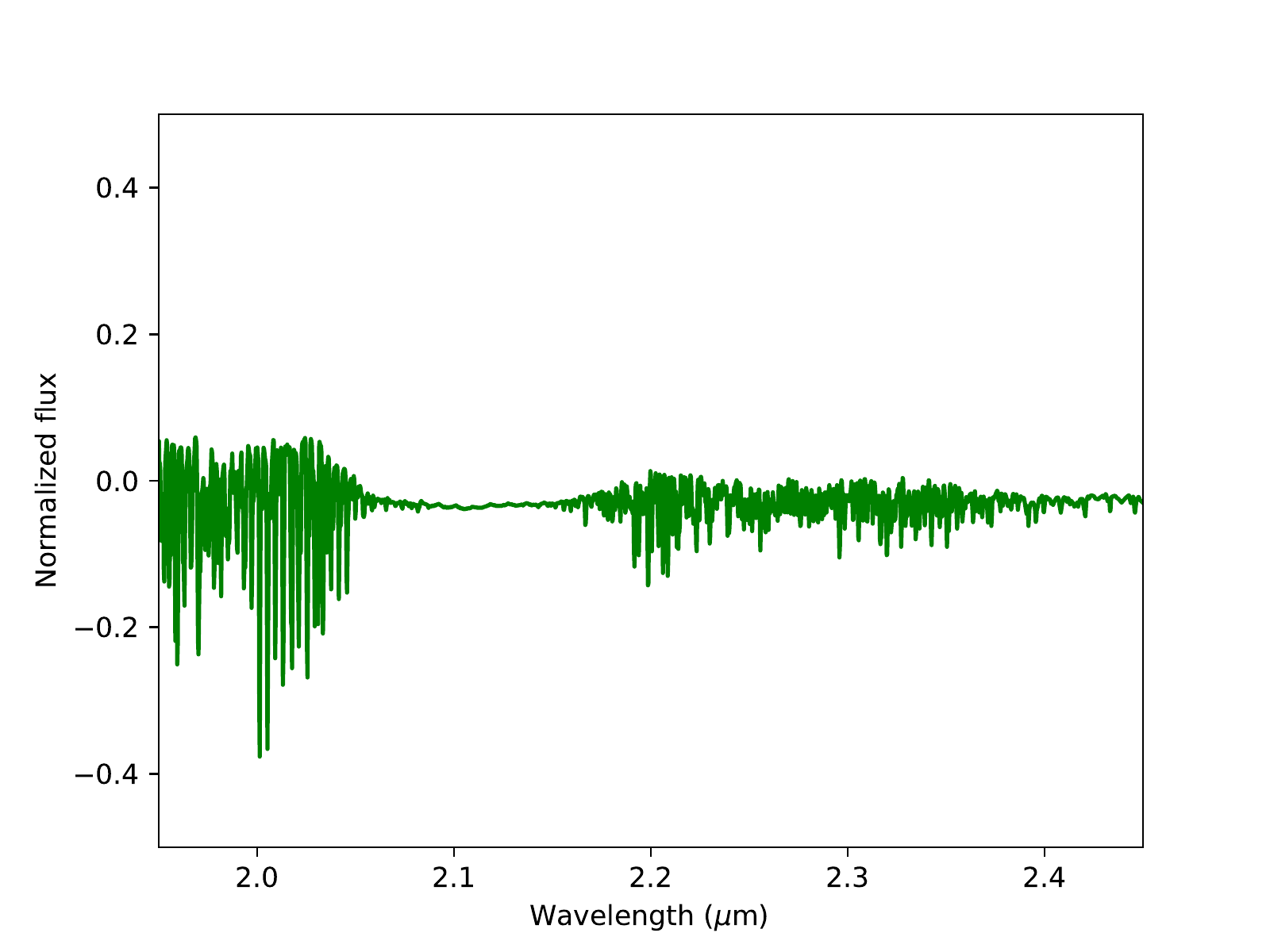}  \\
  \includegraphics[width=5.3cm]{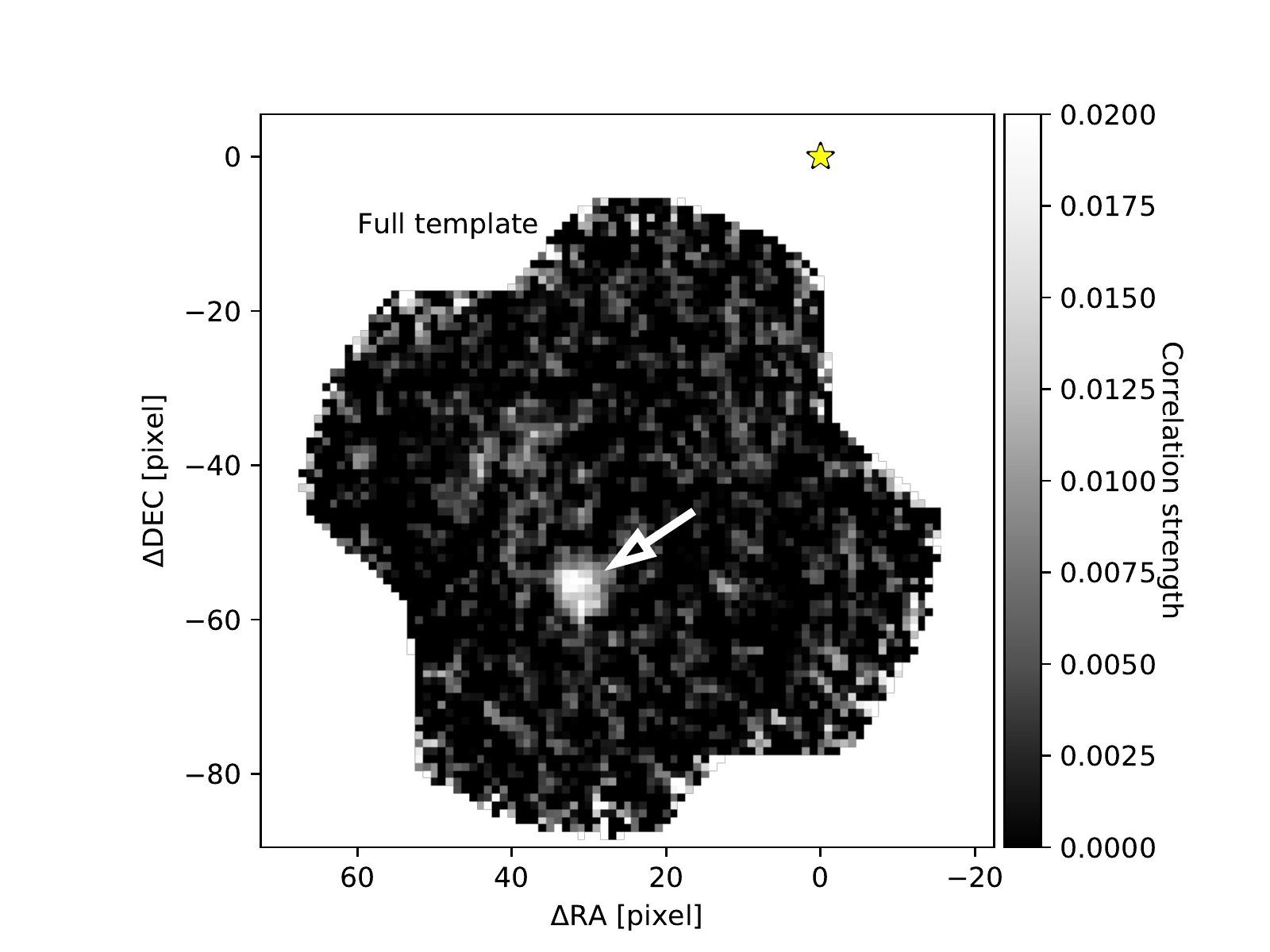}  
  \includegraphics[width=5.3cm]{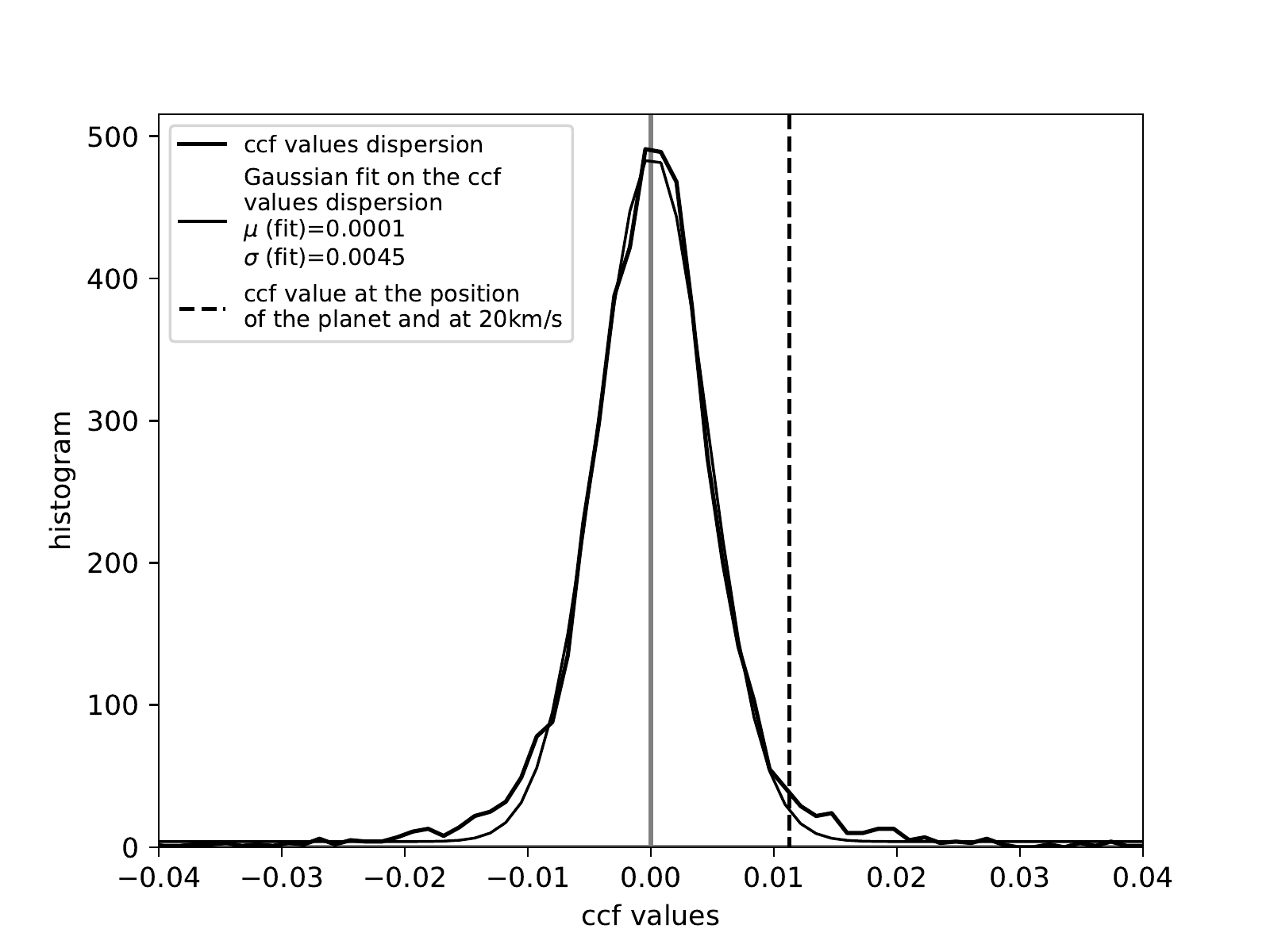} 
  \includegraphics[width=5.3cm]{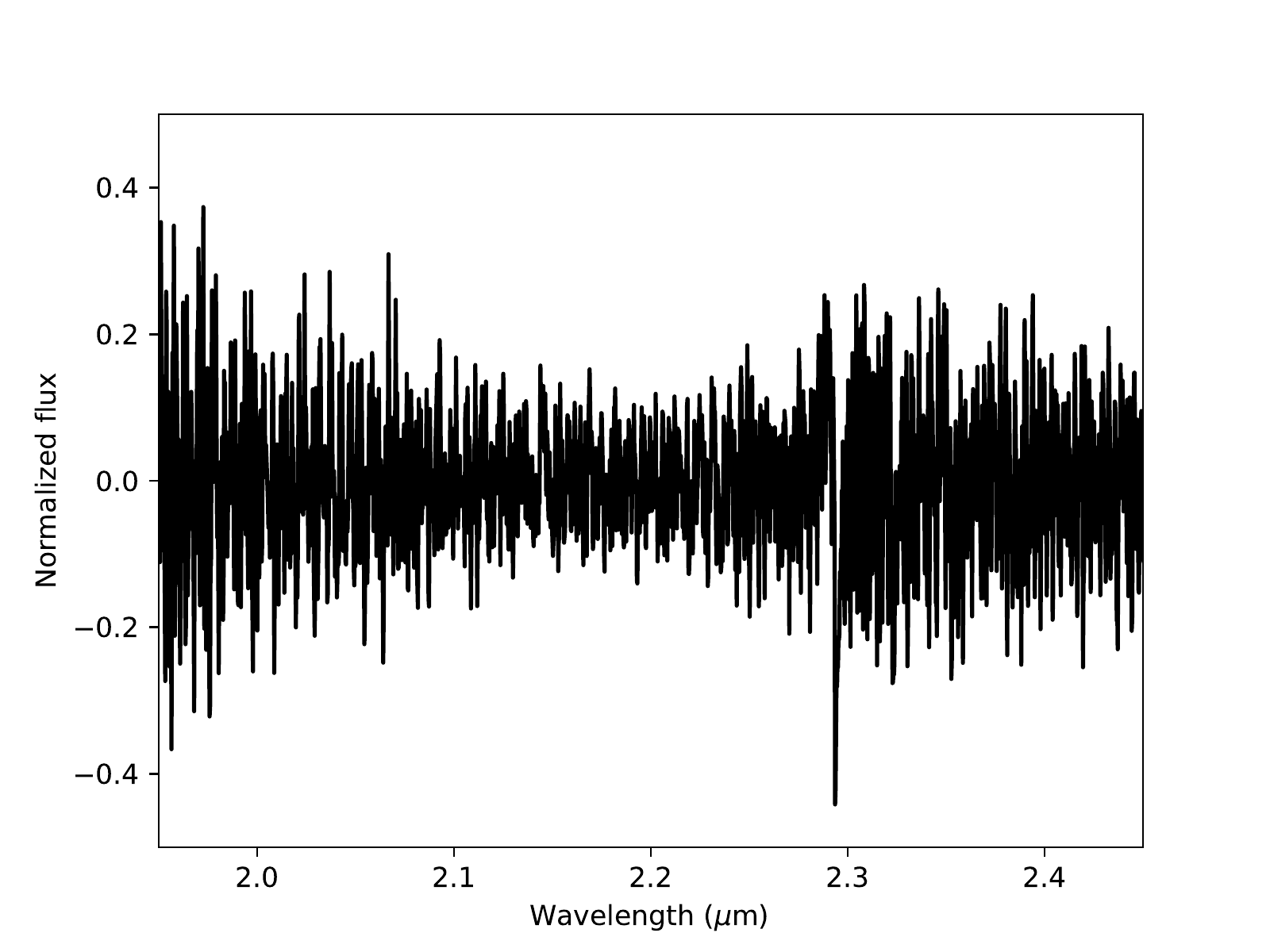}  \\
  \caption{\textit{Left} : Maps of cross-correlation between SINFONI data of HIP~65426\,b and templates of molecules $^{12}$CO, H$_{2}$O, CH$_{4}$, NH$_{3}$ and full molecule template (\textit{top} to \textit{bottom} respectively) at a velocity of 20 \kms. The star represents the position of the host star which is out of field. We detect the companion with the full molecule template (SNR=2.5), H$_{2}$O (SNR=2.2) and possibly with the $^{12}$CO (SNR=1.1). \textit{Center} : Distributions of cross-correlation values in the integrated cross-correlation cube for each position. The correlation signal is significant for the full molecule template, H$_{2}$O and $^{12}$CO  but not for CH$_{4}$ and NH$_{3}$. \textit{Right}: molecular template.}
  \label{Fig:molmaptemplate}
\end{figure*}

\subsection{Characterization with the molecular mapping}
\label{subsec:molmapcarac}

\subsubsection{Extraction of the planet's correlation signal}
\label{subsec:planetcorsig}
We ran the molecular mapping steps with the \texttt{TExTRIS} package (Section \ref{subsubsec:molmap}) on the molecular templates (see Section \ref{subsubsec:Moleculartemplate}) and the grids of \texttt{Exo-REM} and \texttt{BT-SETTL15} model spectra (see Section \ref{subsubsec:Moleculargrids}). The tool was parallelized to gain computation time and produced as many cross-correlation maps as input model spectra. The planet is detected with a radial velocity of 20$\pm$8\kms\, (see Figures\,1).  We extracted the cross-correlation signal in the correlation maps by averaging the cross-correlation function over the FWHM of the auto-correlation, centered on the planet radial velocity to produce averaged correlation maps and within a circular aperture centered on the planet position of 4 pixels radius ($\sim$FWHM of the PSF) to extract the correlation signal of the planet. We detail below our exploitation of the correlation signal to characterize the properties of the planet.

\subsubsection{Molecular template analysis}
\label{subsubsec:Moleculartemplate}
We used pre-computed templates of individual molecules ($^{13}$CO, $^{12}$CO, CO$_{2}$, CH$_{4}$, FeH, H$_{2}$O, HDO, K, NH$_{3}$, Na, PH$_{3}$, TiO and VO) as input of \texttt{TExTRIS} to evaluate whether these species produce a strong set of absorption in the planet spectrum.  The templates were computed with the \texttt{Exo-REM} model assuming the pressure-temperature profile of a solar-metallicity atmosphere with \Teff=1700\,K, log(g)=4.0\,dex (e.g, a generic model close to the physical properties of HIP~65426\,b). We show  in Figure \ref{Fig:molmaptemplate} (\textit{left} column) the averaged correlation maps for the CO, H$_{2}$O, CH$_{4}$ and NH$_{3}$ templates, and with the full molecule template (\Teff=1700 K, log(g)=4.0 dex, [M/H]=0.0, C/O=0.50). The planet is detected  with the full molecule template, the H$_{2}$O template and marginally with the $^{12}$CO template. The distribution of correlation signals (see Figure \ref{Fig:molmaptemplate}, \textit{right} column) shows a Gaussian distribution and indicates a detection at a signal-to-noise of 2.5, 2.2 and 1.2 for the full molecule template, the H$_{2}$O and $^{12}$CO templates, respectively. This is consistent with the mid-L spectral type of the object. 

The lack of a clear detection of the $^{12}$CO is unexpected for a L6 object and is discussed in the Section \ref{Sec: MolecMap_noise}. 



\subsubsection{Grids of synthetic spectra exploration}
\label{subsubsec:Moleculargrids}

  \begin{figure}[t]
  \centering
  \includegraphics[width=\columnwidth]{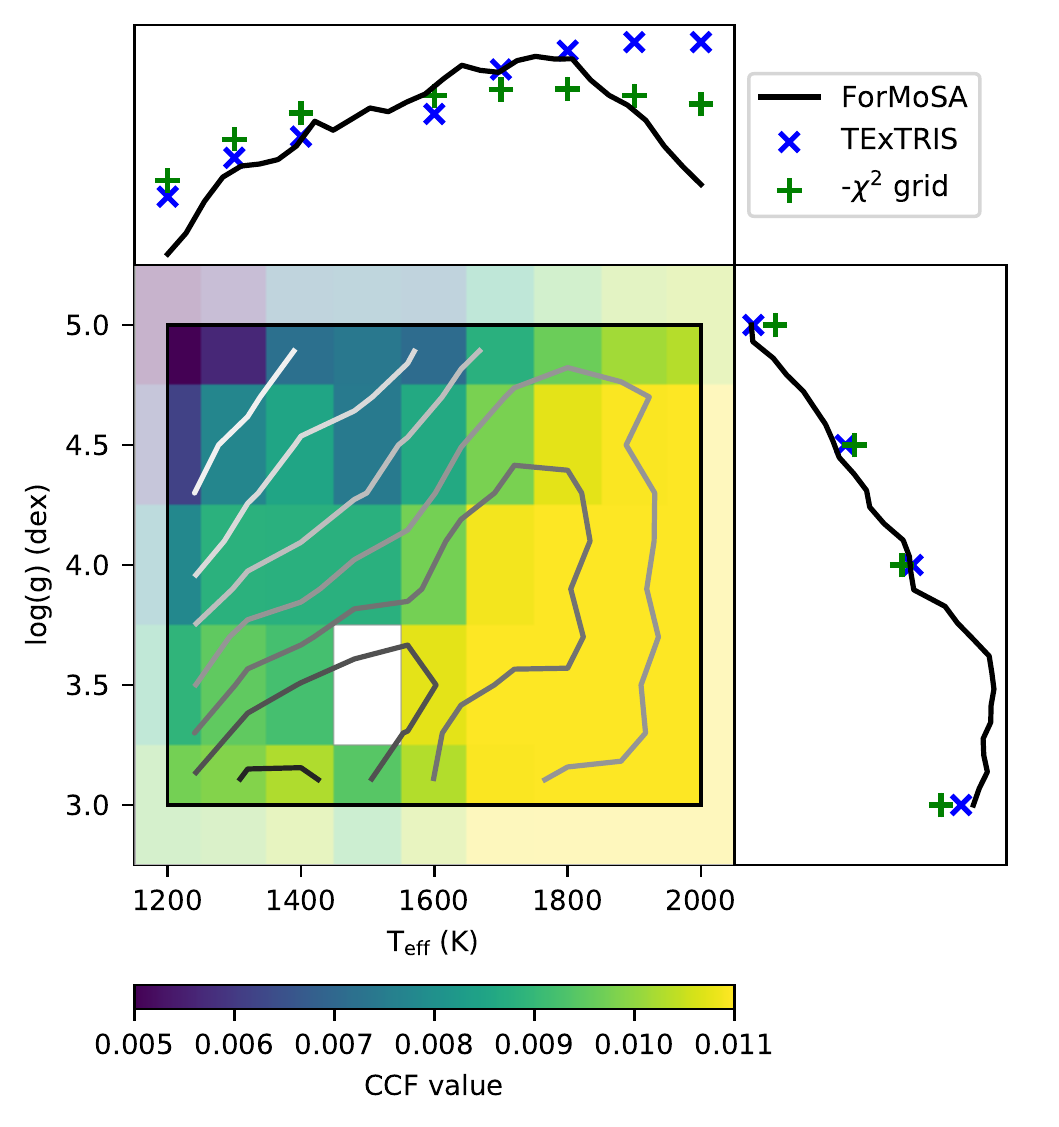} 
  \caption{Comparison between the correlation signal of each spectrum of the grid \texttt{BT-SETTL15} obtained with \texttt{TExTRIS} (color map) and the 2D-posterior obtained with \texttt{ForMoSA} (\textit{grey} contours). We also compare the normalized 1D-posterior obtained with \texttt{ForMoSA} on the K band without the continuum (black line) with the normalized 1D-pseudo-posteriors extracted from the correlation signal map (\textit{blue} cross) and a $\chi^{2}$ map (\textit{green} cross) for the \Teff\, and the log(g) (\textit{top} and \textit{right} panel respectively). The \textit{white} rectangle corresponds to an non-converged synthetic spectra.}
  \label{Fig:textrixVSForMoSA_btsettl}
\end{figure}

As a following step, we explored the evolution of the cross-correlation signal over the synthetic grid of spectra from the \texttt{BT-SETTL15} and \texttt{Exo-REM} models. Synthetic spectra with anomalous spectral slopes indicative of non-converged cloud models were excluded and are reported as white rectangles. The application with the \texttt{BT-SETTL15} grid is shown in Figure\,\ref{Fig:textrixVSForMoSA_btsettl} giving the correlation signal evolution with the effective temperature and the surface gravity. The 2D posterior distribution obtained with the \texttt{ForMoSA} code on the continuum-removed SINFONI spectrum (see Section \ref{Sec:Bayse}) is overlaid. We also show the corresponding 1D posteriors along with the averaged $\chi^{2}$ and mean correlation signals (``pseudo-1D-posteriors''). The $\chi^{2}$ pseudo-1D-posteriors starts from the original grid of models. We note that this alternative approach to the forward modeling one also identifies two distinct modes of physical parameters for HIP~65426\,b at ``low-\Teff'' (\Teff$\simeq$1400\,K) and ``high-\Teff'' (\Teff$\simeq$1800\,K).
 
\begin{figure*}
  \centering
  \includegraphics[width=7.5cm]{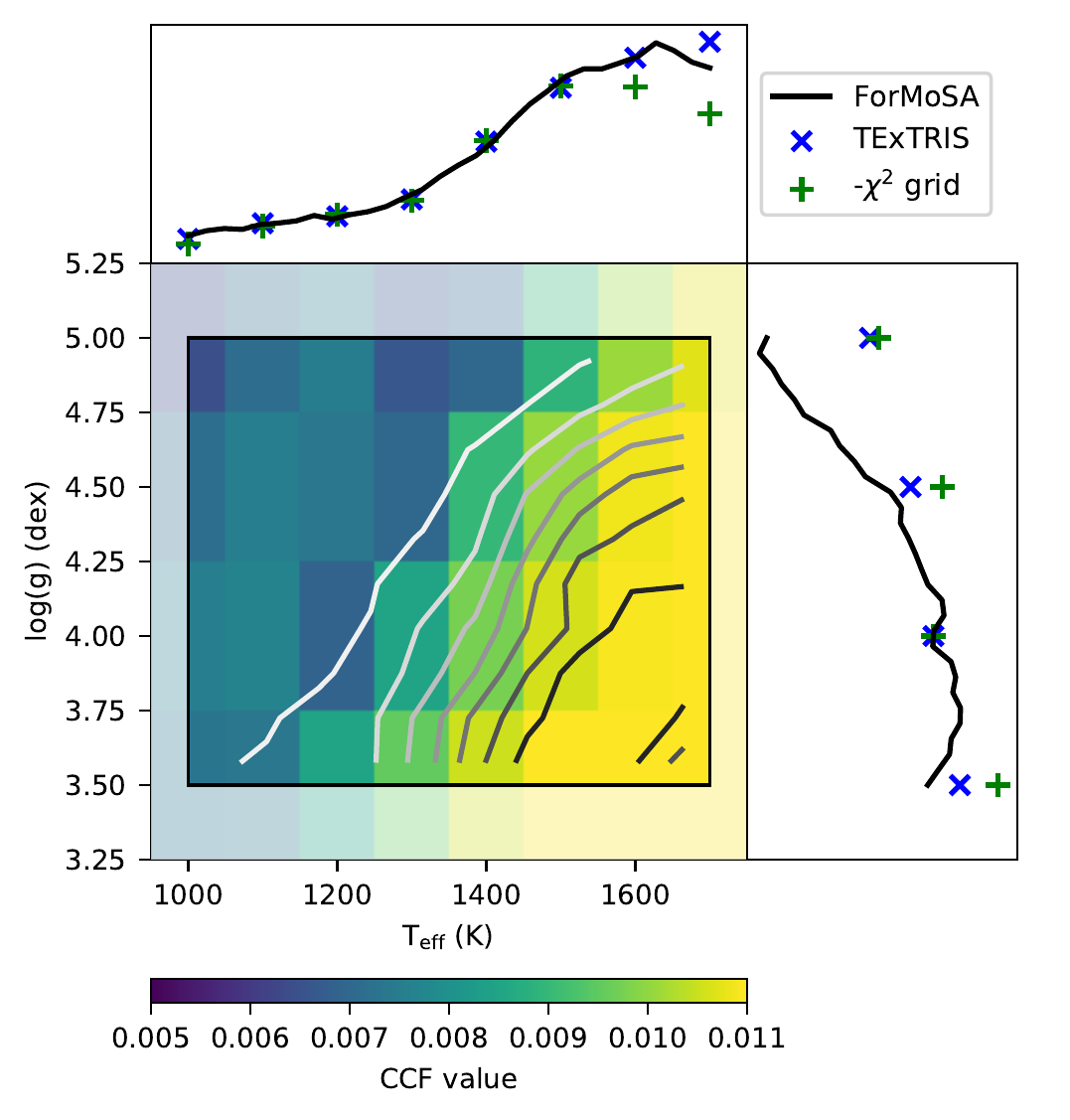} 
  \includegraphics[width=7.5cm]{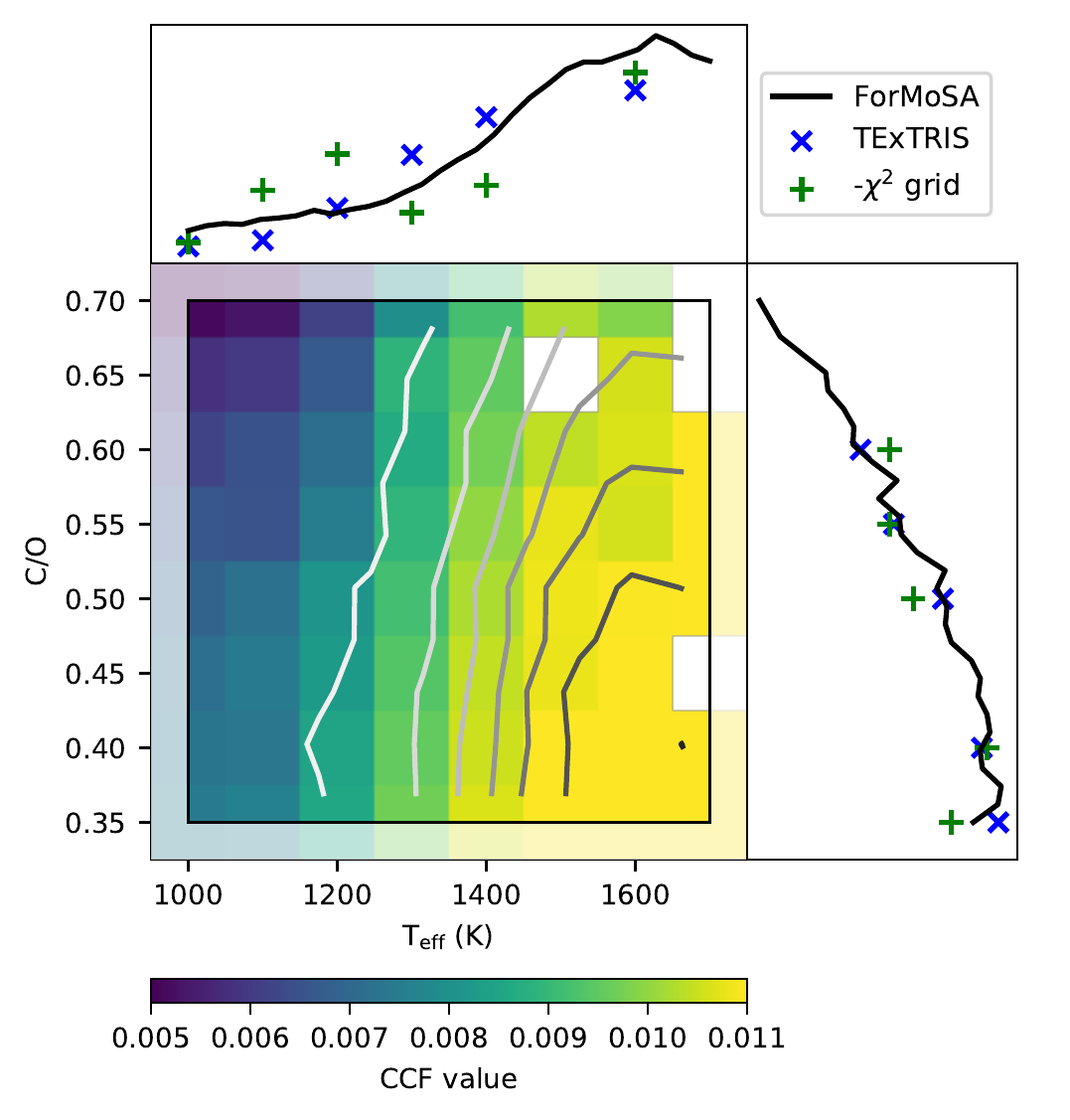} 
  \includegraphics[width=7.5cm]{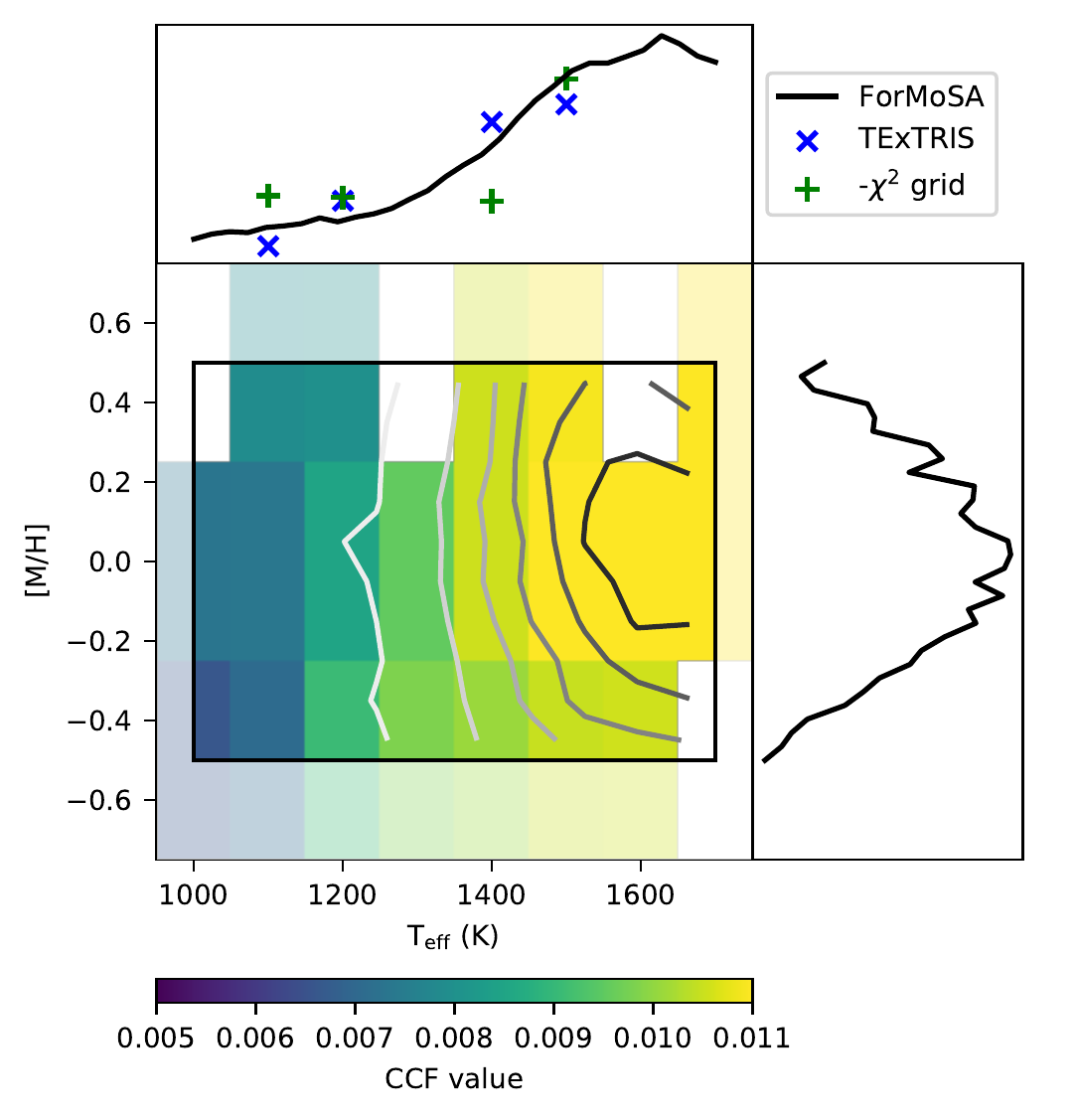} 
  \includegraphics[width=7.5cm]{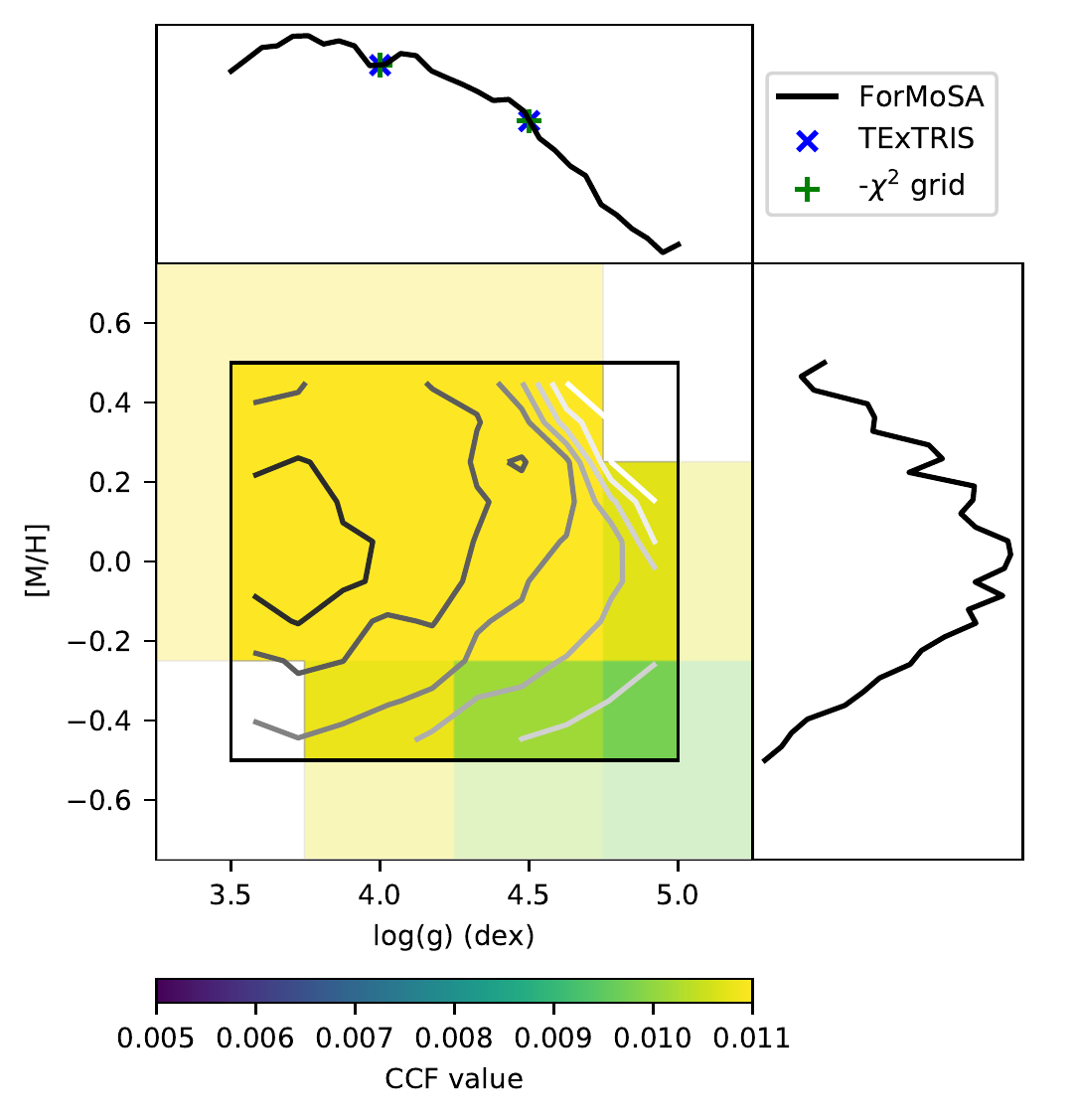} 
  \includegraphics[width=7.5cm]{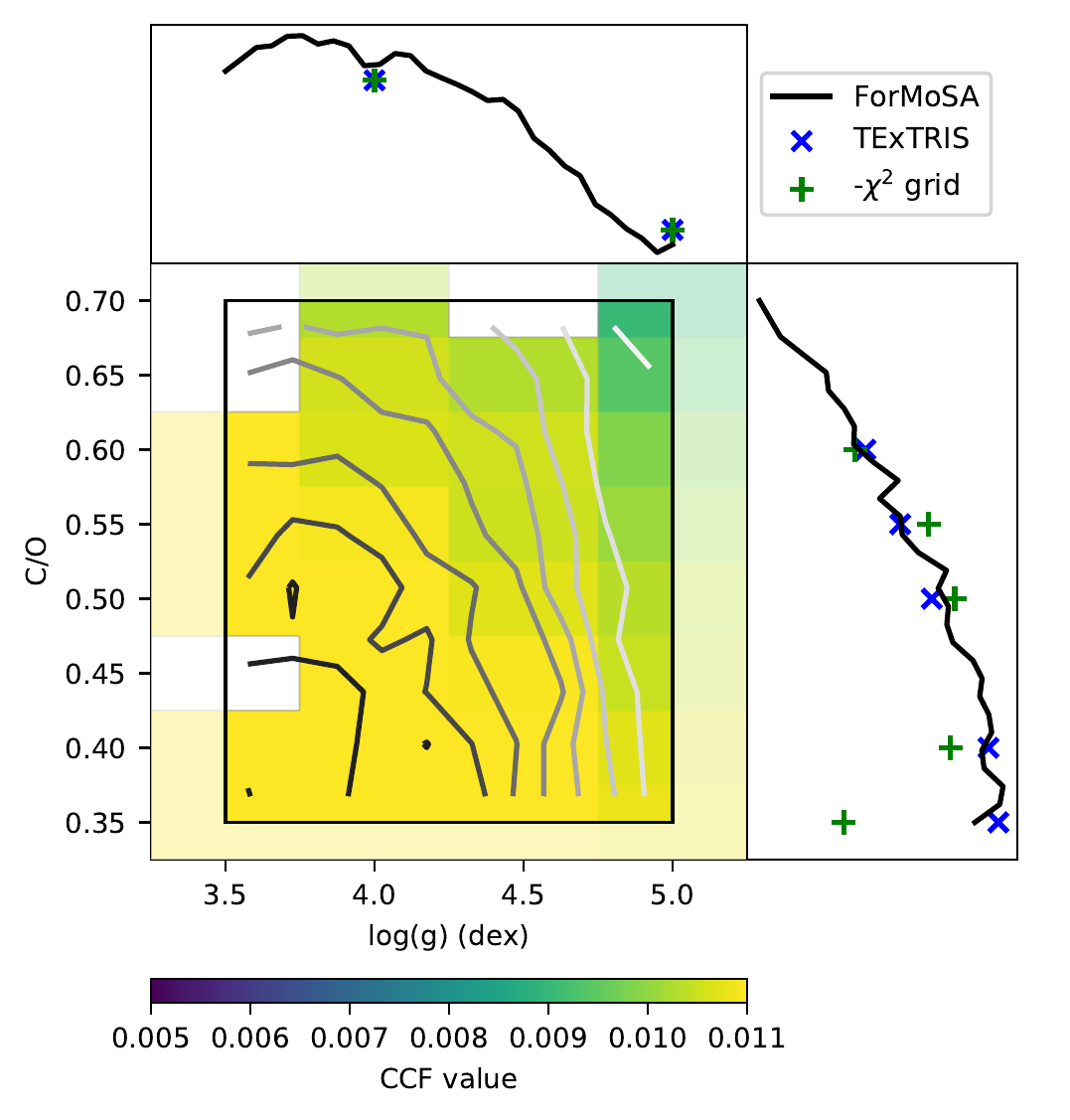} 
  \includegraphics[width=7.5cm]{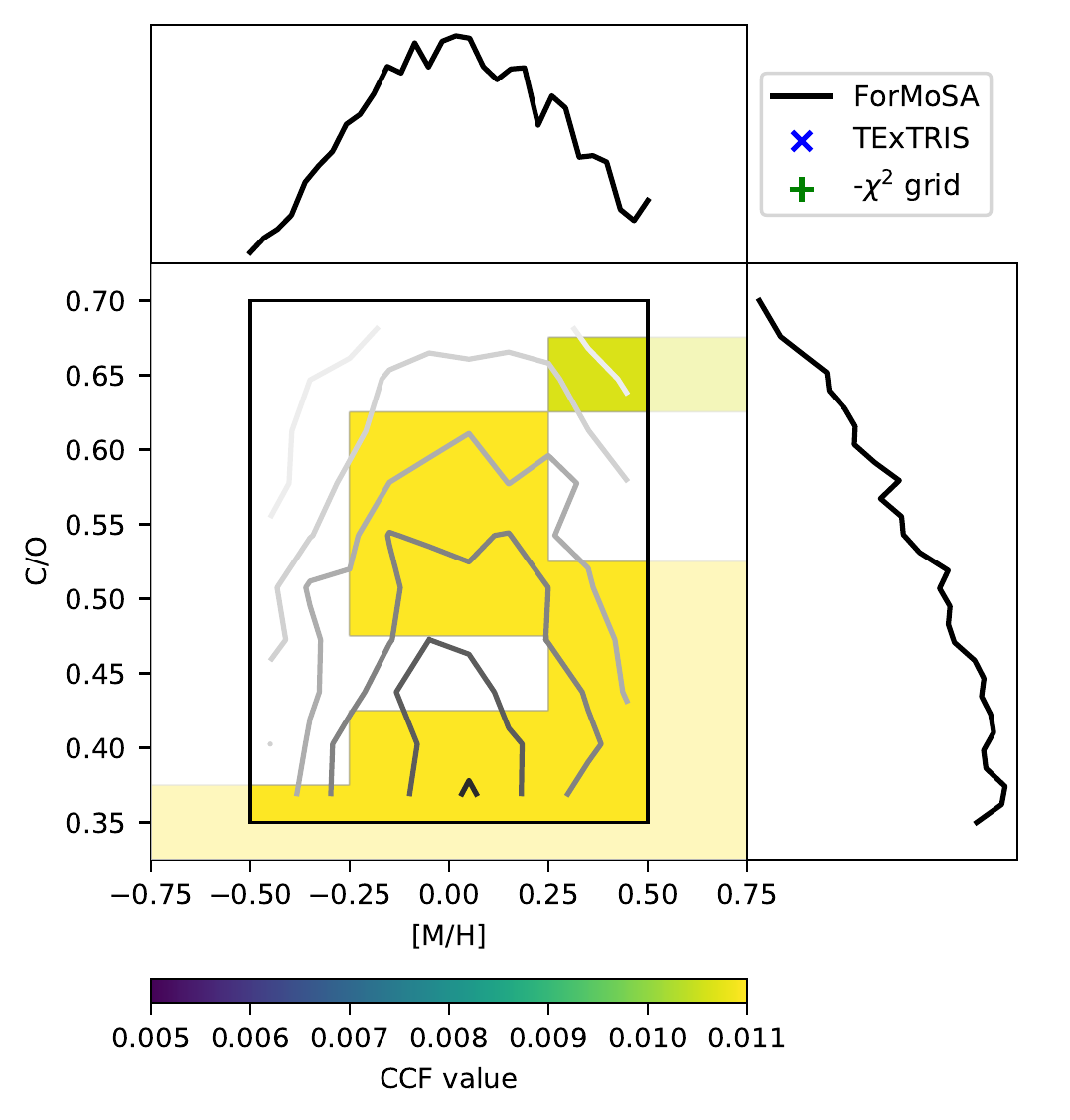} 
  \caption{Same as in the Figure \ref{Fig:textrixVSForMoSA_btsettl} but with the grid \texttt{Exo-REM}. We considered different layouts around the set of parameters that maximise the CCF in each direction of the parameter space. Models at \Teff=1800K are unconverged at the given [M/H] and/or C/O, and/or log (g) so that the corresponding CCF values and posteriors stop at \Teff=1700K.}
  \label{Fig:textrixVSForMoSA_exorem}
\end{figure*}

A similar analysis with the \texttt{Exo-REM} model grid are shown in Figure \ref{Fig:textrixVSForMoSA_exorem} considering different layouts in the parameter space explored. 
Their exploitation also shows a similar behavior than previously and favors one family of solutions with \Teff=1700\,K, log(g)=3.5, [M/H]=0.0 and C/O=0.40. For both models, the detailed set of physical parameters maximizing the correlation signal are reported in Table \ref{Tab:ParamSETTL} for the \texttt{BT-SETTL15} and \texttt{Exo-REM} models. We notice that the maximum of the CCF value obtained with both model are similar ($\sim$ 0.011). The CCF signal using the \texttt{Exo-REM} models is maximized at solar-metallicity, which correspond "de-facto"  to the one of the \texttt{BT-SETTL15} models.

The comparison of both approaches (forward modeling versus molecular mapping) for the characterization of planetary signals is discussed below, together with the physical properties and possible origin of HIP~65426\,b.



\section{Discussion}
\label{Sec:Discussion}
\subsection{Performances and limitations of our data analysis}
\subsubsection{The molecular mapping on noisy data}
\label{Sec: MolecMap_noise}
The exposure time calculator  (ETC- of SINFONI estimates a theoretical maximum S/N=15.2 from 2.1 to 2.15 $\mu$m for the first night of observations. We measure a S/N of 9.7 on the extracted spectrum over a similar wavelength interval (see Appendix \ref{Appendix:extraction}). The difference may be due to wavelength-to-wavelength noise introduced by the extraction process. The data are read-out noise limited because of the small integration time used for avoiding the persistence on the  detector. This advocates for the  use of specific high-contrast modes on such medium-resolution IFUs \citep[e.g., GTC/FRIDA,  ELT/HARMONI;][]{2018SPIE10703E..3EN, 2018SPIE10702E..9NC}.  

The observations of HIP~65426\,b offer an interesting test-case of the molecular mapping method when applied in the low S/N regime. In such a case, only the strongest molecular absorptions will contribute to the correlation signal. The  H$_{2}$O tend to produce a larger amount of strong lines compared to $^{12}$CO. The $^{12}$CO lines are spread in a narrow range of wavelength where the theoretical S/N predicted by the ETC is lower (10.5 for wavelengths longward of 2.29 $\mu$m) due to the numerous telluric lines and the lower transmission of the K-band filter of the instrument. Conversely, the set of narrow water absorption are found as a set of partly blended features at the resolution of SINFONI and spread over the entire K-band (Figure \ref{Fig:molmaptemplate}). This explains why the  $^{12}$CO does not produce a strong correlation signal as would naively be expected for a mid-L type exoplanet. We investigated the detection capability of the molecular mapping using the CO template further in the Appendix \ref{Appendix:C} injecting fake planets at different C/O ratio into our datacube. The tests reveal the ability of the method for recovering a planet with a noiseless spectral signature  close to the one of HIP~65426b at a low S/N. We show in addition that the significance of the detection drops for synthetic planets with low C/O ratio for which the strongest CO overtone is shallower.

\begin{figure}
  \centering
  \includegraphics[width=\columnwidth]{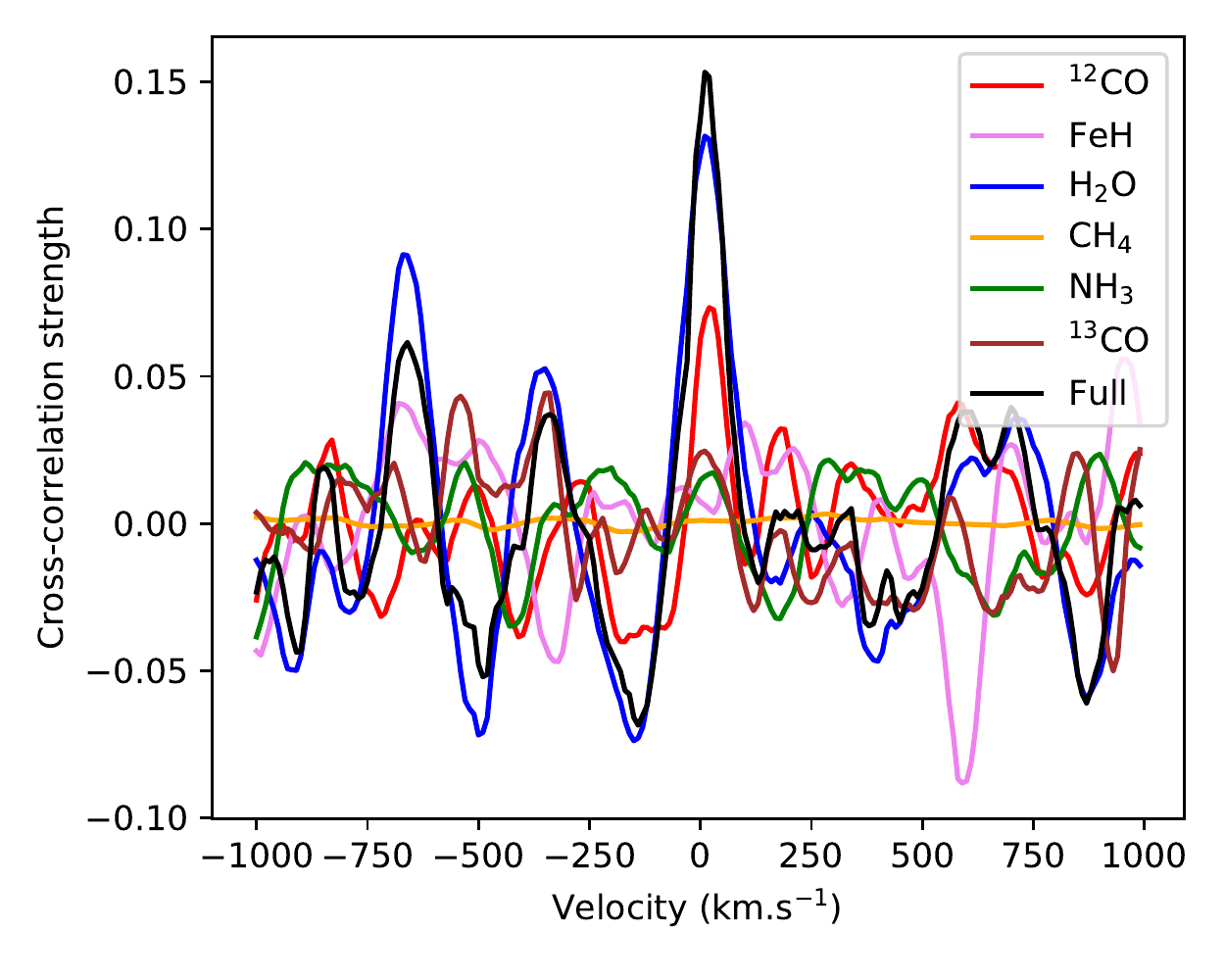} 
  \caption{Cross-correlation signals obtained when using the continuum-subtracted spectrum of HIP~65426\,b and the molecular templates (color) or a full Exo-REM spectrum at \Teff=1700K, log g=4.0 and solar abundances (black).}
 \label{Fig:CCFsimues}
\end{figure}

We show in Figure \ref{Fig:CCFsimues} the correlation signal between the extracted SINFONI spectrum of the planet (continuum-subtracted) and the molecular templates. We confirm the conclusions from Figure \ref{Fig:molmaptemplate} and in particular the faintness of the correlation signal of $^{12}$CO with respect to H$_{2}$O. 
To conclude, the figure also reveals additional correlation and anti-correlation peaks over the velocity span. As noted by \cite{2018A&A...617A.144H}, some of these peaks are likely to be caused by the regular spacing of  molecular absorption both in the template and exoplanet spectra (overtones). This was checked by computing the auto-correlation of the templates and justifies the use of the spatial distribution of CCF values for computing a signal-to-noise rather than the velocity dimension. The overtones appear at different velocities in the various molecules and therefore the correlation of full synthetic spectra combining all the molecular absorption offer to reduce these parasitic signals with respect to the main correlation peak in the velocity space. 

These features are slightly reduced and the central correlation peak is enhanced when using a full synthetic spectrum corresponding to the \Teff\, log g, and composition assumed for the  templates of individual molecules. For all these reasons, the use of full synthetic spectra  with respect to templates of individual molecules offers a more robust way of characterizing any object and  achieving a convincing detection of new ones. 

\subsubsection{Comparison of the characterization approaches}
In this study, we compare three different approaches to determine the physical parameters of HIP~65426\,b: a standard $\chi^{2}$ minimization, a Bayesian inference with the code \texttt{ForMoSA}, and the molecular mapping with the code \texttt{TExTRIS}. We show in the Figures \ref{Fig:textrixVSForMoSA_btsettl} and \ref{Fig:textrixVSForMoSA_exorem}, and Table \ref{Tab:param_ForMoSA} that all three methods yield consistent results.


The coherency between the $\chi^{2}$ minimization and the Bayesian inference stems from our assumption of Gaussian distribution of uncertainties on the observed spectra and the lack of correlations between the data-points. The likelihood and the cross-correlation function are related assuming the same underlying hypothesis on the data \citep{2019AJ....157..114B, 2019AJ....158..200R}. That can explain the overall similar trends between the results from the Bayesian inference (and  the $\chi^{2}$ minimization) and the molecular mapping. The remaining differences, evidenced in Figures \ref{Fig:textrixVSForMoSA_btsettl} and \ref{Fig:textrixVSForMoSA_exorem}, may have several origins:
\begin{itemize}
    \item The likelihood and the CCF  are related by the model variance and data variance. The model variances vary accross the grid. The data variance is expected to be constant for a given spaxel, which might not be the case as the planet signal is sampled by different spaxels (with their own noise distribution) along the sequence because of the field rotation.
    \item The removal of the halo  and the continuum-subtraction method  is not performed at the same stage in \texttt{TExTRIS} and \texttt{ForMoSA} and may result in the introduction of different correlated noises. 
    \item The remaining differences may arise from correlations in the CCF signals. This can make the use of molecular mapping for the characterization more perilous. It further advocates the direct extraction of spectra of companions whenever they are detected or more robust approaches such as the forward modelling of the joint speckles and planetary signals for  objects blurred into the stellar halo \citep[e.g.,][]{2019AJ....158..200R}.
\end{itemize}

\begin{figure}[t]
  \centering
  \includegraphics[width=\columnwidth]{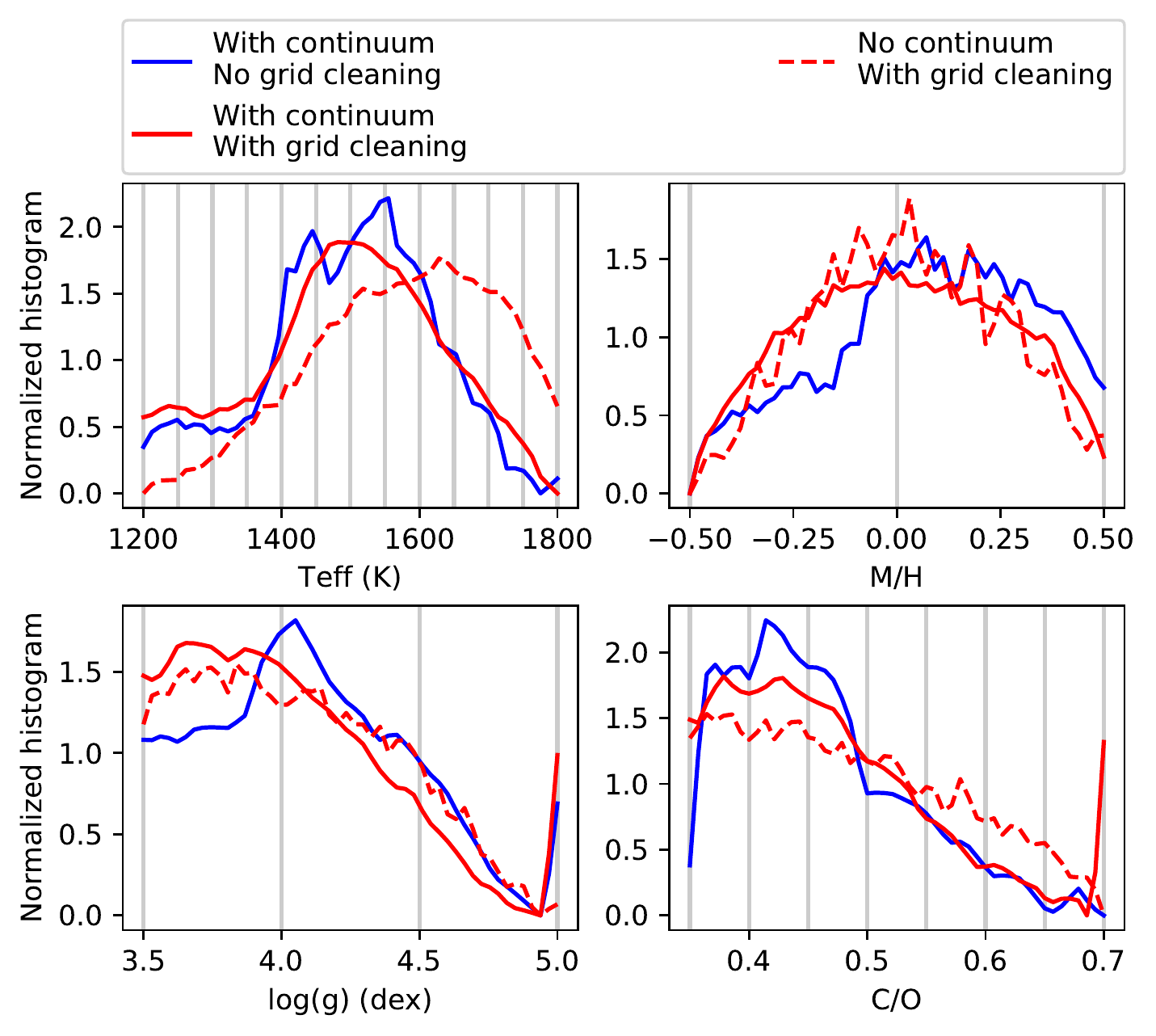} 
  \caption{Comparison of posteriors from \texttt{ForMoSA} with different data and models configurations with the grid \texttt{Exo-REM}. In \textit{blue} : The grid has been taken on the shelf and the continuum has been kept. In \textit{red} (\textit{solid} line) : The grid has been cleaned of its non-converged models and the continuum has been kept. In \textit{red} (\textit{dashed} line) : The grid has been cleaned of its non-converged models and the continuum has been removed. In \textit{grey} : The step of the grid for each parameter.}
  \label{Fig:comp_posteriors}
\end{figure}

Yet, the evolution of the CCF across the grids remains very similar to the one of the $\chi^{2}$ values and the main differences with the posteriors (\texttt{ForMoSA}) appear at the grid edges. This may stem from the interpolation steps performed by \texttt{ForMoSA}. The grid generation can originally produce non-converged synthetic spectra which will propagate at the interpolation steps and impact the posterior shape. 
The Figure \ref{Fig:comp_posteriors} shows the difference between the posteriors obtained with the original grid \texttt{Exo-REM} spectra and with a grid where we removed all the obvious non-converged models. When no cleaning of the grid is performed beforehand, local maxima of likelihood appear due to the non-converged models and produce  patterns in the posteriors. We have been taking a particular care to that issue and  removed any non-converged models before running the code. However, it is still possible that some models may be slightly non-converged at the edges of the grids  especially in the low gravity range where dust is expected to play a prominent role in the atmospheric balance or at high  \Teff\,where the \texttt{Exo-REM} models are less suited.  

These appear as inherent limitations of Bayesian inference with pre-computed grids of forward models. These biases could be lifted when a fully integrated Bayesian scheme coupled to \texttt{Exo-REM} (retrieval) becomes available or by using extended and input grids of models with a refined meshing. 

  


\subsubsection{Losing the pseudo-continuum information}

\begin{figure*}[t]
  \centering
  \includegraphics[width=8cm]{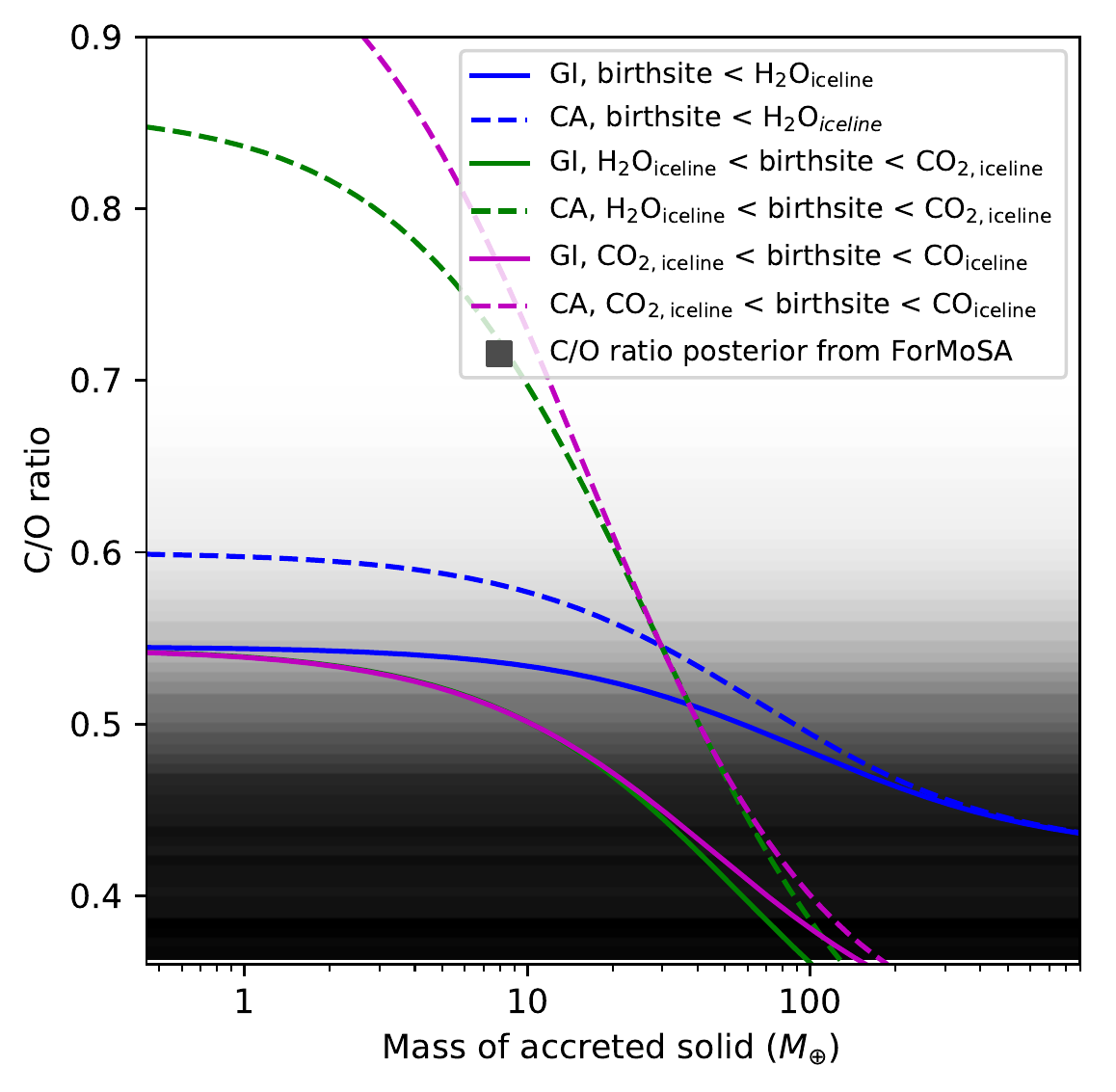} 
  \includegraphics[width=8cm]{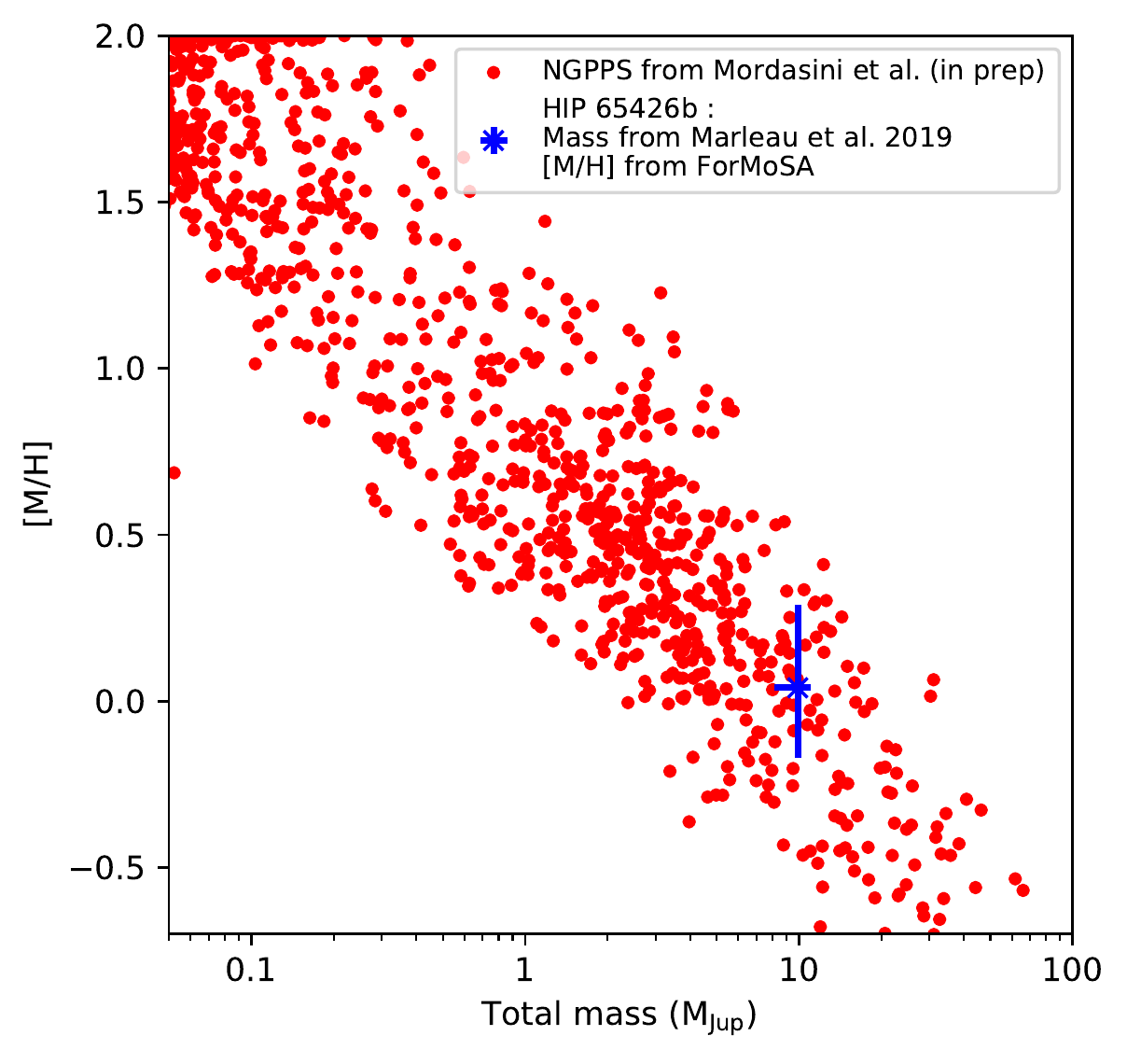} 
  \caption{\textit{Left}: Comparison between the posterior of the C/O ratio given by \texttt{ForMoSA} (\textit{gradient; black}: high likelihood) and different scenario of formation as a function of the accreted solid mass. The \textit{solid} lines correspond to the gravitational instability (GI) scenario while the \textit{dashed} lines correspond to the core accretion (CA) scenario. The \textit{blue}, \textit{green} and \textit{purple} curves represent a formation within the water iceline , between the water and CO$_{2}$ icelines and between CO$_{2}$ and CO icelines) respectively.  \textit{Right}: Comparison between the bulk enrichment of the NGPPS (\textit{red} dots) and the estimated properties of HIP~65426\,b (\textit{blue} dot). We show here the NGPPS with a mass higher than 0.05 \MJup\, ($\sim$ Neptune's mass).}
  \label{Fig:formation_pathway}
 \end{figure*}

The molecular mapping implies a removal of the information contained in the planet spectral continuum.  The Figure \ref{Fig:comp_posteriors} compares Bayesian posteriors obtained using the grid \texttt{Exo-REM} on the band K with and without the continuum (\textit{solid} and \textit{dashed} \textit{green} line respectively), and the Table \ref{Tab:param_ForMoSA} summarizes the values of the parameters extracted with error bars. Both approaches yield results in agreement within error bars (1$\sigma$). Nevertheless, the subtraction of the continuum shifts and flattens the \Teff\, posterior (factor of $\simeq$2.0 between the error bars). The constraint on the other physical parameters is less impacted by the continuum removal, which indicates that the related information is encoded in the line intensities.


These differences clearly point out the limitations of the molecular mapping for constraining the \Teff~of objects in  the L-type regime from observations at K-band. 
We still need to investigate if the method, when applied to a different band (J or H-band)  or at higher spectral resolution (VLT/ERIS K-band mode at R$\sim$8000, ELT/HARMONI), could retrieve that information.


\subsection{Interpreting the radial velocity of HIP~65426\,b}

An interesting result coming from the SINFONI K-band spectrum of HIP~65426\,b is the determination of the radial velocity of the planet itself. As described in Section\,\ref{sec:obs}, great care has been taken to calibrate the wavelength dispersion of the cubes using the various telluric absorption lines, and  cross-checked at the end with solution found for the OH lines present in the data (centered on zero \kms, see Figure\,\ref{Fig:recalOH}). Using both independent approaches, the Bayesian inference and the molecular mapping to measure the radial velocity of HIP~65426\,b in the SINFONI datacubes, we find compatible values that give a radial velocity estimate of $21\pm7$\kms. Similar results are obtained with the molecular mapping approach using the observation acquired on May 25$\rm^{th}$, 2018 ($20\pm8$\kms). We used the molecular mapping technique to estimate the radial velocity on the set of observations acquired on May 26$\rm^{th}$, 2018 and we found $31\pm8$\kms. The radial velocity of both night are consistent at 2-$\mathrm{sigma}$ and we chose to define our final estimate of the radial velocity as RV = $26\pm15$\kms\, to be the most conservative.  

\begin{figure}[t]
  \centering
  \includegraphics[width=\columnwidth]{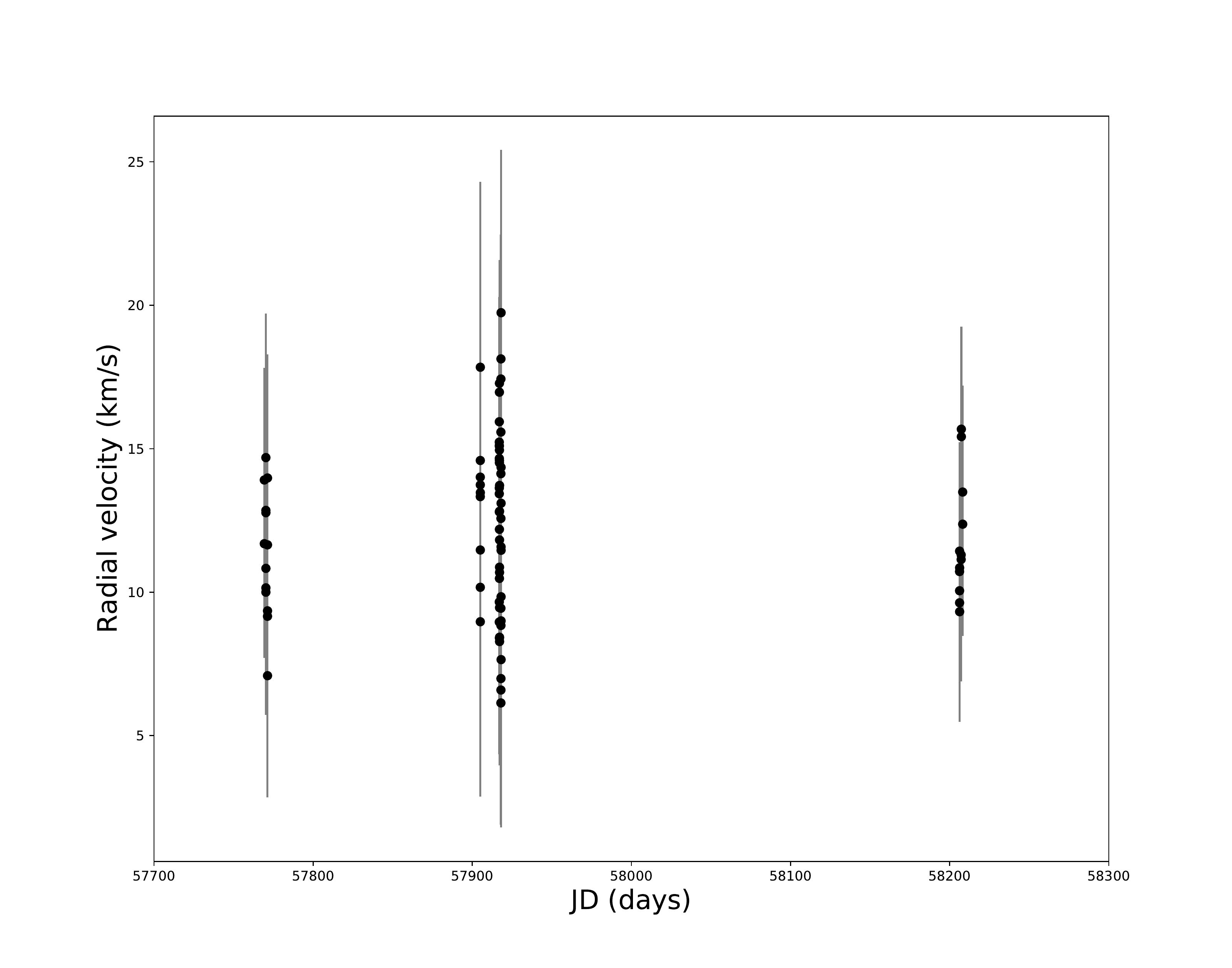} 
  \caption{Radial velocity measurements of HIP\,65426 using a total of 78 HARPS spectra obtained between 2017 and 2018 at La Silla/3.6m telescope and weak metallic lines less sensitive to the continuum tracing.}
  \label{Fig:rv}
\end{figure}

A direct application of this result is to check the consistency with the expected Keplerian motion of a bound companion orbiting HIP~65426\,A. Considering the orbital solution found for HIP~65426\,b by \cite{2019A&A...622A..80C}, one might expect a maximum difference of $\pm4.4$\kms\, relative to the primary star absolute radial velocity. In the discovery paper of HIP~65426\,b, \cite{2017A&A...605L...9C} reported for HIP~65426\,A a value of $5.2\pm1.3$\kms, and a very high projected rotational velocity of $v$sin$i=299\pm9$\kms. They considered for this analysis a set of 14 HARPS spectra acquired during the nights of January 16$\rm^{th}$, 17$\rm^{th}$ and 18$\rm^{th}$, 2017, and a two sets of lines: (i) Six strong lines (Ca II K and five H lines from H$\beta$ to H9, excluding H$\epsilon$; using the rotationally broadened core of the  lines) and (ii) 35 atomic metallic lines. More details can be found in Appendix\,B of \cite{2017A&A...605L...9C}. Given the primary spectral type and its rotational velocity, a precise measurement of the absolute radial velocity is challenging. It motivates to revisit this first analysis mainly focused on the extraction of the stellar rotational velocity.  We used this time 78 HARPS spectra\footnote{Program IDs: 098.C-0739(A) and 099.C-0205(A)} obtained during 10 nights spanning a period of time between January, 2017 to March, 2018. We first followed a similar analysis to extract the radial and rotational velocities using both sets of lines (i) and (ii). Although H lines are adapted for the extraction of rotational velocities and the search for variability given their strength, their extended wings makes them less reliable to derive absolute radial velocity as they are more sensitive on the continuum tracing. The revisited absolute radial velocity of HIP\,65426 was therefore extracted using only the second set of lines (ii), i.e. the weak metallic lines that are essentially only broadened by the atmospheric velocity field. The resulting radial velocity measurements are shown in Figure\,\ref{Fig:rv}. A mean radial velocity of $12.2\pm0.3$ \kms\, (2.9\kms\, for the standard deviation of the RV measurements) is derived, which is significantly different from the value derived by \cite{2017A&A...605L...9C}, but more reliable given the number of spectra used, the temporal baseline, and the selection of metallic lines more adpated for absolute radial velocity measurements. Moreover, high frequency variation probably related the existence of pulsation are clearly identified. The average value of $v$sin$i$ is lower than originally derived with a value of $261\pm2$\kms\, (16\kms\, for the standard deviation of the RV measurements) and no sign of Spectral Binary 1 (SB1) or Spectral Binary 2 (SB2) binary are detected.

From this, we conclude that the difference of radial velocities between HIP~65426\,b and A is compatible with a physical object in Keplerian motion (considering the uncertainties on the radial velocities of A and b, and the predicted Keplerian motion for b). We note that the new value of radial velocity measurement of HIP~65426\,A is in better agreement with a membership to the LCC subgroup considering the probability predictions of the \texttt{BANYAN $\Sigma$} code by  \cite{2018ApJ...856...23G}. Finally, the direct determination of the radial velocity of imaged exoplanets with medium (to high) resolution spectrographs, as already obtained for $\beta$ Pic\,b \citep{2014Natur.509...63S}, HR\,8799\,b and c \citep{2019AJ....158..200R} and now HIP~65426\,b, highlights the rich perspective of this technique in synergy with astrometric monitoring to: i/ test the companionship of the newly imaged exoplanetary candidates, ii/ refine their orbital parameters when confirmed, and iii/ measure, with very high spectral spectra, their projected rotational velocity and potentially their spin.



\subsection{HIP~65426\,b formation pathway}

The  broad constraints on the C/O ratio and metallicity of HIP~65426\,b derived in Section \ref{Sec:Physical_properties} give the opportunity of discussing the planet formation mode. We caution that the C/O ratio estimate depends on the correct assumptions on the cloud prescriptions where grains are efficient oxygen carriers and can  impact the composition of the gas phase from which the molecular lines arise \citep[e.g.,][]{2014Life....4..142H}. 

The interpretation relies on several additional hypothesis. The original composition of the putative circumstellar disk where HIP~65426\,b may have formed is obviously unknown and the C/O and metallicity of HIP~65426\,A, which is a fast rotator \citep{2017A&A...605L...9C}, have not been determined either, to our knowledge.  \cite{1980A&A....84..115K} report the C/O ratio of 6 bona-fide members of LCC and UCL but these estimates show a strong dispersion and would need to be revised with the help of recent stellar model atmospheres.  \cite{2011AJ....142..180B} find on average a sub-solar metallicity ([Fe/H]=-0.12 dex, 0.10 dex rms)  for a sample of UCL and LCC solar-type stars. \cite{2009A&A...501..965V} report metallicities of $0.02\pm0.05$ dex and  $-0.02\pm0.09$ dex for 6 and 10 stars from UCL and Upper-Scorpius, respectively. \cite{2013A&A...552A..73N} reports a correlation between the C/O ratio and the Fe/H of stars with planets with a mean C/O=0.58 at Fe/H=0 (0.06 rms scatter). Therefore, we can expect both HIP~65426A metallicity and C/O ratio to be close to solar values. Yet, the knowledge of the host star composition would be important for strengthening the conclusions on the formation pathway based on the exoplanet atmospheric composition.


\begin{table*}[t]
\centering
\caption{Imaged planets around intermediate-mass stars with reported abundance measurements}
\label{Tab:abundances}
\small
\begin{tabular}{cccccc}
\hline
Planet  &  $a$ &   Mass    &   [M/H] &   C/O & Refs.\\
        &   (au)    &   (\MJup) &   (dex)   &   &   \\    
\hline
$\beta$\,Pic\,b  & $11^{+0.3}_{-0.4}$ &  $12.7 \pm 2.2$ &  $0.68^{+0.11}_{-0.08}$  & $0.43^{+0.04}_{-0.03}$   &   1 \\
HR\,8799\,e & $15.3^{+1.4}_{-1.1}$ &  $7.2^{+0.6}_{-0.7}$ &  $0.48^{+0.25}_{−0.29}$  &  $0.60^{+0.07}_{-0.08}$ & 2, 3 \\
HR\,8799\,c &   $37.6^{+2.2}_{-1.7}$ &  $7.2^{+0.6}_{-0.7}$ & \dots & $0.65^{+0.10}_{-0.05}$  & 2, 4 \\
HR\,8799\,b &    $69.5^{+9.3}_{-7.0}$ & $5.8\pm0.5$ & \dots &  $0.61^{+0.09}_{-0.03}$ & 2, 5 \\
HIP~65426\,b & $110^{+90}_{-30}$  & $9.9^{+1.1}_{-1.8}$  &  0.05$_{-0.22}^{+0.24}$ &  $\leq$0.55 &  6, 7, 8 \\ 
\hline
\end{tabular}
\tablefoot{References: [1] - \cite{2020A&A...633A.110G},  [2] -  \cite{2018AJ....156..192W} , [3] - \cite{2020arXiv200609394M},  [4] - \cite{2013Sci...339.1398K}, [5] - \cite{2015ApJ...804...61B}, [6] - \cite{2019A&A...622A..80C}, [7] - \cite{2019A&A...624A..20M} , [8] - this work.}
\end{table*}
     The \cite{2020A&A...633A.110G} have proposed a simple analytical method for interpreting the C/O ratio measured for imaged exoplanets. From the relative abundances of CO, CO$_{2}$, H$_{2}$O, C (grains) and O (silicates) given in \cite{2011ApJ...743L..16O}, they estimated the relative abundance of the C and O in both the solid (\ncs, \nos) and gaseous phases (\ncg, \nog)  and for two possible planet birth locations : within the water iceline and between the water and CO$_{2}$ icelines.  We considered in addition here the case of a formation between the  CO$_{2}$ and CO icelines given the projected physical separation of 92\,au for HIP~65426\,b.      
     The planet birthsite defines \Msolid\, and \Mgas\, as the amount of solid and gas in the atmosphere of the planet and \fsg\, the solid/gas ratio in the initial disk to calculate the C/O ratio :

\begin{equation}
\label{Eq:C_O}
\mathrm{C}/\mathrm{O} = \frac{\ncs\,\fsg\,^{-1}\,\Msolid+\ncg\,(1-\fsg)^{-1}\,\Mgas}{\nos\,\fsg\,^{-1}\,\Msolid+\nog\,(1-\fsg)^{-1}\,\Mgas}
\end{equation}

In the case of \Msolid\,=\,\fsg\,\Mplanet\, and \Mgas\,=\,(1-\fsg)\,\Mplanet\, where \Mplanet\, is the total mass of the planet and \fsg\, is assumed to be equal to the standard solid-to-gas ratio of circumstellar disks (0.01), the Equation \ref{Eq:C_O} gives a  C/O$\sim$0.55  corresponding to the solar value. During planet formation, this C/O ratio is modified by the accretion of solids and gas. In the case of the gravitational instability scenario, solids are accreted during the sudden collapse (\Msolidini) and later on through planetesimal accretion (\Maccreted). Assuming that the total mass accreted during the collapse \Mini\, $\gg$ \Maccreted, the total mass of solids  in the planet then writes: 

\begin{equation}
\Msolid\, = \fsg\,\Mplanet\,+\,\Maccreted    
\end{equation}

In the case of a formation following a core accretion, \Msolid\ is collected through the slower core accretion process while the gas is brought mostly at the runaway accretion step (\Msolid\, = \Maccreted). For both scenario, the fraction of the mass of gas in the planet write \Mgas\, = (1 - \fsg)\,\Mplanet \,and that mix is assumed to be representative of the atmospheric composition.

We show in Figure \ref{Fig:formation_pathway} (\textit{left})  the predictions of the C/O ratio for different values of \Msolid\, and both formation paradigms for a \Mplanet\,=\,10 \MJup~\citep{2019A&A...624A..20M} planet and birthsites within the water iceline (\textit{blue} lines), between the water and CO$_{2}$ icelines (\textit{green} lines) and between CO$_{2}$ and CO icelines (\textit{purple} lines). The predictions are compared with the estimate of HIP~65426\,b C/O ratio. Both scenarii are consistent with our estimate of C/O ratio. 


%

\cite{2008ApJ...673..502K} estimate the water iceline to be originally at $\sim$7\,au around a $\sim2M_{\odot}$ star before moving inward to $\sim$2.5\,au at 3\,Myr.  \cite{2019ApJ...872..112V} locate the CO$_{2}$ iceline at $\sim10$\,au in the mid-plane of the disks around the $\sim2M_{\odot}$ young stars V1247\,Ori and HD\,163296. They also predict CO icelines at $\sim$70\,au and $\sim$140\,au for these two stars.  \cite{2015ApJ...813..128Q} report  observations of the CO snowline at 75\ au around HD\ 163296 (NB: we revised the published value using the GAIA-DR2 distance of the star) e.g. at shorter separation than the projected physical separation of HIP~65426\,b. Therefore, if formed \textit{in-situ} possibly beyond the CO snowline, the planet must have accreted a substantial amount of solids to lower its C/O ratio. 

We caution that these conclusions remain based on a simple approach of planet formation processes and snow lines in disks.  The \fsg \,might  be different than the interstellar value adopted here and is expected to decrease with age due to the growth from micron-sized particles to planetesimals and planets.  Ionisation process in the disk can modify drastically the chemical abundances of elements in the region between the H$_{2}$O and CO$_{2}$ icelines \citep{2018A&A...613A..14E}. Furthermore, our 1D approach neglects the variation of the snowline radii at different scale height in the disk \citep[e.g.,][]{2020A&A...633A.137D} and their complex evolution in time due to episodic accretion  \citep[e.g.,][]{2016Natur.535..258C}. \\

The population synthesis approach offers to track down the subtle effects involved in planet formation. The ``New Generation Planetary Population Synthesis'' (hereafter NGPPS) from the Bern group \citep{2019A&A...624A..20M, 2020arXiv200705563S} predict the bulk enrichment of planets formed by core accretion. They account for type I and II migration, the formation and dynamical interactions of multiple planet embryos per disk. The population has not yet been computed for a host-star more massive than 1.5 \Msun.~However, the predictions of bulk enrichment for the populations of planets around 1.0 and 1.5 \Msun\ show not significant change and we therefore assume here that it remains valid for a 2 \Msun\ star such as HIP~65426\,A. 

At an age of 10\,Myr, the simulation contains $\sim$ 24500 bodies with masses up to 65.9 \MJup\ and semi-major axis out to $\sim800$~au. The solar-metallicity of HIP~65426\,b is well compatible with the bulk metallicity of the NGPPS planets in the same mass range (Figure \ref{Fig:formation_pathway}) but the simulated planets are found on much tighter orbits (semi-major axis from 0.57 to 11.16 au for planets in the same ranges of mass and metallicity than HIP~65426\,b) than the projected separation of HIP~65426\,b (92\,au). The conclusions remain unchanged at 20\,Myr. This comparison of the atmospheric metallicity and bulk enrichment assumes that the metals acquired at formation get diluted in the planet envelop.



To conclude, the abundances of HIP~65426\,b can be compared to those of imaged companions around the intermediate-mass stars with similar measurements (Table \ref{Tab:abundances})\footnote{We did not included 51 Eri\,b because of discrepancies between the published GPI and SPHERE data and the inferred atmospheric composition.}. The C/O ratio and mass of HIP~65426\,b are reminiscent of $\beta$\,Pictoris\,b located much closer to its host star at  11\,au. This supports the above hypothesis on the formation of the planet closer in, then scattered at larger separation via planet-planet interactions (see \citealt{marleau2019a}). Such hypothesis could also explain the different separations observed between HIP~65426\,b and the NGPPS population at the same mass, age, and metallicity ranges. A future refinement of the orbital parameters with GRAVITY will help clarifying the dynamical history of the planet and the interpretation of the planet composition derived here.

\section{Summary and conclusions}
We analysed new K-band medium-resolution IFS data obtained with SINFONI at VLT of the HIP~65426\,b exoplanet, discovered by SPHERE, to further characterize its atmosphere. Our \texttt{TExTRIS} python analysis toolkit for IFS data was able to retrieve the star position outside the field of view and to optimise the data processing allowing for the extraction of the planet emission spectrum. We interpreted this spectrum following a Bayesian inference with self-consistent atmospheric forward models with the \texttt{ForMoSA} code \citep{2020A&A...633A.124P} and compared the inferred atmospheric properties of the planet to those obtained from the recent ``molecular mapping'' technique based on  cross-correlation between the IFS data elements and grid of models. We can summarize the main results as follows:

\begin{enumerate}
\item We find a \Teff=1560$\pm100$K, log g$\leq$4.40 dex, [M/H]=$0.05
^{+0.24}_{-0.22}$ dex, and C/O $\leq$0.55 from \texttt{ForMoSA}. The accuracy is  limited by the systematic uncertainties induced by the use of different models and by the bi-modal distribution of solutions found with the \texttt{BT-SETTL15} models. That composition is consistent with a formation of HIP~65426\,b closer to the host-star by core-accretion. We can not exclude a formation by gravitational instability at large separation, but the latter would imply the accretion of a substantial amount of solids. 

\item The molecular mapping as performed with \texttt{TExTRIS} yields results consistent with the ones from \texttt{ForMoSA} that validates the ability of the molecular mapping to characterize atmospheres. Nevertheless, this alternative method is limited by the loss of the spectral continuum information in the data that can shift the estimates of \Teff, and degrades the constraints on the other parameters (log(g), [M/H], C/O).

\item We estimated a radial velocity of 26$\pm$15 \kms\, for the planet  compatible with the revised radial velocity of the host-star. This highlights the usefulness of medium (to high) resolution spectrographs to test companionship.
\end{enumerate}

GRAVITY observations could soon drastically improve the knowledge of the orbital parameter of the planet and potentially draw a consistent picture of its formation. Observations with ERIS and CRIRES+ would also allow for improving the knowledge of the radial velocity and consolidate the constraints on the atmospheric parameters derived from our K-band data.

A new generation of integral field spectrographs successor of SINFONI (VLT-ERIS, GTC-FRIDA, \textit{JWST}-NIRSpec and MIRI, ELT-HARMONI and METIS) will soon observe at high-Strehls,  higher spectral resolution (up to $R_\lambda\sim$100\,000 for METIS) and  high-contrast  (GTC-FRIDA, HARMONI-high contrast arm; METIS).  All these instruments  belong to the same class of slicer-based IFS and present similar challenges in terms of data analysis. The techniques developed and explored as part of this study could be adapted to process these observations. 

\begin{acknowledgements}
 We dedicate this work to the memory of Dr. France Allard, a great source of inspiration for our community. We hope that this paper will honor her career, her research and her huge  contribution in the domain of the physics of atmospheres of stars, brown dwarfs and exoplanets. S.Pe, M.Bo, G.Ch, B.Ch., A-M.La, P.De and A.Bo, acknowledge support of the \textit{Programme National de Planétologie} through project grants ``ISEP'' and ``EXO-SPEC'' and the \textit{Agence Nationale de la Recherche} through grant ANR-14-CE33-0018. This research has benefited from the support of the IDEX cross-disciplinary project \textit{Origin of Life}, and the visiting program of the French Chilean Lab for Astronomy (FCLA, UMI-3886). A.Vi and M.Ho acknowledge funding from the European Research Council (ERC) under the European Union's Horizon 2020 research and innovation programme (grant agreement No. 757561). A.Wy acknowledges the financial support of the SNSF by grant number P400P2\_186765. G.-D.Ma acknowledges the support of the DFG priority program SPP~1992 ``Exploring the Diversity of Extrasolar Planets'' (KU~2849/7-1) and from the Swiss National Science Foundation under grant BSSGI0\_155816 ``PlanetsInTime''. Part of this work has been carried out within the framework of the National Centre of Competence in Research PlanetS supported by the Swiss National Science Foundation. F. Al acknowledges financial support from the {\sl ``Programme National de Physique Stellaire''} (PNPS) and the {\sl ``Programme National de Planétologie''} of CNRS/INSU, France. The computations of brown dwarf and exoplanet models were performed at the {\sl P\^ole Scientifique de Mod\'elisation Num\'erique} (PSMN) at the {\sl \'Ecole Normale Sup\'erieure} (ENS) in Lyon. The results presented in this research paper were obtained using the \texttt{matplotlib} and \texttt{astropy} libraries \citep{Hunter:2007, astropy:2018}.
\end{acknowledgements}

\begin{appendix}
\section{Robustness of the wavelength solution at the companion location}
\label{Appendix:A}
\begin{figure}[t]
  \centering
  \includegraphics[width=\columnwidth]{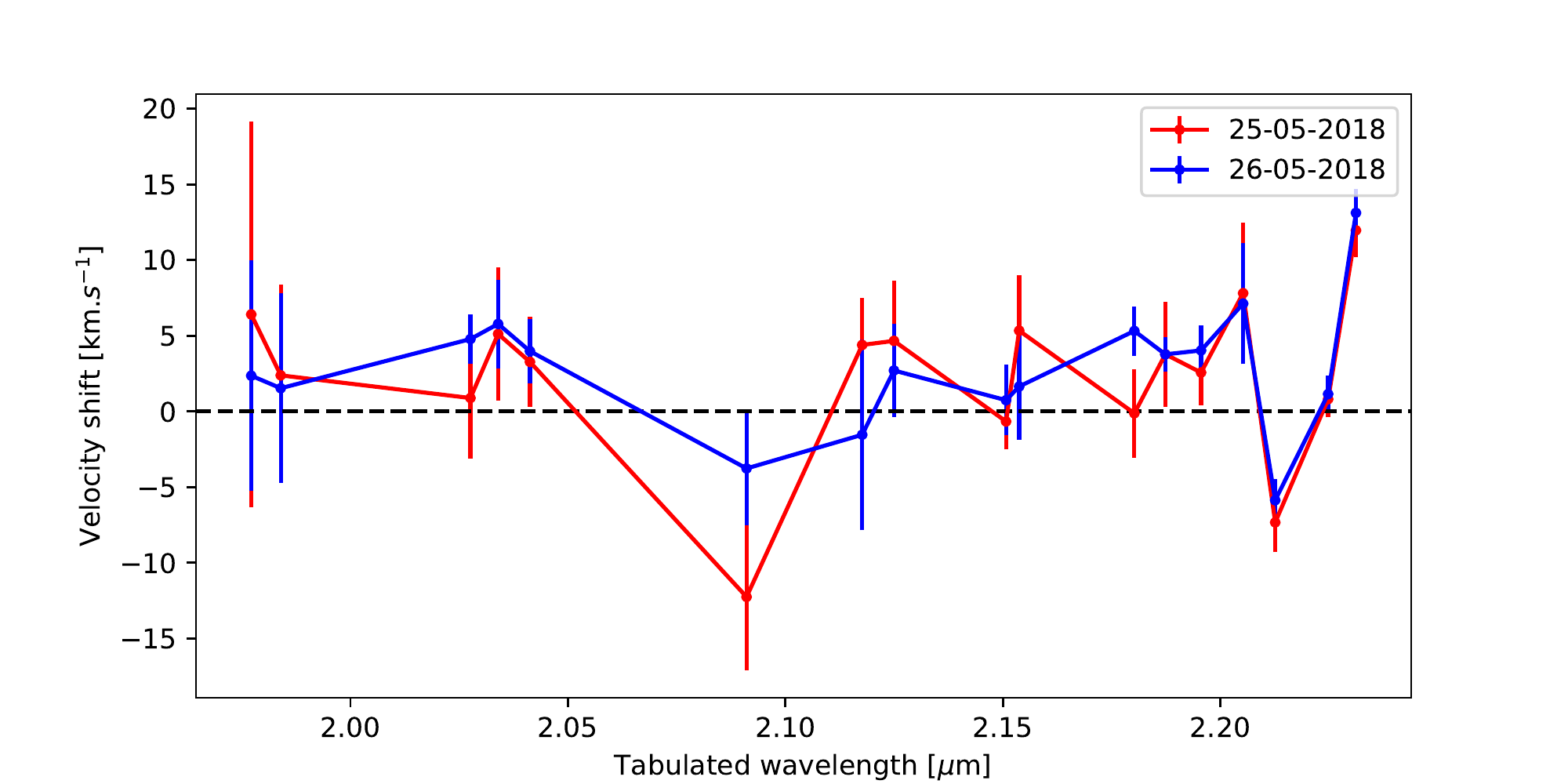} 
  \caption{Velocity shifts between the measured and tabulated positions (vacuum) of OH lines at the position of the companion in the stacked datacubes.}
  \label{Fig:recalOH}
\end{figure}

We checked the accuracy of the correction we applied on the wavelength solution of the instrument using OH$^{-}$ emission lines. Seventeen lines from 1.977 to 2.23 \mic\, are well detected in the stacked cubes (e.g., Figure \ref{Fig:detecSINF}) on both nights.  Their central wavelength at the companion location could therefore be determined with an average accuracy of 3.5 \kms\, using a Gaussian fit and be compared to reference values from \cite{2000A&A...354.1134R}. The residual shifts are reported in Figure \ref{Fig:recalOH} and oscillates around means of 2.3 and 2.8 \kms, for the May 25 and 26 data, respectively. These values and the measurement accuracy indicate no significant residual wavelength shift and validate locally our re-calibration based on telluric absorption. 

\section{Validating the extraction procedure}
\label{Appendix:extraction}
\begin{figure*}[h]
\centering
\begin{tabular}{cc}
\includegraphics[width=9cm]{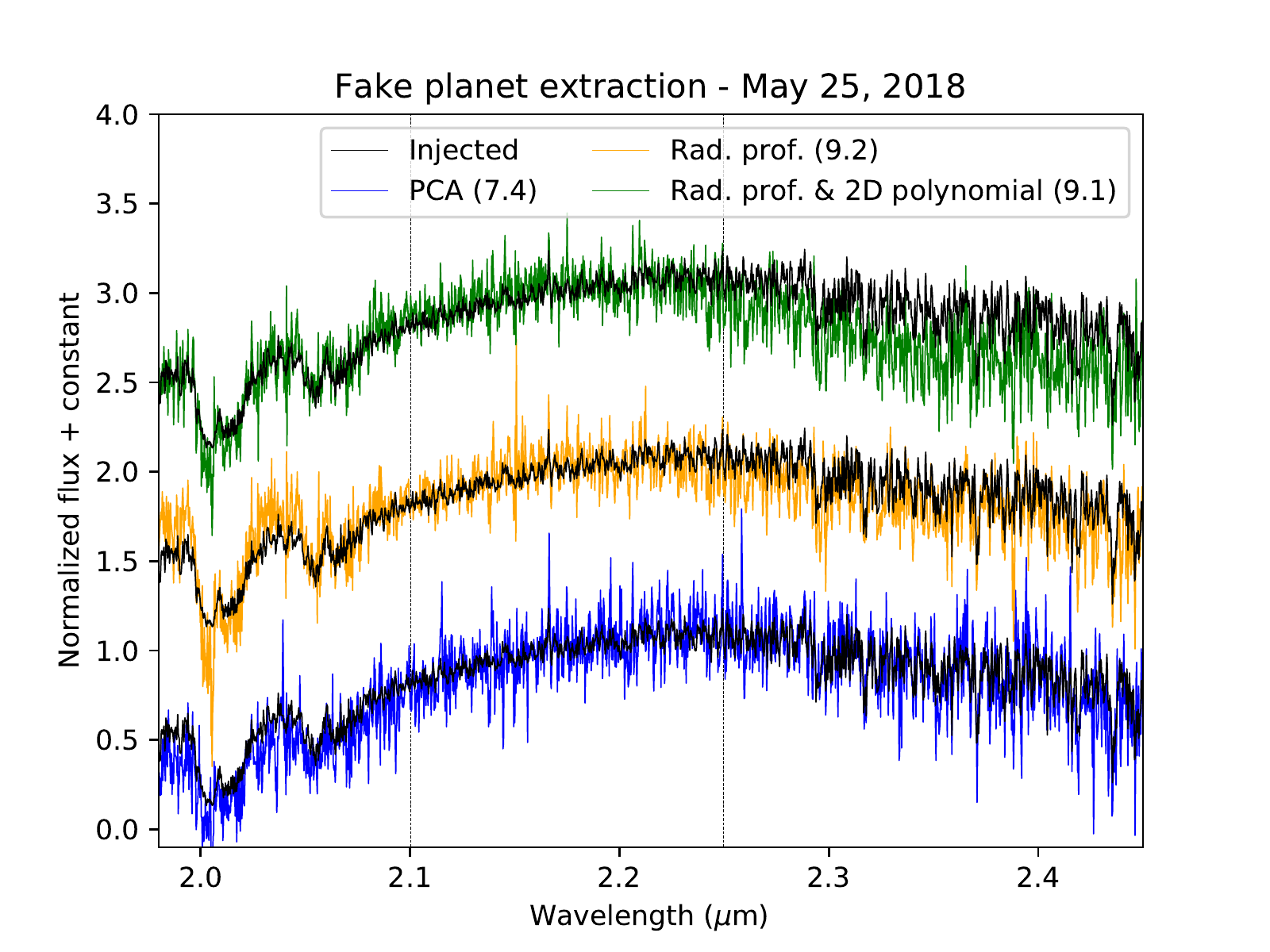} &   \includegraphics[width=9cm]{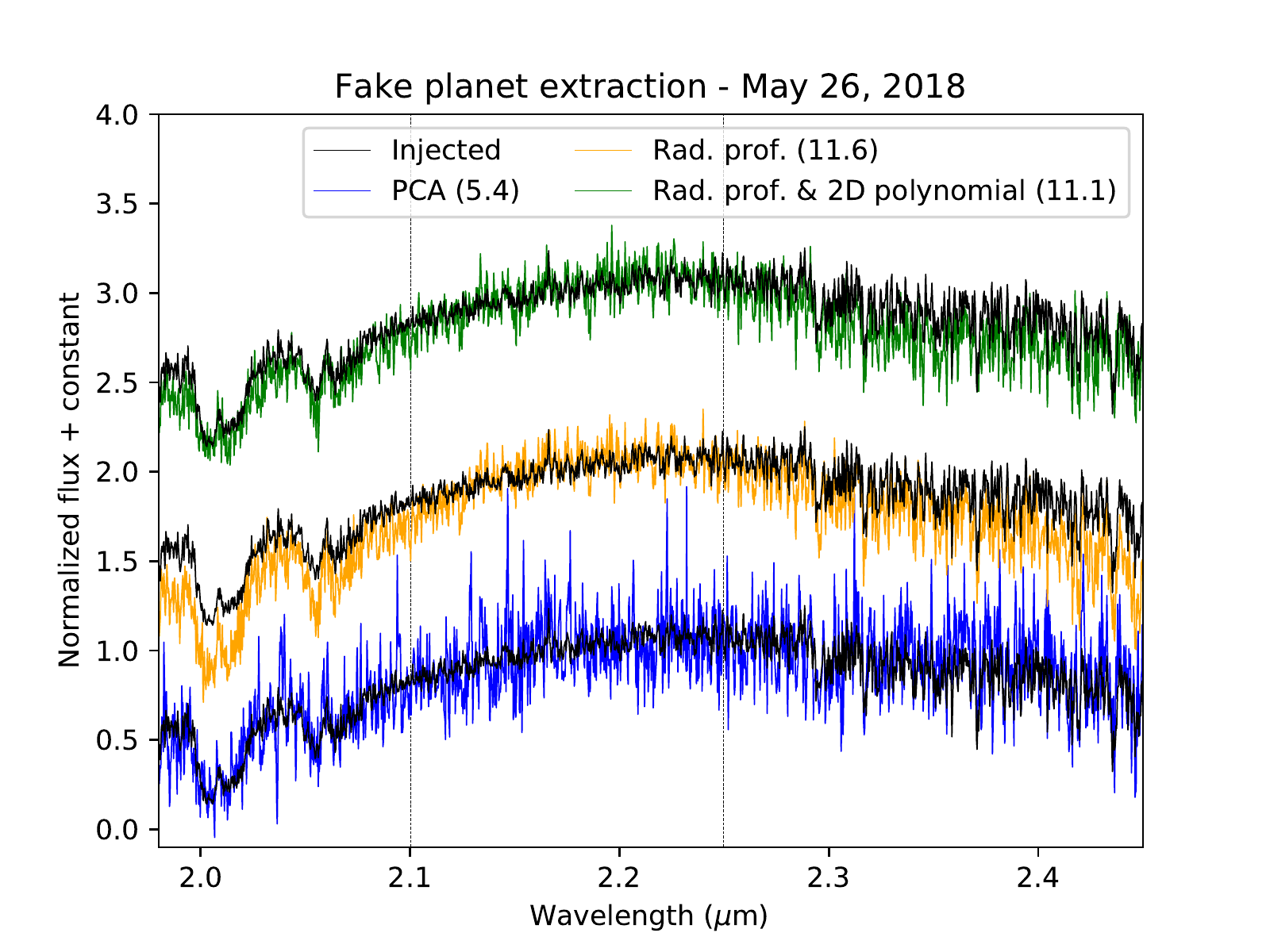} \\
\end{tabular}
\caption{Comparison of the injected (black lines) and extracted fake planet spectra (color lines) using the  methods described in Section \ref{sec:obs}: the principal-component analysis (blue), the removal of a circular profile at each wavelengths (orange), and the removal of a 2D polynomials at each wavelengths at the fake planet position (green). The signal-to-noise ratio of the recovered spectra is computed from 2.1 to 2.25 $\mu$m (dashed vertical lines) and given in parenthesis.} 
\label{Fig:extractionfp}
\end{figure*}

We tested the different extraction methods described in section \ref{sec:obs} injecting a fake planet with a K-band contrast of 10 magnitudes \citep[corresponding to the values reported in][]{2019A&A...622A..80C} and a position angle of 165.2$^{\circ}$. The fake planet spectrum was extracted within a 75mas wide circular aperture and compared to the injected signal. The results are shown for both nights in Figure \ref{Fig:extractionfp}.

The test confirms the ability of the PCA to conserve the  continuum shape. Conversely, the removal of the circular profile in the nADI cubes leaves  large-scale flux halo residuals which introduce a bias of the continuum at the fake planet location. The 2D-polynomials applied on these cubes remove these residuals efficiently at shorter wavelengths, but bias the pseudo-continuum longward of 2.3 \mic. We estimated the signal-to-noise (S/N) from the difference of the normalised extracted and injected spectra. The signal-to-noise is systematically lower using the PCA. We therefore conclude that the PCA conserves the planet continuum information, but is not optimised for accessing the molecular lines.

\begin{figure}[h]
\centering
\includegraphics[width=\columnwidth]{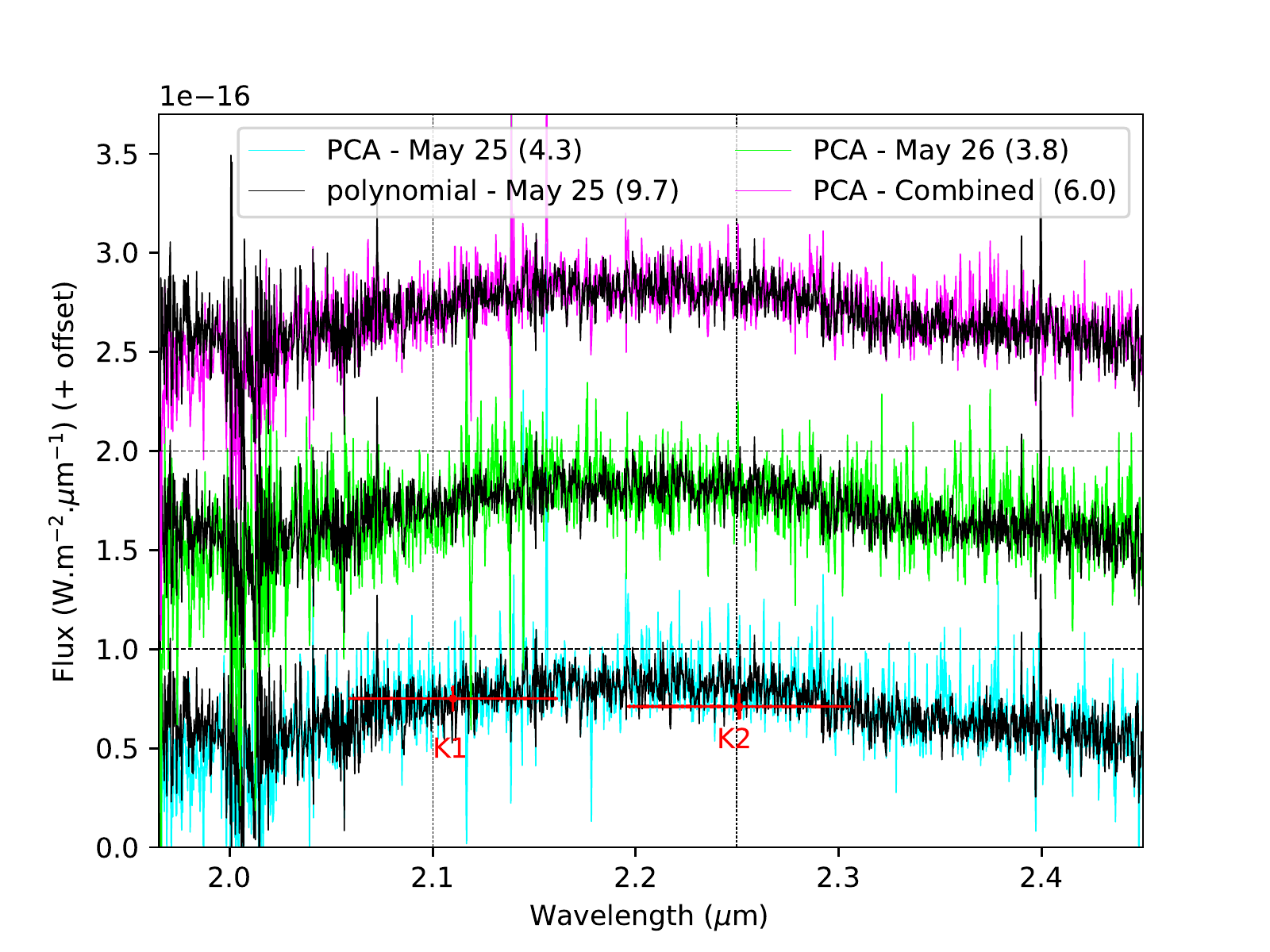}   
\caption{Flux-calibrated spectra of HIP~65426b obtained with the PCA on both nights and compared to the spectrum extracted from the nADI reduction of the May 25, 2018 data followed by a subtraction of a 2D polynomials. The signal-to-noise ratio (S/N) estimated from 2.1 to 2.25  $\mu$m  is given in parenthesis. The red datapoints correspond to the SPHERE K1 and K2 photometry from \cite{2019A&A...622A..80C}.} 
\label{Fig:extractionplanet}
\end{figure}

We show in Figure \ref{Fig:extractionplanet} the flux-calibrated spectrum of HIP~65426b. The PCA-extracted spectra obtained from the May 25 and 26 datacubes and the one extracted  from the May 25 nADI cubes show identical continua, confirming the later method works better at the planet position. The PCA spectrum is noisier than the nADI spectrum, in agreement with the conclusions based on fake planets. The combination of the two PCA-extracted spectra of HIP~65426b does not  reach the signal-to-noise obtained from the nADI/polynomial fit from the May 25 data. We therefore adopted the spectrum extracted from the nADI and polynomial fit.  The final flux-calibrated spectrum of HIP~65426b appears to be consistent with the SPHERE K-band photometry.

\begin{figure*}[h]
\centering
\begin{tabular}{cc}
\includegraphics[width=9cm]{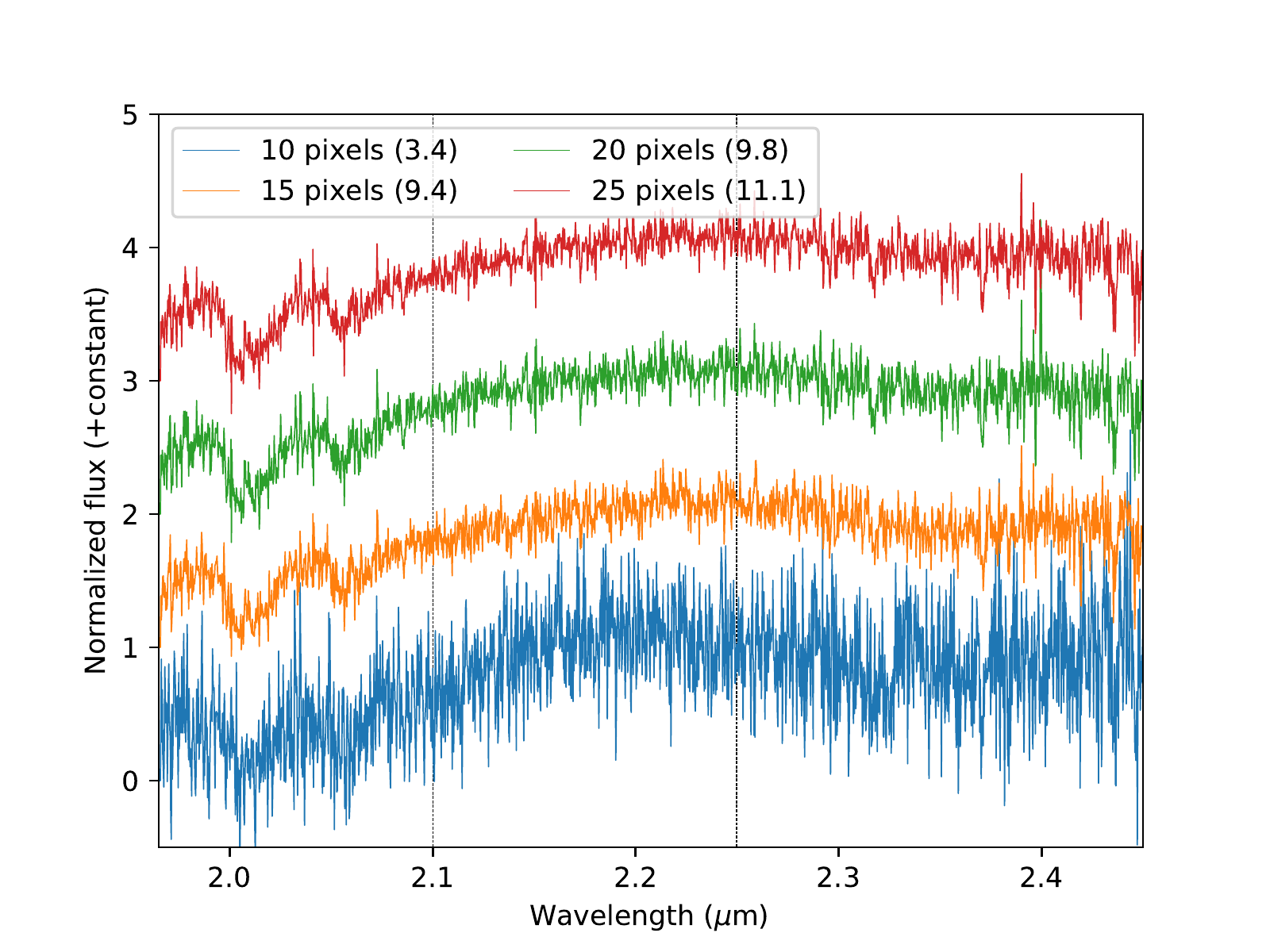} &   \includegraphics[width=9cm]{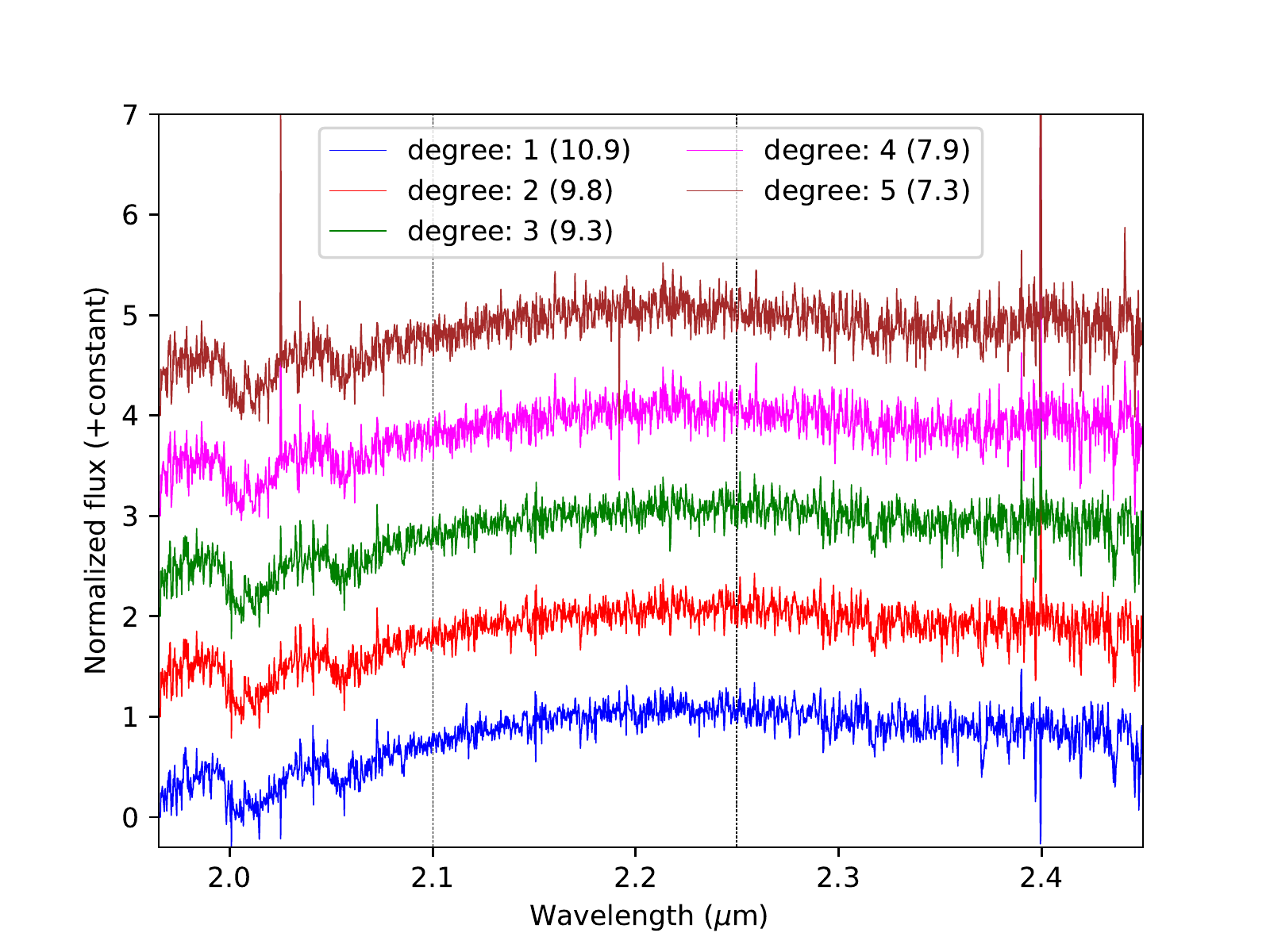} \\
\end{tabular}
\caption{Extracted spectra (with telluric features) of HIP~65426b for different box sizes (left) and degree (right) of the polynomial fit applied to the nADI data. The signal-to-noise ratio (S/N) of the spectra estimated from 2.1 to 2.25  $\mu$m  is given in parenthesis.}
\label{Fig:polyparams}
\end{figure*}

To conclude, we tested different square box sizes (10, 15, 20, 25 pixels width) as well as different degrees (1 to 5) of polynomials when applying the background correction to the nADI cubes. The results are shown in Figure \ref{Fig:polyparams} for the May 25, 2018 observations. Boxes smaller than 15 pixels do not contain enough spaxels to evaluate the background and lead to noisier spectra. Boxes larger than 20 pixels encompass artefacts at the edges of the field of view of SINFONI and were discarded. Square boxes of 15 and 20 pixels width yield similar results. We chose the later to allow for estimating the level of residuals within an annuli of 53mas width (e.g., a resolution element) centered on the planet. Polynomials of degree 2 to 5 produce spectra with identical continua. The S/N decreases when higher degrees are considered. Polynomials of degree 1 seem to optimise the S/N ratio. However, it leaves a higher level of residuals around the planet and does not subtract the continuum properly at shorter wavelengths. Therefore, we concluded that a box-size of 20 pixels and a polynomials of degree 2 is the best trade-off on these data.

\section{Detection of planets with different C/O ratio with the CO molecular template.}
\label{Appendix:C}


  \begin{figure*}[t]
  \centering
\begin{tabular}{cc}
  \includegraphics[width=8cm]{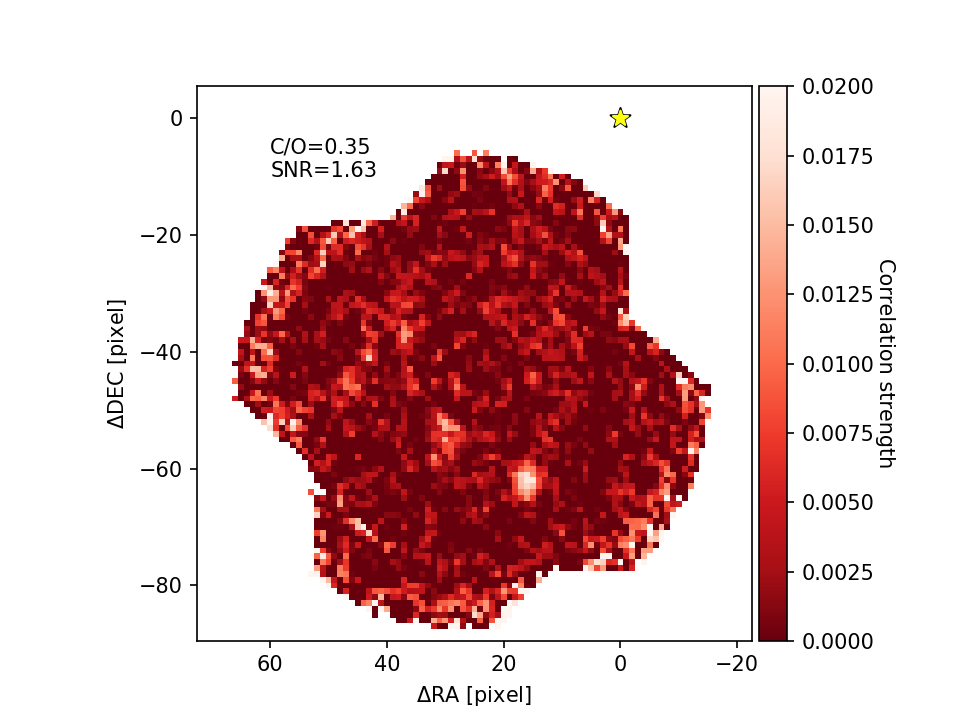}  &
  \includegraphics[width=8cm]{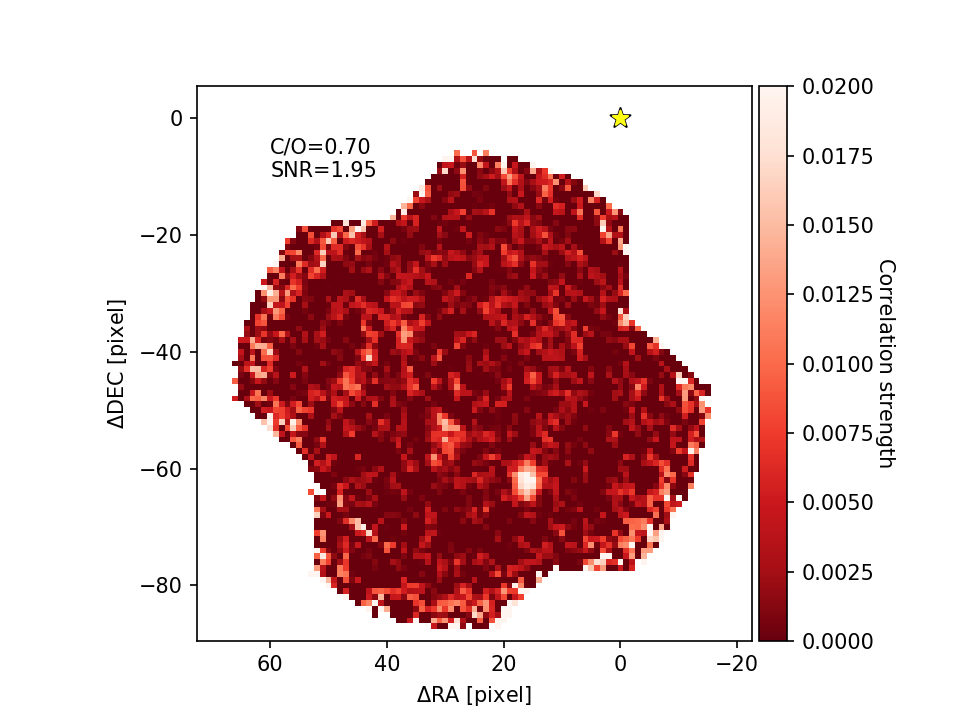} \\
  \includegraphics[width=8cm]{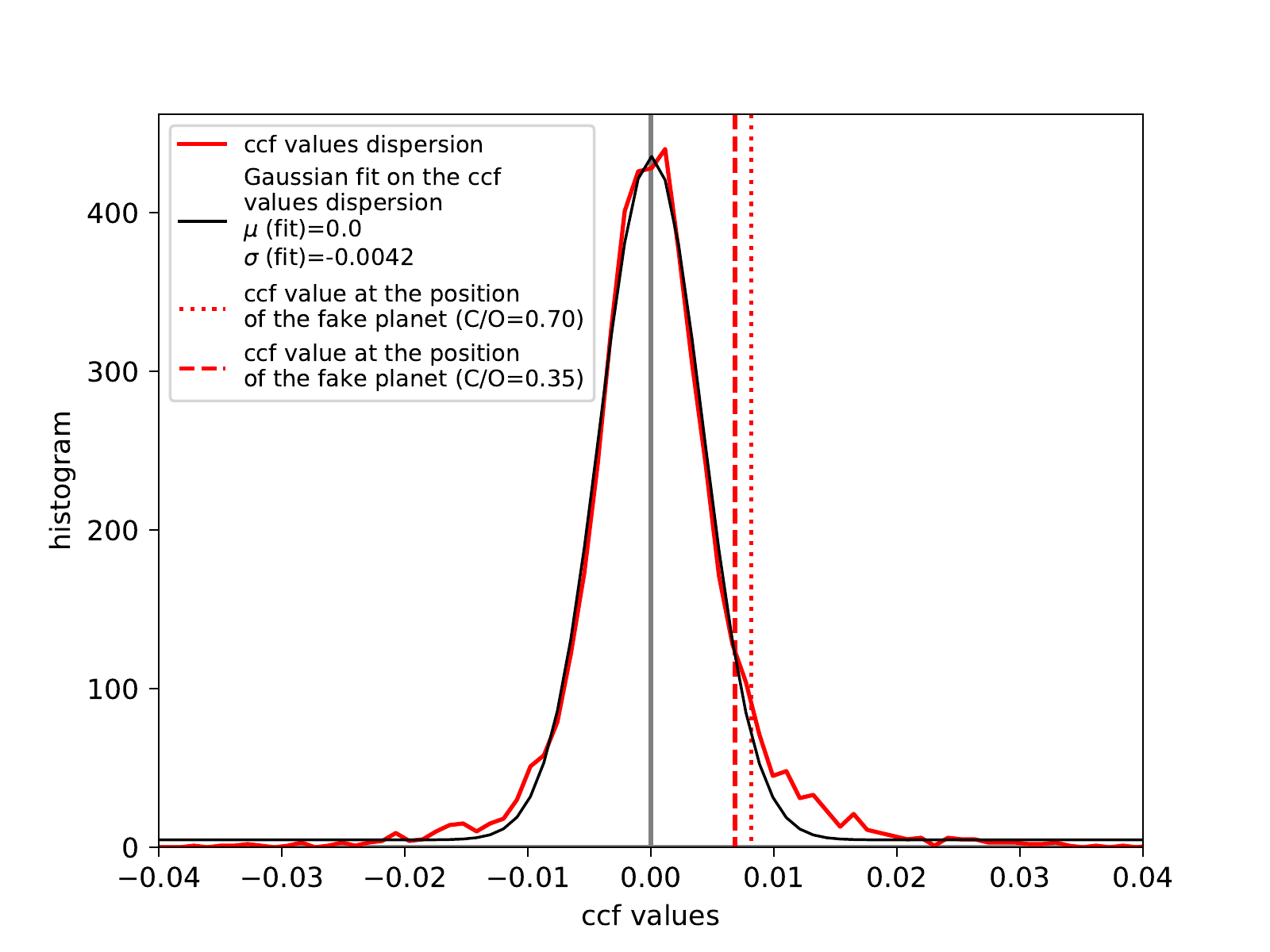} &  \includegraphics[width=8cm]{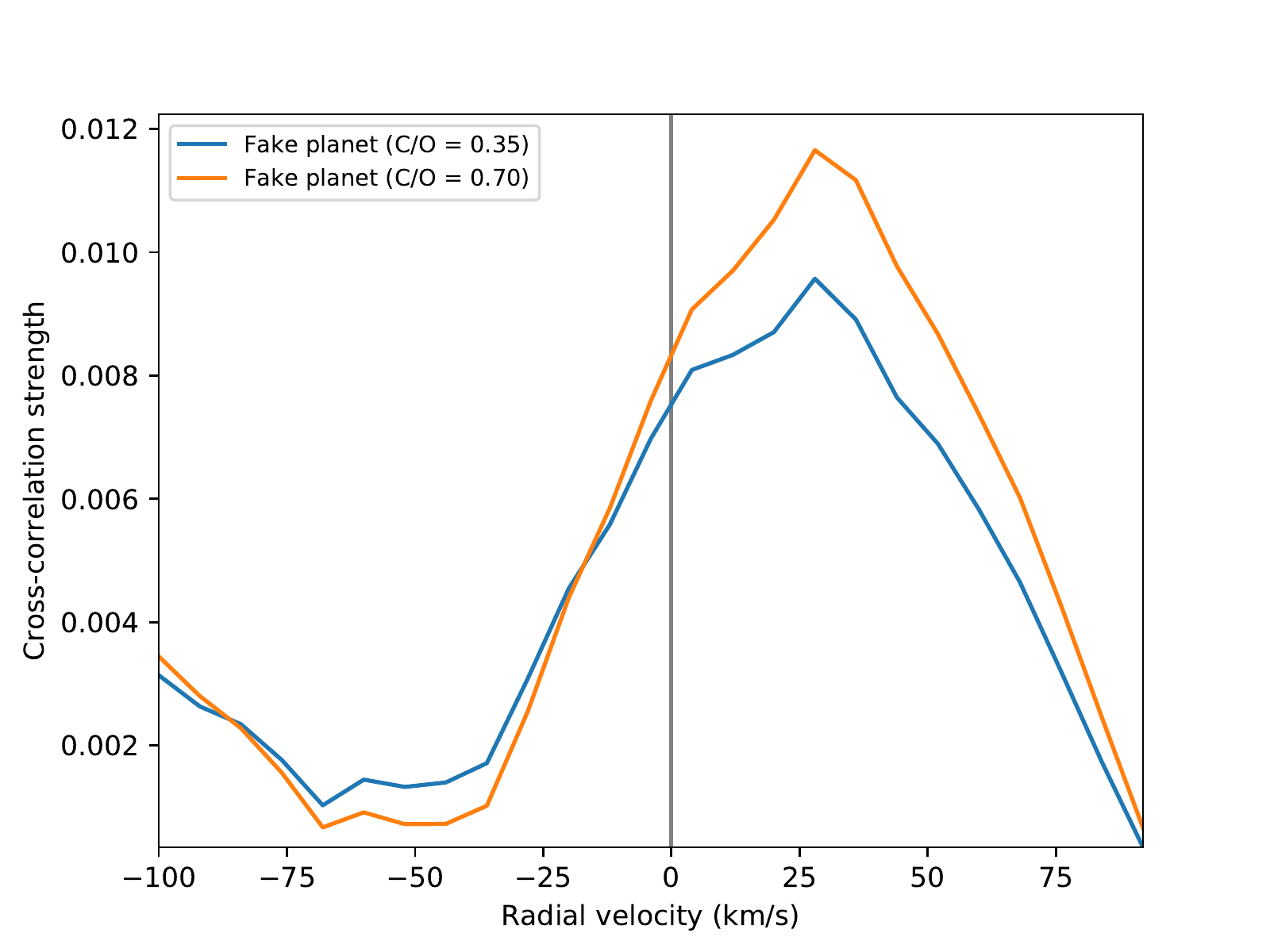}\\
  \end{tabular}
  \caption{\textit{Upper-Left and Right}: CO molecular maps obtained following the injection of a fake planet at $\rho$=812 mas and PA=165.2$^{\circ}$ in the original data-cubes using \texttt{Exo-REM} models  at \Teff=1650 K, log(g)=4.0 dex, [M/H]=0.0, and C/O=0.35 and 0.70 at a velocity of 30 \kms. \textit{Bottom-left} Histograms of CCF values at a velicity of 30 \kms. \textit{Bottom-right} Cross-correlation function obtained at the fake planet position.}
  \label{Fig:C_O_SNR_recap}
\end{figure*}

  \begin{figure*}[t]
  \centering
  \includegraphics[width=9cm]{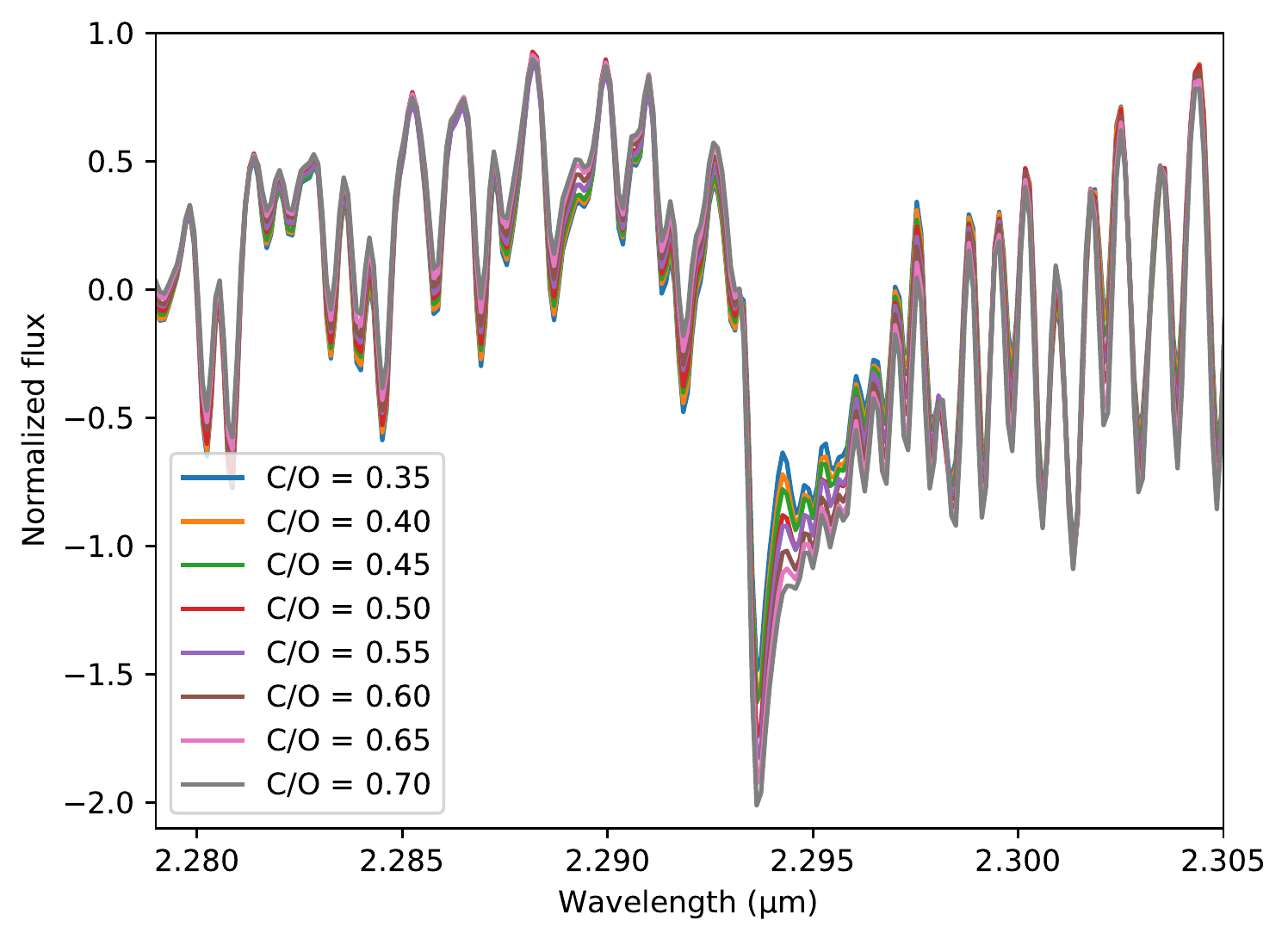} 
  \includegraphics[width=9cm]{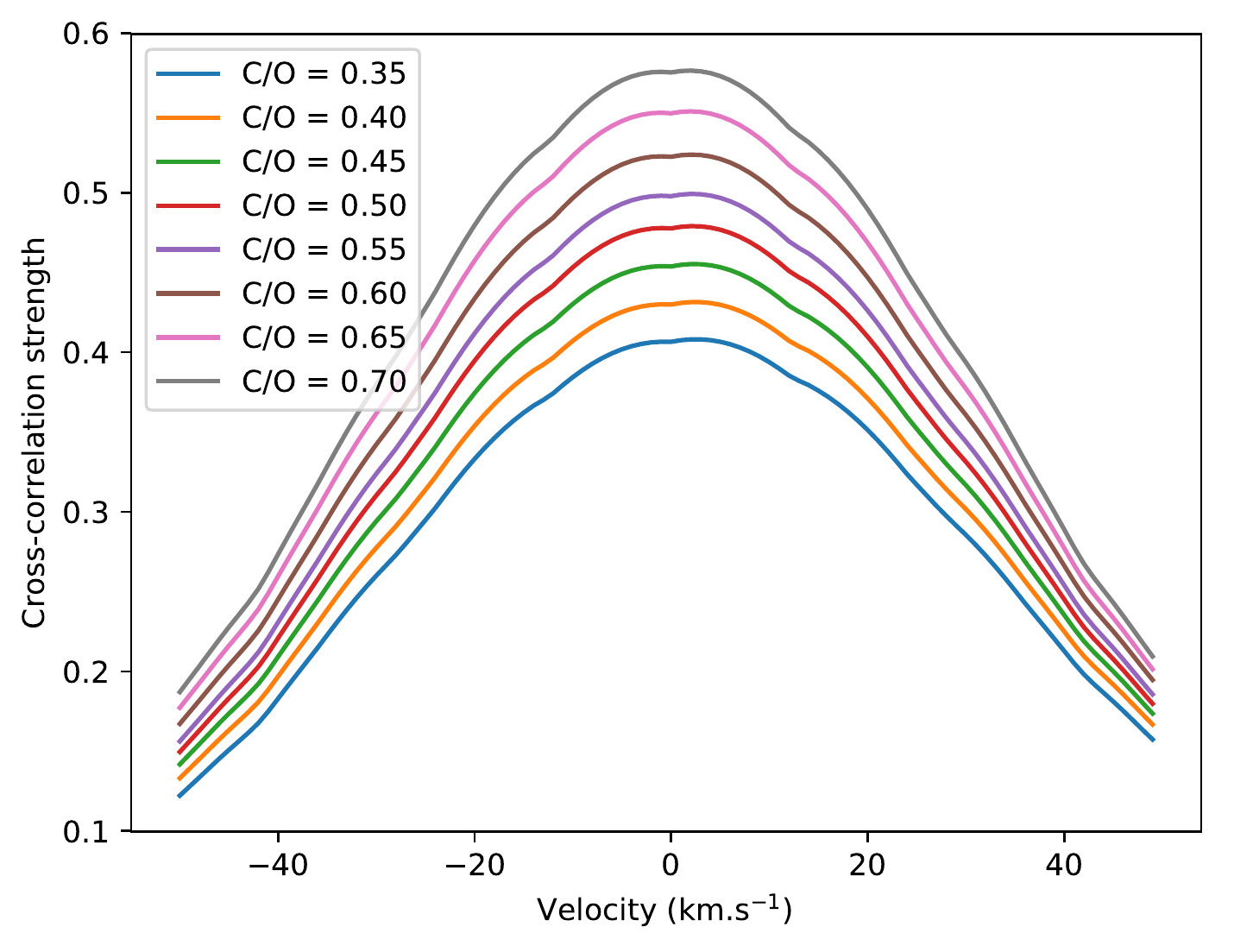} 
  \caption{\textit{Left} : Zoom on the $\nu=$2--0 CO overtone at 2.29 \mic~ of normalized continuum-subtracted \texttt{Exo-REM} models at different C/O ratios (\Teff=1650 K, log(g)=0.40 dex, [M/H]=0.0 and C/O=0.35-0.70). \textit{Right} : cross-correlation functions between the \texttt{Exo-REM} models and the CO template shown in Figure \ref{Fig:molmaptemplate}.}
  \label{Fig:C_O_overtones}
\end{figure*}

We investigated in the following the faint detection of HIP~65426b using the CO molecular template. We  injected fake planets (FP hereafter) into our datacube and compared the detection S/N into the CCF signal maps generated for each cases. The fake planets were injected at $\rho$=812 mas, PA=165.2$^{\circ}$, a broad-band contrast $\Delta_{K}$=10 mag, and a radial velocity RV=30 \kms. We used a noiseless synthetic spectrum generated with the \texttt{Exo-REM} model  corresponding to a \Teff=1650 K, a log(g)=4.0 dex, a [M/H]=0.0 and a C/O=0.35 and 0.70 for each FP respectively. The Figure \ref{Fig:C_O_SNR_recap} shows the cross-correlation function for each FP obtain following the method described in Section \ref{subsec:planetcorsig}. The signal is detected at a larger S/N than on for the real planet which might be due to the lack of noise in the injected spectral signature. The comparison however shows that planets with low C/O such as HIP~65426b are detected at a lower S/N than planets with super-solar C/O. Indeed, the \texttt{Exo-REM} models predict a less pronounced $\nu=$2--0 overtone, which is the strongest spectroscopic signature of $^{12}$CO in the spectrum (see also Fig. \ref{Fig:molmaptemplate}). This translates to a reduced cross-correlation signal (Figure \ref{Fig:C_O_overtones}). This may therefore explain in part the marginal detection of HIP~65426b with the CO molecular template.


\end{appendix}

\clearpage
\newpage
\bibliographystyle{aa}
\bibliography{ms}

\begin{thebibliography}{125}
\expandafter\ifx\csname natexlab\endcsname\relax\def\natexlab#1{#1}\fi

\bibitem[{{Abuter} {et~al.}(2006){Abuter}, {Schreiber}, {Eisenhauer}, {Ott},
  {Horrobin}, \& {Gillesen}}]{2006NewAR..50..398A}
{Abuter}, R., {Schreiber}, J., {Eisenhauer}, F., {et~al.} 2006, \nar, 50, 398

\bibitem[{{Ackerman} \& {Marley}(2001)}]{2001ApJ...556..872A}
{Ackerman}, A.~S. \& {Marley}, M.~S. 2001, \apj, 556, 872

\bibitem[{{Allard} {et~al.}(2001){Allard}, {Hauschildt}, {Alexander},
  {Tamanai}, \& {Schweitzer}}]{2001ApJ...556..357A}
{Allard}, F., {Hauschildt}, P.~H., {Alexander}, D.~R., {Tamanai}, A., \&
  {Schweitzer}, A. 2001, \apj, 556, 357

\bibitem[{{Allard} {et~al.}(2012){Allard}, {Homeier}, \&
  {Freytag}}]{2012RSPTA.370.2765A}
{Allard}, F., {Homeier}, D., \& {Freytag}, B. 2012, Philosophical Transactions
  of the Royal Society of London Series A, 370, 2765

\bibitem[{{Astropy Collaboration} {et~al.}(2018){Astropy Collaboration},
  {Price-Whelan}, {Sip{H{o}}cz}, {G{"u}nther}, {Lim}, {Crawford}, {Conseil},
  {Shupe}, {Craig}, {Dencheva}, {Ginsburg}, {Vand erPlas}, {Bradley},
  {P{'e}rez-Su{'a}rez}, {de Val-Borro}, {Aldcroft}, {Cruz}, {Robitaille},
  {Tollerud}, {Ardelean}, {Babej}, {Bach}, {Bachetti}, {Bakanov}, {Bamford},
  {Barentsen}, {Barmby}, {Baumbach}, {Berry}, {Biscani}, {Boquien}, {Bostroem},
  {Bouma}, {Brammer}, {Bray}, {Breytenbach}, {Buddelmeijer}, {Burke},
  {Calderone}, {Cano Rodr{'i}guez}, {Cara}, {Cardoso}, {Cheedella}, {Copin},
  {Corrales}, {Crichton}, {D'Avella}, {Deil}, {Depagne}, {Dietrich}, {Donath},
  {Droettboom}, {Earl}, {Erben}, {Fabbro}, {Ferreira}, {Finethy}, {Fox},
  {Garrison}, {Gibbons}, {Goldstein}, {Gommers}, {Greco}, {Greenfield},
  {Groener}, {Grollier}, {Hagen}, {Hirst}, {Homeier}, {Horton}, {Hosseinzadeh},
  {Hu}, {Hunkeler}, {Ivezi{'c}}, {Jain}, {Jenness}, {Kanarek}, {Kendrew},
  {Kern}, {Kerzendorf}, {Khvalko}, {King}, {Kirkby}, {Kulkarni}, {Kumar},
  {Lee}, {Lenz}, {Littlefair}, {Ma}, {Macleod}, {Mastropietro}, {McCully},
  {Montagnac}, {Morris}, {Mueller}, {Mumford}, {Muna}, {Murphy}, {Nelson},
  {Nguyen}, {Ninan}, {N{"o}the}, {Ogaz}, {Oh}, {Parejko}, {Parley}, {Pascual},
  {Patil}, {Patil}, {Plunkett}, {Prochaska}, {Rastogi}, {Reddy Janga},
  {Sabater}, {Sakurikar}, {Seifert}, {Sherbert}, {Sherwood-Taylor}, {Shih},
  {Sick}, {Silbiger}, {Singanamalla}, {Singer}, {Sladen}, {Sooley},
  {Sornarajah}, {Streicher}, {Teuben}, {Thomas}, {Tremblay}, {Turner},
  {Terr{'o}n}, {van Kerkwijk}, {de la Vega}, {Watkins}, {Weaver}, {Whitmore},
  {Woillez}, {Zabalza}, \& {Astropy Contributors}}]{astropy:2018}
{Astropy Collaboration}, {Price-Whelan}, A.~M., {Sip{H{o}}cz}, B.~M., {et~al.}
  2018, aj, 156, 123

\bibitem[{{Bailey} {et~al.}(2014){Bailey}, {Meshkat}, {Reiter}, {Morzinski},
  {Males}, {Su}, {Hinz}, {Kenworthy}, {Stark}, {Mamajek}, {Briguglio}, {Close},
  {Follette}, {Puglisi}, {Rodigas}, {Weinberger}, \&
  {Xompero}}]{2014ApJ...780L...4B}
{Bailey}, V., {Meshkat}, T., {Reiter}, M., {et~al.} 2014, \apjl, 780, L4

\bibitem[{{Baraffe} {et~al.}(2003){Baraffe}, {Chabrier}, {Barman}, {Allard}, \&
  {Hauschildt}}]{2003A&A...402..701B}
{Baraffe}, I., {Chabrier}, G., {Barman}, T.~S., {Allard}, F., \& {Hauschildt},
  P.~H. 2003, \aap, 402, 701

\bibitem[{{Barman} {et~al.}(2015){Barman}, {Konopacky}, {Macintosh}, \&
  {Marois}}]{2015ApJ...804...61B}
{Barman}, T.~S., {Konopacky}, Q.~M., {Macintosh}, B., \& {Marois}, C. 2015,
  \apj, 804, 61

\bibitem[{{Barman} {et~al.}(2011){Barman}, {Macintosh}, {Konopacky}, \&
  {Marois}}]{2011ApJ...733...65B}
{Barman}, T.~S., {Macintosh}, B., {Konopacky}, Q.~M., \& {Marois}, C. 2011,
  \apj, 733, 65

\bibitem[{{Bayo} {et~al.}(2008){Bayo}, {Rodrigo}, {Barrado Y Navascu{\'e}s},
  {Solano}, {Guti{\'e}rrez}, {Morales-Calder{\'o}n}, \&
  {Allard}}]{2008A&A...492..277B}
{Bayo}, A., {Rodrigo}, C., {Barrado Y Navascu{\'e}s}, D., {et~al.} 2008, \aap,
  492, 277

\bibitem[{{Becklin} \& {Zuckerman}(1988)}]{1988Natur.336..656B}
{Becklin}, E.~E. \& {Zuckerman}, B. 1988, \nat, 336, 656

\bibitem[{{Beuzit} {et~al.}(2019){Beuzit}, {Vigan}, {Mouillet}, {Dohlen},
  {Gratton}, {Boccaletti}, {Sauvage}, {Schmid}, {Langlois}, {Petit},
  {Baruffolo}, {Feldt}, {Milli}, {Wahhaj}, {Abe}, {Anselmi}, {Antichi},
  {Barette}, {Baudrand}, {Baudoz}, {Bazzon}, {Bernardi}, {Blanchard}, {Brast},
  {Bruno}, {Buey}, {Carbillet}, {Carle}, {Cascone}, {Chapron}, {Charton},
  {Chauvin}, {Claudi}, {Costille}, {De Caprio}, {de Boer}, {Delboulb{\'e}},
  {Desidera}, {Dominik}, {Downing}, {Dupuis}, {Fabron}, {Fantinel}, {Farisato},
  {Feautrier}, {Fedrigo}, {Fusco}, {Gigan}, {Ginski}, {Girard}, {Giro},
  {Gisler}, {Gluck}, {Gry}, {Henning}, {Hubin}, {Hugot}, {Incorvaia}, {Jaquet},
  {Kasper}, {Lagadec}, {Lagrange}, {Le Coroller}, {Le Mignant}, {Le Ruyet},
  {Lessio}, {Lizon}, {Llored}, {Lundin}, {Madec}, {Magnard}, {Marteaud},
  {Martinez}, {Maurel}, {M{\'e}nard}, {Mesa}, {M{\"o}ller-Nilsson}, {Moulin},
  {Moutou}, {Orign{\'e}}, {Parisot}, {Pavlov}, {Perret}, {Pragt}, {Puget},
  {Rabou}, {Ramos}, {Reess}, {Rigal}, {Rochat}, {Roelfsema}, {Rousset}, {Roux},
  {Saisse}, {Salasnich}, {Santambrogio}, {Scuderi}, {Segransan}, {Sevin},
  {Siebenmorgen}, {Soenke}, {Stadler}, {Suarez}, {Tiph{\`e}ne}, {Turatto},
  {Udry}, {Vakili}, {Waters}, {Weber}, {Wildi}, {Zins}, \&
  {Zurlo}}]{2019A&A...631A.155B}
{Beuzit}, J.~L., {Vigan}, A., {Mouillet}, D., {et~al.} 2019, \aap, 631, A155

\bibitem[{{Boley} {et~al.}(2011){Boley}, {Helled}, \&
  {Payne}}]{2011ApJ...735...30B}
{Boley}, A.~C., {Helled}, R., \& {Payne}, M.~J. 2011, \apj, 735, 30

\bibitem[{{Bonnefoy} {et~al.}(2013){Bonnefoy}, {Boccaletti}, {Lagrange},
  {Allard}, {Mordasini}, {Beust}, {Chauvin}, {Girard}, {Homeier}, {Apai},
  {Lacour}, \& {Rouan}}]{2013A&A...555A.107B}
{Bonnefoy}, M., {Boccaletti}, A., {Lagrange}, A.~M., {et~al.} 2013, \aap, 555,
  A107

\bibitem[{{Bonnefoy} {et~al.}(2014){Bonnefoy}, {Chauvin}, {Lagrange}, {Rojo},
  {Allard}, {Pinte}, {Dumas}, \& {Homeier}}]{2014A&A...562A.127B}
{Bonnefoy}, M., {Chauvin}, G., {Lagrange}, A.~M., {et~al.} 2014, \aap, 562,
  A127

\bibitem[{{Bonnefoy} {et~al.}(2016){Bonnefoy}, {Zurlo}, {Baudino}, {Lucas},
  {Mesa}, {Maire}, {Vigan}, {Galicher}, {Homeier}, {Marocco}, {Gratton},
  {Chauvin}, {Allard}, {Desidera}, {Kasper}, {Moutou}, {Lagrange}, {Antichi},
  {Baruffolo}, {Baudrand }, {Beuzit}, {Boccaletti}, {Cantalloube}, {Carbillet},
  {Charton}, {Claudi}, {Costille}, {Dohlen}, {Dominik}, {Fantinel},
  {Feautrier}, {Feldt}, {Fusco}, {Gigan}, {Girard}, {Gluck}, {Gry}, {Henning},
  {Janson}, {Langlois}, {Madec}, {Magnard}, {Maurel}, {Mawet}, {Meyer},
  {Milli}, {Moeller-Nilsson}, {Mouillet}, {Pavlov}, {Perret}, {Pujet}, {Quanz},
  {Rochat}, {Rousset}, {Roux}, {Salasnich}, {Salter}, {Sauvage}, {Schmid},
  {Sevin}, {Soenke}, {Stadler}, {Turatto}, {Udry}, {Vakili}, {Wahhaj}, \&
  {Wildi}}]{2016A&A...587A..58B}
{Bonnefoy}, M., {Zurlo}, A., {Baudino}, J.~L., {et~al.} 2016, \aap, 587, A58

\bibitem[{{Bonnet} {et~al.}(2004){Bonnet}, {Abuter}, {Baker}, {Bornemann},
  {Brown}, {Castillo}, {Conzelmann}, {Damster}, {Davies}, {Delabre},
  {Donaldson}, {Dumas}, {Eisenhauer}, {Elswijk}, {Fedrigo}, {Finger},
  {Gemperlein}, {Genzel}, {Gilbert}, {Gillet}, {Goldbrunner}, {Horrobin}, {Ter
  Horst}, {Huber}, {Hubin}, {Iserlohe}, {Kaufer}, {Kissler-Patig}, {Kragt},
  {Kroes}, {Lehnert}, {Lieb}, {Liske}, {Lizon}, {Lutz}, {Modigliani}, {Monnet},
  {Nesvadba}, {Patig}, {Pragt}, {Reunanen}, {R{\"o}hrle}, {Rossi}, {Schmutzer},
  {Schoenmaker}, {Schreiber}, {Stroebele}, {Szeifert}, {Tacconi}, {Tecza},
  {Thatte}, {Tordo}, {van der Werf}, \& {Weisz}}]{2004Msngr.117...17B}
{Bonnet}, H., {Abuter}, R., {Baker}, A., {et~al.} 2004, The Messenger, 117, 17

\bibitem[{{Bowler} \& {Hillenbrand}(2015)}]{2015ApJ...811L..30B}
{Bowler}, B.~P. \& {Hillenbrand}, L.~A. 2015, \apjl, 811, L30

\bibitem[{{Bowler} {et~al.}(2017){Bowler}, {Kraus}, {Bryan}, {Knutson},
  {Brogi}, {Rizzuto}, {Mace}, {Vanderburg}, {Liu}, {Hillenbrand}, \&
  {Cieza}}]{2017AJ....154..165B}
{Bowler}, B.~P., {Kraus}, A.~L., {Bryan}, M.~L., {et~al.} 2017, \aj, 154, 165

\bibitem[{{Bowler} {et~al.}(2010){Bowler}, {Liu}, {Dupuy}, \&
  {Cushing}}]{2010ApJ...723..850B}
{Bowler}, B.~P., {Liu}, M.~C., {Dupuy}, T.~J., \& {Cushing}, M.~C. 2010, \apj,
  723, 850

\bibitem[{{Bowler} {et~al.}(2011){Bowler}, {Liu}, {Kraus}, {Mann}, \&
  {Ireland}}]{2011ApJ...743..148B}
{Bowler}, B.~P., {Liu}, M.~C., {Kraus}, A.~L., {Mann}, A.~W., \& {Ireland},
  M.~J. 2011, \apj, 743, 148

\bibitem[{Briechle \& Hanebeck(2001)}]{10.1117/12.421129}
Briechle, K. \& Hanebeck, U.~D. 2001, in Optical Pattern Recognition XII, ed.
  D.~P. Casasent \& T.-H. Chao, Vol. 4387, International Society for Optics and
  Photonics (SPIE), 95 -- 102

\bibitem[{{Brogi} \& {Line}(2019)}]{2019AJ....157..114B}
{Brogi}, M. \& {Line}, M.~R. 2019, \aj, 157, 114

\bibitem[{{Bubar} {et~al.}(2011){Bubar}, {Schaeuble}, {King}, {Mamajek}, \&
  {Stauffer}}]{2011AJ....142..180B}
{Bubar}, E.~J., {Schaeuble}, M., {King}, J.~R., {Mamajek}, E.~E., \&
  {Stauffer}, J.~R. 2011, \aj, 142, 180

\bibitem[{{Carlotti} {et~al.}(2018){Carlotti}, {H{\'e}nault}, {Dohlen},
  {Sauvage}, {Rabou}, {Magnard}, {Vigan}, {Mouillet}, {Chauvin}, {Vola},
  {Bonnefoy}, {Fusco}, {El Hadi}, {Thatte}, {Clarke}, {Tecza}, {Bryson},
  {Schnetler}, \& {V{\'e}rinaud}}]{2018SPIE10702E..9NC}
{Carlotti}, A., {H{\'e}nault}, F., {Dohlen}, K., {et~al.} 2018, in Society of
  Photo-Optical Instrumentation Engineers (SPIE) Conference Series, Vol. 10702,
  \procspie, 107029N

\bibitem[{{Chabrier} {et~al.}(2000){Chabrier}, {Baraffe}, {Allard}, \&
  {Hauschildt}}]{2000ApJ...542..464C}
{Chabrier}, G., {Baraffe}, I., {Allard}, F., \& {Hauschildt}, P. 2000, \apj,
  542, 464

\bibitem[{{Charnay} {et~al.}(2018){Charnay}, {B{\'e}zard}, {Baudino},
  {Bonnefoy}, {Boccaletti}, \& {Galicher}}]{2018ApJ...854..172C}
{Charnay}, B., {B{\'e}zard}, B., {Baudino}, J.~L., {et~al.} 2018, \apj, 854,
  172

\bibitem[{{Chauvin} {et~al.}(2017){Chauvin}, {Desidera}, {Lagrange}, {Vigan},
  {Gratton}, {Langlois}, {Bonnefoy}, {Beuzit}, {Feldt}, {Mouillet}, {Meyer},
  {Cheetham}, {Biller}, {Boccaletti}, {D'Orazi}, {Galicher}, {Hagelberg},
  {Maire}, {Mesa}, {Olofsson}, {Samland}, {Schmidt}, {Sissa}, {Bonavita},
  {Charnay}, {Cudel}, {Daemgen}, {Delorme}, {Janin-Potiron}, {Janson},
  {Keppler}, {Le Coroller}, {Ligi}, {Marleau}, {Messina}, {Molli{\`e}re},
  {Mordasini}, {M{\"u}ller}, {Peretti}, {Perrot}, {Rodet}, {Rouan}, {Zurlo},
  {Dominik}, {Henning}, {Menard}, {Schmid}, {Turatto}, {Udry}, {Vakili}, {Abe},
  {Antichi}, {Baruffolo}, {Baudoz}, {Baudrand}, {Blanchard}, {Bazzon}, {Buey},
  {Carbillet}, {Carle}, {Charton}, {Cascone}, {Claudi}, {Costille}, {Deboulbe},
  {De Caprio}, {Dohlen}, {Fantinel}, {Feautrier}, {Fusco}, {Gigan}, {Giro},
  {Gisler}, {Gluck}, {Hubin}, {Hugot}, {Jaquet}, {Kasper}, {Madec}, {Magnard},
  {Martinez}, {Maurel}, {Le Mignant}, {M{\"o}ller-Nilsson}, {Llored}, {Moulin},
  {Orign{\'e}}, {Pavlov}, {Perret}, {Petit}, {Pragt}, {Puget}, {Rabou},
  {Ramos}, {Rigal}, {Rochat}, {Roelfsema}, {Rousset}, {Roux}, {Salasnich},
  {Sauvage}, {Sevin}, {Soenke}, {Stadler}, {Suarez}, {Weber}, {Wildi},
  {Antoniucci}, {Augereau}, {Baudino}, {Brandner}, {Engler}, {Girard}, {Gry},
  {Kral}, {Kopytova}, {Lagadec}, {Milli}, {Moutou}, {Schlieder},
  {Szul{\'a}gyi}, {Thalmann}, \& {Wahhaj}}]{2017A&A...605L...9C}
{Chauvin}, G., {Desidera}, S., {Lagrange}, A.~M., {et~al.} 2017, \aap, 605, L9

\bibitem[{{Chauvin} {et~al.}(2018){Chauvin}, {Gratton}, {Bonnefoy}, {Lagrange},
  {de Boer}, {Vigan}, {Beust}, {Lazzoni}, {Boccaletti}, {Galicher}, {Desidera},
  {Delorme}, {Keppler}, {Lannier}, {Maire}, {Mesa}, {Meunier}, {Kral},
  {Henning}, {Menard}, {Moor}, {Avenhaus}, {Bazzon}, {Janson}, {Beuzit},
  {Bhowmik}, {Bonavita}, {Borgniet}, {Brandner}, {Cheetham}, {Cudel}, {Feldt},
  {Fontanive}, {Ginski}, {Hagelberg}, {Janin-Potiron}, {Lagadec}, {Langlois},
  {Le Coroller}, {Messina}, {Meyer}, {Mouillet}, {Peretti}, {Perrot}, {Rodet},
  {Samland}, {Sissa}, {Olofsson}, {Salter}, {Schmidt}, {Zurlo}, {Milli}, {van
  Boekel}, {Quanz}, {Feautrier}, {Le Mignant}, {Perret}, {Ramos}, \&
  {Rochat}}]{2018A&A...617A..76C}
{Chauvin}, G., {Gratton}, R., {Bonnefoy}, M., {et~al.} 2018, \aap, 617, A76

\bibitem[{{Chauvin} {et~al.}(2005){Chauvin}, {Lagrange}, {Zuckerman}, {Dumas},
  {Mouillet}, {Song}, {Beuzit}, {Lowrance}, \& {Bessell}}]{2005A&A...438L..29C}
{Chauvin}, G., {Lagrange}, A.~M., {Zuckerman}, B., {et~al.} 2005, \aap, 438,
  L29

\bibitem[{{Cheetham} {et~al.}(2019){Cheetham}, {Samland}, {Brems}, {Launhardt},
  {Chauvin}, {S{\'e}gransan}, {Henning}, {Quirrenbach}, {Avenhaus}, {Cugno},
  {Girard}, {Godoy}, {Kennedy}, {Maire}, {Metchev}, {M{\"u}ller}, {Musso
  Barcucci}, {Olofsson}, {Pepe}, {Quanz}, {Queloz}, {Reffert}, {Rickman}, {van
  Boekel}, {Boccaletti}, {Bonnefoy}, {Cantalloube}, {Charnay}, {Delorme},
  {Janson}, {Keppler}, {Lagrange}, {Langlois}, {Lazzoni}, {Menard}, {Mesa},
  {Meyer}, {Schmidt}, {Sissa}, {Udry}, \& {Zurlo}}]{2019A&A...622A..80C}
{Cheetham}, A.~C., {Samland}, M., {Brems}, S.~S., {et~al.} 2019, \aap, 622, A80

\bibitem[{{Chilcote} {et~al.}(2017){Chilcote}, {Pueyo}, {De Rosa}, {Vargas},
  {Macintosh}, {Bailey}, {Barman}, {Bauman}, {Bruzzone}, {Bulger}, {Burrows},
  {Cardwell}, {Chen}, {Cotten}, {Dillon}, {Doyon}, {Draper}, {Duch{\^e}ne},
  {Dunn}, {Erikson}, {Fitzgerald}, {Follette}, {Gavel}, {Goodsell}, {Graham},
  {Greenbaum}, {Hartung}, {Hibon}, {Hung}, {Ingraham}, {Kalas}, {Konopacky},
  {Larkin}, {Maire}, {Marchis}, {Marley}, {Marois}, {Metchev},
  {Millar-Blanchaer}, {Morzinski}, {Nielsen}, {Norton}, {Oppenheimer},
  {Palmer}, {Patience}, {Perrin}, {Poyneer}, {Rajan}, {Rameau},
  {Rantakyr{\"o}}, {Sadakuni}, {Saddlemyer}, {Savransky}, {Schneider}, {Serio},
  {Sivaramakrishnan}, {Song}, {Soummer}, {Thomas}, {Wallace}, {Wang},
  {Ward-Duong}, {Wiktorowicz}, \& {Wolff}}]{2017AJ....153..182C}
{Chilcote}, J., {Pueyo}, L., {De Rosa}, R.~J., {et~al.} 2017, \aj, 153, 182

\bibitem[{{Christiaens} {et~al.}(2019){Christiaens}, {Cantalloube}, {Casassus},
  {Price}, {Absil}, {Pinte}, {Girard}, \& {Montesinos}}]{2019ApJ...877L..33C}
{Christiaens}, V., {Cantalloube}, F., {Casassus}, S., {et~al.} 2019, \apjl,
  877, L33

\bibitem[{{Cieza} {et~al.}(2016){Cieza}, {Casassus}, {Tobin}, {Bos},
  {Williams}, {Perez}, {Zhu}, {Caceres}, {Canovas}, {Dunham}, {Hales},
  {Prieto}, {Principe}, {Schreiber}, {Ruiz-Rodriguez}, \&
  {Zurlo}}]{2016Natur.535..258C}
{Cieza}, L.~A., {Casassus}, S., {Tobin}, J., {et~al.} 2016, \nat, 535, 258

\bibitem[{{Cutri} \& {et al.}(2012)}]{2012yCat.2311....0C}
{Cutri}, R.~M. \& {et al.} 2012, VizieR Online Data Catalog, II/311

\bibitem[{{Cutri} {et~al.}(2003){Cutri}, {Skrutskie}, {van Dyk}, {Beichman},
  {Carpenter}, {Chester}, {Cambresy}, {Evans}, {Fowler}, {Gizis}, {Howard},
  {Huchra}, {Jarrett}, {Kopan}, {Kirkpatrick}, {Light}, {Marsh}, {McCallon},
  {Schneider}, {Stiening}, {Sykes}, {Weinberg}, {Wheaton}, {Wheelock}, \&
  {Zacarias}}]{2003yCat.2246....0C}
{Cutri}, R.~M., {Skrutskie}, M.~F., {van Dyk}, S., {et~al.} 2003, VizieR Online
  Data Catalog, II/246

\bibitem[{{Daemgen} {et~al.}(2017){Daemgen}, {Todorov}, {Quanz}, {Meyer},
  {Mordasini}, {Marleau}, \& {Fortney}}]{2017A&A...608A..71D}
{Daemgen}, S., {Todorov}, K., {Quanz}, S.~P., {et~al.} 2017, \aap, 608, A71

\bibitem[{{Davies}(2007)}]{2007MNRAS.375.1099D}
{Davies}, R.~I. 2007, \mnras, 375, 1099

\bibitem[{{De Rosa} {et~al.}(2016){De Rosa}, {Rameau}, {Patience}, {Graham},
  {Doyon}, {Lafreni{\`e}re}, {Macintosh}, {Pueyo}, {Rajan}, {Wang},
  {Ward-Duong}, {Hung}, {Maire}, {Nielsen}, {Ammons}, {Bulger}, {Cardwell},
  {Chilcote}, {Galvez}, {Gerard}, {Goodsell}, {Hartung}, {Hibon}, {Ingraham},
  {Johnson-Groh}, {Kalas}, {Konopacky}, {Marchis}, {Marois}, {Metchev},
  {Morzinski}, {Oppenheimer}, {Perrin}, {Rantakyr{\"o}}, {Savransky}, \&
  {Thomas}}]{2016ApJ...824..121D}
{De Rosa}, R.~J., {Rameau}, J., {Patience}, J., {et~al.} 2016, \apj, 824, 121

\bibitem[{{de Zeeuw} {et~al.}(1999){de Zeeuw}, {Hoogerwerf}, {de Bruijne},
  {Brown}, \& {Blaauw}}]{1999AJ....117..354D}
{de Zeeuw}, P.~T., {Hoogerwerf}, R., {de Bruijne}, J.~H.~J., {Brown}, A.~G.~A.,
  \& {Blaauw}, A. 1999, \aj, 117, 354

\bibitem[{{Delorme} {et~al.}(2017){Delorme}, {Schmidt}, {Bonnefoy}, {Desidera},
  {Ginski}, {Charnay}, {Lazzoni}, {Christiaens}, {Messina}, {D'Orazi}, {Milli},
  {Schlieder}, {Gratton}, {Rodet}, {Lagrange}, {Absil}, {Vigan}, {Galicher},
  {Hagelberg}, {Bonavita}, {Lavie}, {Zurlo}, {Olofsson}, {Boccaletti},
  {Cantalloube}, {Mouillet}, {Chauvin}, {Hambsch}, {Langlois}, {Udry},
  {Henning}, {Beuzit}, {Mordasini}, {Lucas}, {Marocco}, {Biller}, {Carson},
  {Cheetham}, {Covino}, {De Caprio}, {Delboulbe}, {Feldt}, {Girard}, {Hubin},
  {Maire}, {Pavlov}, {Petit}, {Rouan}, {Roelfsema}, \&
  {Wildi}}]{2017A&A...608A..79D}
{Delorme}, P., {Schmidt}, T., {Bonnefoy}, M., {et~al.} 2017, \aap, 608, A79

\bibitem[{{DENIS Consortium}(2005)}]{2005yCat.2263....0D}
{DENIS Consortium}. 2005, VizieR Online Data Catalog, II/263

\bibitem[{{Dullemond} {et~al.}(2020){Dullemond}, {Isella}, {Andrews},
  {Skobleva}, \& {Dzyurkevich}}]{2020A&A...633A.137D}
{Dullemond}, C.~P., {Isella}, A., {Andrews}, S.~M., {Skobleva}, I., \&
  {Dzyurkevich}, N. 2020, \aap, 633, A137

\bibitem[{{Eisenhauer} {et~al.}(2003){Eisenhauer}, {Abuter}, {Bickert},
  {Biancat-Marchet}, {Bonnet}, {Brynnel}, {Conzelmann}, {Delabre}, {Donaldson},
  {Farinato}, {Fedrigo}, {Genzel}, {Hubin}, {Iserlohe}, {Kasper},
  {Kissler-Patig}, {Monnet}, {Roehrle}, {Schreiber}, {Stroebele}, {Tecza},
  {Thatte}, \& {Weisz}}]{2003SPIE.4841.1548E}
{Eisenhauer}, F., {Abuter}, R., {Bickert}, K., {et~al.} 2003, Society of
  Photo-Optical Instrumentation Engineers (SPIE) Conference Series, Vol. 4841,
  {SINFONI - Integral field spectroscopy at 50 milli-arcsecond resolution with
  the ESO VLT}, ed. M.~{Iye} \& A.~F.~M. {Moorwood}, 1548--1561

\bibitem[{{Eistrup} {et~al.}(2018){Eistrup}, {Walsh}, \& {van
  Dishoeck}}]{2018A&A...613A..14E}
{Eistrup}, C., {Walsh}, C., \& {van Dishoeck}, E.~F. 2018, \aap, 613, A14

\bibitem[{{Freudling} {et~al.}(2013){Freudling}, {Romaniello}, {Bramich},
  {Ballester}, {Forchi}, {Garc{\'{\i}}a-Dabl{\'o}}, {Moehler}, \&
  {Neeser}}]{2013A&A...559A..96F}
{Freudling}, W., {Romaniello}, M., {Bramich}, D.~M., {et~al.} 2013, \aap, 559,
  A96

\bibitem[{{Gagn{\'e}} {et~al.}(2018){Gagn{\'e}}, {Mamajek}, {Malo}, {Riedel},
  {Rodriguez}, {Lafreni{\`e}re}, {Faherty}, {Roy-Loubier}, {Pueyo}, {Robin}, \&
  {Doyon}}]{2018ApJ...856...23G}
{Gagn{\'e}}, J., {Mamajek}, E.~E., {Malo}, L., {et~al.} 2018, \apj, 856, 23

\bibitem[{{Gaia Collaboration} {et~al.}(2016){Gaia Collaboration}, {Brown},
  {Vallenari}, {Prusti}, {de Bruijne}, {Mignard}, {Drimmel}, {Babusiaux},
  {Bailer-Jones}, {Bastian}, {Biermann}, {Evans}, {Eyer}, {Jansen}, {Jordi},
  {Katz}, {Klioner}, {Lammers}, {Lindegren}, {Luri}, {O'Mullane}, {Panem},
  {Pourbaix}, {Randich}, {Sartoretti}, {Siddiqui}, {Soubiran}, {Valette}, {van
  Leeuwen}, {Walton}, {Aerts}, {Arenou}, {Cropper}, {H{\o}g}, {Lattanzi},
  {Grebel}, {Holland}, {Huc}, {Passot}, {Perryman}, {Bramante}, {Cacciari},
  {Casta{\~n}eda}, {Chaoul}, {Cheek}, {De Angeli}, {Fabricius}, {Guerra},
  {Hern{\'a}ndez}, {Jean-Antoine-Piccolo}, {Masana}, {Messineo}, {Mowlavi},
  {Nienartowicz}, {Ord{\'o}{\~n}ez-Blanco}, {Panuzzo}, {Portell}, {Richards},
  {Riello}, {Seabroke}, {Tanga}, {Th{\'e}venin}, {Torra}, {Els},
  {Gracia-Abril}, {Comoretto}, {Garcia-Reinaldos}, {Lock}, {Mercier},
  {Altmann}, {Andrae}, {Astraatmadja}, {Bellas-Velidis}, {Benson}, {Berthier},
  {Blomme}, {Busso}, {Carry}, {Cellino}, {Clementini}, {Cowell}, {Creevey},
  {Cuypers}, {Davidson}, {De Ridder}, {de Torres}, {Delchambre}, {Dell'Oro},
  {Ducourant}, {Fr{\'e}mat}, {Garc{\'\i}a-Torres}, {Gosset}, {Halbwachs},
  {Hambly}, {Harrison}, {Hauser}, {Hestroffer}, {Hodgkin}, {Huckle}, {Hutton},
  {Jasniewicz}, {Jordan}, {Kontizas}, {Korn}, {Lanzafame}, {Manteiga},
  {Moitinho}, {Muinonen}, {Osinde}, {Pancino}, {Pauwels}, {Petit},
  {Recio-Blanco}, {Robin}, {Sarro}, {Siopis}, {Smith}, {Smith}, {Sozzetti},
  {Thuillot}, {van Reeven}, {Viala}, {Abbas}, {Abreu Aramburu}, {Accart},
  {Aguado}, {Allan}, {Allasia}, {Altavilla}, {{\'A}lvarez}, {Alves},
  {Anderson}, {Andrei}, {Anglada Varela}, {Antiche}, {Antoja}, {Ant{\'o}n},
  {Arcay}, {Bach}, {Baker}, {Balaguer-N{\'u}{\~n}ez}, {Barache}, {Barata},
  {Barbier}, {Barblan}, {Barrado y Navascu{\'e}s}, {Barros}, {Barstow},
  {Becciani}, {Bellazzini}, {Bello Garc{\'\i}a}, {Belokurov}, {Bendjoya},
  {Berihuete}, {Bianchi}, {Bienaym{\'e}}, {Billebaud}, {Blagorodnova},
  {Blanco-Cuaresma}, {Boch}, {Bombrun}, {Borrachero}, {Bouquillon}, {Bourda},
  {Bouy}, {Bragaglia}, {Breddels}, {Brouillet}, {Br{\"u}semeister},
  {Bucciarelli}, {Burgess}, {Burgon}, {Burlacu}, {Busonero}, {Buzzi}, {Caffau},
  {Cambras}, {Campbell}, {Cancelliere}, {Cantat-Gaudin}, {Carlucci},
  {Carrasco}, {Castellani}, {Charlot}, {Charnas}, {Chiavassa}, {Clotet},
  {Cocozza}, {Collins}, {Costigan}, {Crifo}, {Cross}, {Crosta}, {Crowley},
  {Dafonte}, {Damerdji}, {Dapergolas}, {David}, {David}, {De Cat}, {de Felice},
  {de Laverny}, {De Luise}, {De March}, {de Martino}, {de Souza}, {Debosscher},
  {del Pozo}, {Delbo}, {Delgado}, {Delgado}, {Di Matteo}, {Diakite},
  {Distefano}, {Dolding}, {Dos Anjos}, {Drazinos}, {Duran}, {Dzigan},
  {Edvardsson}, {Enke}, {Evans}, {Eynard Bontemps}, {Fabre}, {Fabrizio},
  {Faigler}, {Falc{\~a}o}, {Farr{\`a}s Casas}, {Federici}, {Fedorets},
  {Fern{\'a}ndez-Hern{\'a}ndez}, {Fernique}, {Fienga}, {Figueras}, {Filippi},
  {Findeisen}, {Fonti}, {Fouesneau}, {Fraile}, {Fraser}, {Fuchs}, {Gai},
  {Galleti}, {Galluccio}, {Garabato}, {Garc{\'\i}a-Sedano}, {Garofalo},
  {Garralda}, {Gavras}, {Gerssen}, {Geyer}, {Gilmore}, {Girona}, {Giuffrida},
  {Gomes}, {Gonz{\'a}lez-Marcos}, {Gonz{\'a}lez-N{\'u}{\~n}ez},
  {Gonz{\'a}lez-Vidal}, {Granvik}, {Guerrier}, {Guillout}, {Guiraud},
  {G{\'u}rpide}, {Guti{\'e}rrez-S{\'a}nchez}, {Guy}, {Haigron},
  {Hatzidimitriou}, {Haywood}, {Heiter}, {Helmi}, {Hobbs}, {Hofmann}, {Holl},
  {Holland }, {Hunt}, {Hypki}, {Icardi}, {Irwin}, {Jevardat de Fombelle},
  {Jofr{\'e}}, {Jonker}, {Jorissen}, {Julbe}, {Karampelas}, {Kochoska},
  {Kohley}, {Kolenberg}, {Kontizas}, {Koposov}, {Kordopatis}, {Koubsky},
  {Krone-Martins}, {Kudryashova}, {Kull}, {Bachchan}, {Lacoste-Seris}, {Lanza},
  {Lavigne}, {Le Poncin-Lafitte}, {Lebreton}, {Lebzelter}, {Leccia}, {Leclerc},
  {Lecoeur-Taibi}, {Lemaitre}, {Lenhardt}, {Leroux}, {Liao}, {Licata},
  {Lindstr{\o}m}, {Lister}, {Livanou}, {Lobel}, {L{\"o}ffler}, {L{\'o}pez},
  {Lorenz}, {MacDonald}, {Magalh{\~a}es Fernandes}, {Managau}, {Mann},
  {Mantelet}, {Marchal}, {Marchant}, {Marconi}, {Marinoni}, {Marrese},
  {Marschalk{\'o}}, {Marshall}, {Mart{\'\i}n-Fleitas}, {Martino}, {Mary},
  {Matijevi{\v{c}}}, {Mazeh}, {McMillan}, {Messina}, {Michalik}, {Millar},
  {Mirand a}, {Molina}, {Molinaro}, {Molinaro}, {Moln{\'a}r}, {Moniez},
  {Montegriffo}, {Mor}, {Mora}, {Morbidelli}, {Morel}, {Morgenthaler},
  {Morris}, {Mulone}, {Muraveva}, {Musella}, {Narbonne}, {Nelemans},
  {Nicastro}, {Noval}, {Ord{\'e}novic}, {Ordieres-Mer{\'e}}, {Osborne},
  {Pagani}, {Pagano}, {Pailler}, {Palacin}, {Palaversa}, {Parsons}, {Pecoraro},
  {Pedrosa}, {Pentik{\"a}inen}, {Pichon}, {Piersimoni}, {Pineau}, {Plachy},
  {Plum}, {Poujoulet}, {Pr{\v{s}}a}, {Pulone}, {Ragaini}, {Rago}, {Rambaux},
  {Ramos-Lerate}, {Ranalli}, {Rauw}, {Read}, {Regibo}, {Reyl{\'e}}, {Ribeiro},
  {Rimoldini}, {Ripepi}, {Riva}, {Rixon}, {Roelens}, {Romero-G{\'o}mez},
  {Rowell}, {Royer}, {Ruiz-Dern}, {Sadowski}, {Sagrist{\`a} Sell{\'e}s},
  {Sahlmann}, {Salgado}, {Salguero}, {Sarasso}, {Savietto}, {Schultheis},
  {Sciacca}, {Segol}, {Segovia}, {Segransan}, {Shih}, {Smareglia}, {Smart},
  {Solano}, {Solitro}, {Sordo}, {Soria Nieto}, {Souchay}, {Spagna}, {Spoto},
  {Stampa}, {Steele}, {Steidelm{\"u}ller}, {Stephenson}, {Stoev}, {Suess},
  {S{\"u}veges}, {Surdej}, {Szabados}, {Szegedi-Elek}, {Tapiador}, {Taris},
  {Tauran}, {Taylor}, {Teixeira}, {Terrett}, {Tingley}, {Trager}, {Turon},
  {Ulla}, {Utrilla}, {Valentini}, {van Elteren}, {Van Hemelryck}, {van
  Leeuwen}, {Varadi}, {Vecchiato}, {Veljanoski}, {Via}, {Vicente}, {Vogt},
  {Voss}, {Votruba}, {Voutsinas}, {Walmsley}, {Weiler}, {Weingrill}, {Wevers},
  {Wyrzykowski}, {Yoldas}, {{\v{Z}}erjal}, {Zucker}, {Zurbach}, {Zwitter},
  {Alecu}, {Allen}, {Allende Prieto}, {Amorim}, {Anglada-Escud{\'e}},
  {Arsenijevic}, {Azaz}, {Balm}, {Beck}, {Bernstein}, {Bigot}, {Bijaoui},
  {Blasco}, {Bonfigli}, {Bono}, {Boudreault}, {Bressan}, {Brown}, {Brunet},
  {Bunclark}, {Buonanno}, {Butkevich}, {Carret}, {Carrion}, {Chemin},
  {Ch{\'e}reau}, {Corcione}, {Darmigny}, {de Boer}, {de Teodoro}, {de Zeeuw},
  {Delle Luche}, {Domingues}, {Dubath}, {Fodor}, {Fr{\'e}zouls}, {Fries},
  {Fustes}, {Fyfe}, {Gallardo}, {Gallegos}, {Gardiol}, {Gebran}, {Gomboc},
  {G{\'o}mez}, {Grux}, {Gueguen}, {Heyrovsky}, {Hoar}, {Iannicola}, {Isasi
  Parache}, {Janotto}, {Joliet}, {Jonckheere}, {Keil}, {Kim}, {Klagyivik},
  {Klar}, {Knude}, {Kochukhov}, {Kolka}, {Kos}, {Kutka}, {Lainey}, {LeBouquin},
  {Liu}, {Loreggia}, {Makarov}, {Marseille}, {Martayan}, {Martinez-Rubi},
  {Massart}, {Meynadier}, {Mignot}, {Munari}, {Nguyen}, {Nordlander}, {Ocvirk},
  {O'Flaherty}, {Olias Sanz}, {Ortiz}, {Osorio}, {Oszkiewicz}, {Ouzounis},
  {Palmer}, {Park}, {Pasquato}, {Peltzer}, {Peralta}, {P{\'e}turaud},
  {Pieniluoma}, {Pigozzi}, {Poels}, {Prat}, {Prod'homme}, {Raison}, {Rebordao},
  {Risquez}, {Rocca-Volmerange}, {Rosen}, {Ruiz-Fuertes}, {Russo}, {Sembay},
  {Serraller Vizcaino}, {Short}, {Siebert}, {Silva}, {Sinachopoulos}, {Slezak},
  {Soffel}, {Sosnowska}, {Strai{\v{z}}ys}, {ter Linden}, {Terrell}, {Theil},
  {Tiede}, {Troisi}, {Tsalmantza}, {Tur}, {Vaccari}, {Vachier}, {Valles}, {Van
  Hamme}, {Veltz}, {Virtanen}, {Wallut}, {Wichmann}, {Wilkinson}, {Ziaeepour},
  \& {Zschocke}}]{2016A&A...595A...2G}
{Gaia Collaboration}, {Brown}, A.~G.~A., {Vallenari}, A., {et~al.} 2016, \aap,
  595, A2

\bibitem[{Gonzalez {et~al.}(2017)Gonzalez, Wertz, Absil, Christiaens,
  Defr{\`{e}}re, Mawet, Milli, Absil, Droogenbroeck, Cantalloube, Hinz, Skemer,
  Karlsson, \& Surdej}]{Gomez_Gonzalez_2017}
Gonzalez, C. A.~G., Wertz, O., Absil, O., {et~al.} 2017, The Astronomical
  Journal, 154, 7

\bibitem[{{Gravity Collaboration} {et~al.}(2020){Gravity Collaboration},
  {Nowak}, {Lacour}, {Molli{\`e}re}, {Wang}, {Charnay}, {van Dishoeck},
  {Abuter}, {Amorim}, {Berger}, {Beust}, {Bonnefoy}, {Bonnet}, {Brandner},
  {Buron}, {Cantalloube}, {Collin}, {Chapron}, {Cl{\'e}net}, {Coud{\'e} Du
  Foresto}, {de Zeeuw}, {Dembet}, {Dexter}, {Duvert}, {Eckart}, {Eisenhauer},
  {F{\"o}rster Schreiber}, {F{\'e}dou}, {Garcia Lopez}, {Gao}, {Gendron},
  {Genzel}, {Gillessen}, {Hau{\ss}mann}, {Henning}, {Hippler}, {Hubert},
  {Jocou}, {Kervella}, {Lagrange}, {Lapeyr{\`e}re}, {Le Bouquin}, {L{\'e}na},
  {Maire}, {Ott}, {Paumard}, {Paladini}, {Perraut}, {Perrin}, {Pueyo}, {Pfuhl},
  {Rabien}, {Rau}, {Rodr{\'\i}guez-Coira}, {Rousset}, {Scheithauer},
  {Shangguan}, {Straub}, {Straubmeier}, {Sturm}, {Tacconi}, {Vincent},
  {Widmann}, {Wieprecht}, {Wiezorrek}, {Woillez}, {Yazici}, \&
  {Ziegler}}]{2020A&A...633A.110G}
{Gravity Collaboration}, {Nowak}, M., {Lacour}, S., {et~al.} 2020, \aap, 633,
  A110

\bibitem[{{Greenbaum} {et~al.}(2018){Greenbaum}, {Pueyo}, {Ruffio}, {Wang}, {De
  Rosa}, {Aguilar}, {Rameau}, {Barman}, {Marois}, {Marley}, {Konopacky},
  {Rajan}, {Macintosh}, {Ansdell}, {Arriaga}, {Bailey}, {Bulger}, {Burrows},
  {Chilcote}, {Cotten}, {Doyon}, {Duch{\^e}ne}, {Fitzgerald}, {Follette},
  {Gerard}, {Goodsell}, {Graham}, {Hibon}, {Hung}, {Ingraham}, {Kalas},
  {Larkin}, {Maire}, {Marchis}, {Metchev}, {Millar-Blanchaer}, {Nielsen},
  {Norton}, {Oppenheimer}, {Palmer}, {Patience}, {Perrin}, {Poyneer},
  {Rantakyr{\"o}}, {Savransky}, {Schneider}, {Sivaramakrishnan}, {Song},
  {Soummer}, {Thomas}, {Wallace}, {Ward-Duong}, {Wiktorowicz}, \&
  {Wolff}}]{2018AJ....155..226G}
{Greenbaum}, A.~Z., {Pueyo}, L., {Ruffio}, J.-B., {et~al.} 2018, \aj, 155, 226

\bibitem[{{Groff} {et~al.}(2015){Groff}, {Kasdin}, {Limbach}, {Galvin}, {Carr},
  {Knapp}, {Brand t}, {Loomis}, {Jarosik}, {Mede}, {McElwain}, {Leviton},
  {Miller}, {Quijada}, {Guyon}, {Jovanovic}, {Takato}, \&
  {Hayashi}}]{2015SPIE.9605E..1CG}
{Groff}, T.~D., {Kasdin}, N.~J., {Limbach}, M.~A., {et~al.} 2015, Society of
  Photo-Optical Instrumentation Engineers (SPIE) Conference Series, Vol. 9605,
  {The CHARIS IFS for high contrast imaging at Subaru}, 96051C

\bibitem[{{Hauschildt} {et~al.}(1997){Hauschildt}, {Baron}, \&
  {Allard}}]{1997ApJ...483..390H}
{Hauschildt}, P.~H., {Baron}, E., \& {Allard}, F. 1997, \apj, 483, 390

\bibitem[{{Helling} {et~al.}(2008){Helling}, {Ackerman}, {Allard}, {Dehn},
  {Hauschildt}, {Homeier}, {Lodders}, {Marley}, {Rietmeijer}, {Tsuji}, \&
  {Woitke}}]{2008MNRAS.391.1854H}
{Helling}, C., {Ackerman}, A., {Allard}, F., {et~al.} 2008, \mnras, 391, 1854

\bibitem[{{Helling} {et~al.}(2014){Helling}, {Woitke}, {Rimmer}, {Kamp}, {Thi},
  \& {Meijerink}}]{2014Life....4..142H}
{Helling}, C., {Woitke}, P., {Rimmer}, P.~B., {et~al.} 2014, Life, 4, 142

\bibitem[{{Hoeijmakers} {et~al.}(2018){Hoeijmakers}, {Schwarz}, {Snellen}, {de
  Kok}, {Bonnefoy}, {Chauvin}, {Lagrange}, \& {Girard}}]{2018A&A...617A.144H}
{Hoeijmakers}, H.~J., {Schwarz}, H., {Snellen}, I.~A.~G., {et~al.} 2018, \aap,
  617, A144

\bibitem[{{H{\o}g} {et~al.}(2000){H{\o}g}, {Fabricius}, {Makarov}, {Urban},
  {Corbin}, {Wycoff}, {Bastian}, {Schwekendiek}, \&
  {Wicenec}}]{2000A&A...355L..27H}
{H{\o}g}, E., {Fabricius}, C., {Makarov}, V.~V., {et~al.} 2000, \aap, 355, L27

\bibitem[{{Hubeny} \& {Burrows}(2007)}]{2007ApJ...669.1248H}
{Hubeny}, I. \& {Burrows}, A. 2007, \apj, 669, 1248

\bibitem[{Hunter(2007)}]{Hunter:2007}
Hunter, J.~D. 2007, Computing in Science \& Engineering, 9, 90

\bibitem[{{Janson} {et~al.}(2010){Janson}, {Bergfors}, {Goto}, {Brandner}, \&
  {Lafreni{\`e}re}}]{2010ApJ...710L..35J}
{Janson}, M., {Bergfors}, C., {Goto}, M., {Brandner}, W., \& {Lafreni{\`e}re},
  D. 2010, \apjl, 710, L35

\bibitem[{{Janson} {et~al.}(2008){Janson}, {Brandner}, \&
  {Henning}}]{2008A&A...478..597J}
{Janson}, M., {Brandner}, W., \& {Henning}, T. 2008, \aap, 478, 597

\bibitem[{{Jones, A.} {et~al.}(2013){Jones, A.}, {Noll, S.}, {Kausch, W.},
  {Szyszka, C.}, \& {Kimeswenger, S.}}]{refId1}
{Jones, A.}, {Noll, S.}, {Kausch, W.}, {Szyszka, C.}, \& {Kimeswenger, S.}
  2013, A\&A, 560, A91

\bibitem[{{Jovanovic} {et~al.}(2015){Jovanovic}, {Martinache}, {Guyon},
  {Clergeon}, {Singh}, {Kudo}, {Garrel}, {Newman}, {Doughty}, {Lozi}, {Males},
  {Minowa}, {Hayano}, {Takato}, {Morino}, {Kuhn}, {Serabyn}, {Norris},
  {Tuthill}, {Schworer}, {Stewart}, {Close}, {Huby}, {Perrin}, {Lacour},
  {Gauchet}, {Vievard}, {Murakami}, {Oshiyama}, {Baba}, {Matsuo}, {Nishikawa},
  {Tamura}, {Lai}, {Marchis}, {Duchene}, {Kotani}, \&
  {Woillez}}]{2015PASP..127..890J}
{Jovanovic}, N., {Martinache}, F., {Guyon}, O., {et~al.} 2015, \pasp, 127, 890

\bibitem[{{Kane} {et~al.}(1980){Kane}, {McKeith}, \&
  {Dufton}}]{1980A&A....84..115K}
{Kane}, L., {McKeith}, C.~D., \& {Dufton}, P.~L. 1980, \aap, 84, 115

\bibitem[{{Kanodia} \& {Wright}(2018{\natexlab{a}})}]{2018RNAAS...2....4K}
{Kanodia}, S. \& {Wright}, J. 2018{\natexlab{a}}, Research Notes of the
  American Astronomical Society, 2, 4

\bibitem[{{Kanodia} \& {Wright}(2018{\natexlab{b}})}]{2018ascl.soft08001K}
{Kanodia}, S. \& {Wright}, J.~T. 2018{\natexlab{b}}, {Barycorrpy: Barycentric
  velocity calculation and leap second management}

\bibitem[{{Kennedy} \& {Kenyon}(2008)}]{2008ApJ...673..502K}
{Kennedy}, G.~M. \& {Kenyon}, S.~J. 2008, \apj, 673, 502

\bibitem[{{Konopacky} {et~al.}(2013){Konopacky}, {Barman}, {Macintosh}, \&
  {Marois}}]{2013Sci...339.1398K}
{Konopacky}, Q.~M., {Barman}, T.~S., {Macintosh}, B.~A., \& {Marois}, C. 2013,
  Science, 339, 1398

\bibitem[{{Lafreni{\`e}re} {et~al.}(2008){Lafreni{\`e}re}, {Jayawardhana}, \&
  {van Kerkwijk}}]{2008ApJ...689L.153L}
{Lafreni{\`e}re}, D., {Jayawardhana}, R., \& {van Kerkwijk}, M.~H. 2008, \apjl,
  689, L153

\bibitem[{{Lavie} {et~al.}(2017){Lavie}, {Mendon{\c{c}}a}, {Mordasini},
  {Malik}, {Bonnefoy}, {Demory}, {Oreshenko}, {Grimm}, {Ehrenreich}, \&
  {Heng}}]{2017AJ....154...91L}
{Lavie}, B., {Mendon{\c{c}}a}, J.~M., {Mordasini}, C., {et~al.} 2017, \aj, 154,
  91

\bibitem[{{Lavigne} {et~al.}(2009){Lavigne}, {Doyon}, {Lafreni{\`e}re},
  {Marois}, \& {Barman}}]{2009ApJ...704.1098L}
{Lavigne}, J.-F., {Doyon}, R., {Lafreni{\`e}re}, D., {Marois}, C., \& {Barman},
  T. 2009, \apj, 704, 1098

\bibitem[{{Lodders}(2010)}]{2010ASSP...16..379L}
{Lodders}, K. 2010, Astrophysics and Space Science Proceedings, 16, 379

\bibitem[{{Lodieu} {et~al.}(2018){Lodieu}, {Zapatero Osorio}, {B{\'e}jar}, \&
  {Pe{\~n}a Ram{\'\i}rez}}]{2018MNRAS.473.2020L}
{Lodieu}, N., {Zapatero Osorio}, M.~R., {B{\'e}jar}, V.~J.~S., \& {Pe{\~n}a
  Ram{\'\i}rez}, K. 2018, \mnras, 473, 2020

\bibitem[{{Macintosh} {et~al.}(2006){Macintosh}, {Graham}, {Palmer}, {Doyon},
  {Gavel}, {Larkin}, {Oppenheimer}, {Saddlemyer}, {Wallace}, {Bauman}, {Evans},
  {Erikson}, {Morzinski}, {Phillion}, {Poyneer}, {Sivaramakrishnan}, {Soummer},
  {Thibault}, \& {Veran}}]{2006SPIE.6272E..0LM}
{Macintosh}, B., {Graham}, J., {Palmer}, D., {et~al.} 2006, Society of
  Photo-Optical Instrumentation Engineers (SPIE) Conference Series, Vol. 6272,
  {The Gemini Planet Imager}, 62720L

\bibitem[{{Madhusudhan} {et~al.}(2011){Madhusudhan}, {Burrows}, \&
  {Currie}}]{2011ApJ...737...34M}
{Madhusudhan}, N., {Burrows}, A., \& {Currie}, T. 2011, \apj, 737, 34

\bibitem[{{Manjavacas} {et~al.}(2014){Manjavacas}, {Bonnefoy}, {Schlieder},
  {Allard}, {Rojo}, {Goldman}, {Chauvin}, {Homeier}, {Lodieu}, \&
  {Henning}}]{2014A&A...564A..55M}
{Manjavacas}, E., {Bonnefoy}, M., {Schlieder}, J.~E., {et~al.} 2014, \aap, 564,
  A55

\bibitem[{{Marleau} {et~al.}(2019{\natexlab{a}}){Marleau}, {Coleman}, {Leleu},
  \& {Mordasini}}]{marleau2019a}
{Marleau}, G.-D., {Coleman}, G.~A.~L., {Leleu}, A., \& {Mordasini}, C.
  2019{\natexlab{a}}, \aap, 624, A20

\bibitem[{{Marleau} {et~al.}(2019{\natexlab{b}}){Marleau}, {Coleman}, {Leleu},
  \& {Mordasini}}]{2019A&A...624A..20M}
{Marleau}, G.-D., {Coleman}, G. A.~L., {Leleu}, A., \& {Mordasini}, C.
  2019{\natexlab{b}}, \aap, 624, A20

\bibitem[{{Marley} {et~al.}(2012){Marley}, {Saumon}, {Cushing}, {Ackerman},
  {Fortney}, \& {Freedman}}]{2012ApJ...754..135M}
{Marley}, M.~S., {Saumon}, D., {Cushing}, M., {et~al.} 2012, \apj, 754, 135

\bibitem[{{Marois} {et~al.}(2006){Marois}, {Lafreni{\`e}re}, {Doyon},
  {Macintosh}, \& {Nadeau}}]{2006ApJ...641..556M}
{Marois}, C., {Lafreni{\`e}re}, D., {Doyon}, R., {Macintosh}, B., \& {Nadeau},
  D. 2006, \apj, 641, 556

\bibitem[{{Marois} {et~al.}(2008){Marois}, {Macintosh}, {Barman}, {Zuckerman},
  {Song}, {Patience}, {Lafreni{\`e}re}, \& {Doyon}}]{2008Sci...322.1348M}
{Marois}, C., {Macintosh}, B., {Barman}, T., {et~al.} 2008, Science, 322, 1348

\bibitem[{{McElwain} {et~al.}(2007){McElwain}, {Metchev}, {Larkin}, {Barczys},
  {Iserlohe}, {Krabbe}, {Quirrenbach}, {Weiss}, \&
  {Wright}}]{2007ApJ...656..505M}
{McElwain}, M.~W., {Metchev}, S.~A., {Larkin}, J.~E., {et~al.} 2007, \apj, 656,
  505

\bibitem[{{Mesa} {et~al.}(2019){Mesa}, {Bonnefoy}, {Gratton}, {Van Der Plas},
  {D'Orazi}, {Sissa}, {Zurlo}, {Rigliaco}, {Schmidt}, {Langlois}, {Vigan},
  {Ubeira Gabellini}, {Desidera}, {Antoniucci}, {Barbieri}, {Benisty},
  {Boccaletti}, {Claudi}, {Fedele}, {Gasparri}, {Henning}, {Kasper},
  {Lagrange}, {Lazzoni}, {Lodato}, {Maire}, {Manara}, {Meyer}, {Reggiani},
  {Samland}, {Van den Ancker}, {Chauvin}, {Cheetham}, {Feldt}, {Hugot},
  {Janson}, {Ligi}, {M{\"o}ller-Nilsson}, {Petit}, {Rickman}, {Rigal}, \&
  {Wildi}}]{2019A&A...624A...4M}
{Mesa}, D., {Bonnefoy}, M., {Gratton}, R., {et~al.} 2019, \aap, 624, A4

\bibitem[{{Mesa} {et~al.}(2020){Mesa}, {D'Orazi}, {Vigan}, {Kitzmann}, {Heng},
  {Gratton}, {Desidera}, {Bonnefoy}, {Lavie}, {Maire}, {Peretti}, \&
  {Boccaletti}}]{2020MNRAS.495.4279M}
{Mesa}, D., {D'Orazi}, V., {Vigan}, A., {et~al.} 2020, \mnras, 495, 4279

\bibitem[{{Meshkat} {et~al.}(2015){Meshkat}, {Bonnefoy}, {Mamajek}, {Quanz},
  {Chauvin}, {Kenworthy}, {Rameau}, {Meyer}, {Lagrange}, {Lannier}, \&
  {Delorme}}]{2015MNRAS.453.2378M}
{Meshkat}, T., {Bonnefoy}, M., {Mamajek}, E.~E., {et~al.} 2015, \mnras, 453,
  2378

\bibitem[{{Mohanty} {et~al.}(2007){Mohanty}, {Jayawardhana}, {Hu{\'e}lamo}, \&
  {Mamajek}}]{2007ApJ...657.1064M}
{Mohanty}, S., {Jayawardhana}, R., {Hu{\'e}lamo}, N., \& {Mamajek}, E. 2007,
  \apj, 657, 1064

\bibitem[{{Molli{\`e}re} {et~al.}(2020){Molli{\`e}re}, {Stolker}, {Lacour},
  {Otten}, {Shangguan}, {Charnay}, {Molyarova}, {Nowak}, {Henning}, {Marleau},
  {Semenov}, {van Dishoeck}, {Eisenhauer}, {Garcia}, {Garcia Lopez}, {Girard},
  {Greenbaum}, {Hinkley}, {Kervella}, {Kreidberg}, {Maire}, {Nasedkin},
  {Pueyo}, {Snellen}, {Vigan}, {Wang}, {de Zeeuw}, \&
  {Zurlo}}]{2020arXiv200609394M}
{Molli{\`e}re}, P., {Stolker}, T., {Lacour}, S., {et~al.} 2020, arXiv e-prints,
  arXiv:2006.09394

\bibitem[{{Molli{\`e}re} {et~al.}(2019){Molli{\`e}re}, {Wardenier}, {van
  Boekel}, {Henning}, {Molaverdikhani}, \& {Snellen}}]{2019A&A...627A..67M}
{Molli{\`e}re}, P., {Wardenier}, J.~P., {van Boekel}, R., {et~al.} 2019, \aap,
  627, A67

\bibitem[{{Mordasini}(2013)}]{2013A&A...558A.113M}
{Mordasini}, C. 2013, \aap, 558, A113

\bibitem[{{Mordasini} {et~al.}(2016){Mordasini}, {van Boekel}, {Molli{\`e}re},
  {Henning}, \& {Benneke}}]{2016ApJ...832...41M}
{Mordasini}, C., {van Boekel}, R., {Molli{\`e}re}, P., {Henning}, T., \&
  {Benneke}, B. 2016, \apj, 832, 41

\bibitem[{{Morley} {et~al.}(2012){Morley}, {Fortney}, {Marley}, {Visscher},
  {Saumon}, \& {Leggett}}]{2012ApJ...756..172M}
{Morley}, C.~V., {Fortney}, J.~J., {Marley}, M.~S., {et~al.} 2012, \apj, 756,
  172

\bibitem[{{Moses} {et~al.}(2016){Moses}, {Marley}, {Zahnle}, {Line}, {Fortney},
  {Barman}, {Visscher}, {Lewis}, \& {Wolff}}]{2016ApJ...829...66M}
{Moses}, J.~I., {Marley}, M.~S., {Zahnle}, K., {et~al.} 2016, \apj, 829, 66

\bibitem[{{Nakajima} {et~al.}(1995){Nakajima}, {Oppenheimer}, {Kulkarni},
  {Golimowski}, {Matthews}, \& {Durrance}}]{1995Natur.378..463N}
{Nakajima}, T., {Oppenheimer}, B.~R., {Kulkarni}, S.~R., {et~al.} 1995, \nat,
  378, 463

\bibitem[{{N'Diaye} {et~al.}(2018){N'Diaye}, {Cuevas}, {S{\'a}nchez},
  {Carlotti}, {Vigan}, {Dohlen}, {Bonnefoy}, \& {Beuzit}}]{2018SPIE10703E..3EN}
{N'Diaye}, M., {Cuevas}, S., {S{\'a}nchez}, B., {et~al.} 2018, in Society of
  Photo-Optical Instrumentation Engineers (SPIE) Conference Series, Vol. 10703,
  \procspie, 107033E

\bibitem[{{Nissen}(2013)}]{2013A&A...552A..73N}
{Nissen}, P.~E. 2013, \aap, 552, A73

\bibitem[{{Noll, S.} {et~al.}(2012){Noll, S.}, {Kausch, W.}, {Barden, M.},
  {Jones, A. M.}, {Szyszka, C.}, {Kimeswenger, S.}, \& {Vinther, J.}}]{refId0}
{Noll, S.}, {Kausch, W.}, {Barden, M.}, {et~al.} 2012, A\&A, 543, A92

\bibitem[{{Nowak} {et~al.}(2019){Nowak}, {Lacour}, {Molli{\`e}re}, {Wang}, \&
  {Charnay}}]{2019ESS.....440407N}
{Nowak}, M., {Lacour}, S., {Molli{\`e}re}, P., {Wang}, J., \& {Charnay}, B.
  2019, in AAS/Division for Extreme Solar Systems Abstracts, Vol.~51,
  AAS/Division for Extreme Solar Systems Abstracts, 404.07

\bibitem[{{{\"O}berg} \& {Bergin}(2016)}]{2016ApJ...831L..19O}
{{\"O}berg}, K.~I. \& {Bergin}, E.~A. 2016, \apjl, 831, L19

\bibitem[{{{\"O}berg} {et~al.}(2011){{\"O}berg}, {Murray-Clay}, \&
  {Bergin}}]{2011ApJ...743L..16O}
{{\"O}berg}, K.~I., {Murray-Clay}, R., \& {Bergin}, E.~A. 2011, \apjl, 743, L16

\bibitem[{{Patience} {et~al.}(2010){Patience}, {King}, {de Rosa}, \&
  {Marois}}]{2010A&A...517A..76P}
{Patience}, J., {King}, R.~R., {de Rosa}, R.~J., \& {Marois}, C. 2010, \aap,
  517, A76

\bibitem[{{Petit dit de la Roche} {et~al.}(2018){Petit dit de la Roche},
  {Hoeijmakers}, \& {Snellen}}]{2018A&A...616A.146P}
{Petit dit de la Roche}, D.~J.~M., {Hoeijmakers}, H.~J., \& {Snellen}, I.~A.~G.
  2018, \aap, 616, A146

\bibitem[{{Petrus} {et~al.}(2020){Petrus}, {Bonnefoy}, {Chauvin}, {Babusiaux},
  {Delorme}, {Lagrange}, {Florent}, {Bayo}, {Janson}, {Biller}, {Manjavacas},
  {Marleau}, \& {Kopytova}}]{2020A&A...633A.124P}
{Petrus}, S., {Bonnefoy}, M., {Chauvin}, G., {et~al.} 2020, \aap, 633, A124

\bibitem[{{Qi} {et~al.}(2015){Qi}, {{\"O}berg}, {Andrews}, {Wilner}, {Bergin},
  {Hughes}, {Hogherheijde}, \& {D'Alessio}}]{2015ApJ...813..128Q}
{Qi}, C., {{\"O}berg}, K.~I., {Andrews}, S.~M., {et~al.} 2015, \apj, 813, 128

\bibitem[{{Rajan} {et~al.}(2017){Rajan}, {Rameau}, {De Rosa}, {Marley},
  {Graham}, {Macintosh}, {Marois}, {Morley}, {Patience}, {Pueyo}, {Saumon},
  {Ward-Duong}, {Ammons}, {Arriaga}, {Bailey}, {Barman}, {Bulger}, {Burrows},
  {Chilcote}, {Cotten}, {Czekala}, {Doyon}, {Duch{\^e}ne}, {Esposito},
  {Fitzgerald}, {Follette}, {Fortney}, {Goodsell}, {Greenbaum}, {Hibon},
  {Hung}, {Ingraham}, {Johnson-Groh}, {Kalas}, {Konopacky}, {Lafreni{\`e}re},
  {Larkin}, {Maire}, {Marchis}, {Metchev}, {Millar-Blanchaer}, {Morzinski},
  {Nielsen}, {Oppenheimer}, {Palmer}, {Patel}, {Perrin}, {Poyneer},
  {Rantakyr{\"o}}, {Ruffio}, {Savransky}, {Schneider}, {Sivaramakrishnan},
  {Song}, {Soummer}, {Thomas}, {Vasisht}, {Wallace}, {Wang}, {Wiktorowicz}, \&
  {Wolff}}]{2017AJ....154...10R}
{Rajan}, A., {Rameau}, J., {De Rosa}, R.~J., {et~al.} 2017, \aj, 154, 10

\bibitem[{{Rajpurohit} {et~al.}(2018){Rajpurohit}, {Allard}, {Rajpurohit},
  {Sharma}, {Teixeira}, {Mousis}, \& {Kamlesh}}]{2018AeA...620A.180R}
{Rajpurohit}, A.~S., {Allard}, F., {Rajpurohit}, S., {et~al.} 2018, \aap, 620,
  A180

\bibitem[{{Rizzuto} {et~al.}(2011){Rizzuto}, {Ireland}, \&
  {Robertson}}]{2011MNRAS.416.3108R}
{Rizzuto}, A.~C., {Ireland}, M.~J., \& {Robertson}, J.~G. 2011, \mnras, 416,
  3108

\bibitem[{{Rousselot} {et~al.}(2000){Rousselot}, {Lidman}, {Cuby}, {Moreels},
  \& {Monnet}}]{2000A&A...354.1134R}
{Rousselot}, P., {Lidman}, C., {Cuby}, J.~G., {Moreels}, G., \& {Monnet}, G.
  2000, \aap, 354, 1134

\bibitem[{{Ruffio} {et~al.}(2019){Ruffio}, {Macintosh}, {Konopacky}, {Barman},
  {De Rosa}, {Wang}, {Wilcomb}, {Czekala}, \& {Marois}}]{2019AJ....158..200R}
{Ruffio}, J.-B., {Macintosh}, B., {Konopacky}, Q.~M., {et~al.} 2019, \aj, 158,
  200

\bibitem[{{Samland} {et~al.}(2017){Samland}, {Molli{\`e}re}, {Bonnefoy},
  {Maire}, {Cantalloube}, {Cheetham}, {Mesa}, {Gratton}, {Biller}, {Wahhaj},
  {Bouwman}, {Brandner}, {Melnick}, {Carson}, {Janson}, {Henning}, {Homeier},
  {Mordasini}, {Langlois}, {Quanz}, {van Boekel}, {Zurlo}, {Schlieder},
  {Avenhaus}, {Beuzit}, {Boccaletti}, {Bonavita}, {Chauvin}, {Claudi}, {Cudel},
  {Desidera}, {Feldt}, {Fusco}, {Galicher}, {Kopytova}, {Lagrange}, {Le
  Coroller}, {Martinez}, {Moeller-Nilsson}, {Mouillet}, {Mugnier}, {Perrot},
  {Sevin}, {Sissa}, {Vigan}, \& {Weber}}]{2017A&A...603A..57S}
{Samland}, M., {Molli{\`e}re}, P., {Bonnefoy}, M., {et~al.} 2017, \aap, 603,
  A57

\bibitem[{{Schlecker} {et~al.}(2020){Schlecker}, {Mordasini}, {Emsenhuber},
  {Klahr}, {Henning}, {Burn}, {Alibert}, \& {Benz}}]{2020arXiv200705563S}
{Schlecker}, M., {Mordasini}, C., {Emsenhuber}, A., {et~al.} 2020, arXiv
  e-prints, arXiv:2007.05563

\bibitem[{{Schmidt} {et~al.}(2014){Schmidt}, {Mugrauer}, {Neuh{\"a}user},
  {Vogt}, {Witte}, {Hauschildt}, {Helling}, \&
  {Seifahrt}}]{2014A&A...566A..85S}
{Schmidt}, T.~O.~B., {Mugrauer}, M., {Neuh{\"a}user}, R., {et~al.} 2014, \aap,
  566, A85

\bibitem[{{Schmidt} {et~al.}(2008){Schmidt}, {Neuh{\"a}user}, {Seifahrt},
  {Vogt}, {Bedalov}, {Helling}, {Witte}, \& {Hauschildt}}]{2008A&A...491..311S}
{Schmidt}, T.~O.~B., {Neuh{\"a}user}, R., {Seifahrt}, A., {et~al.} 2008, \aap,
  491, 311

\bibitem[{{Seifahrt} {et~al.}(2007){Seifahrt}, {Neuh{\"a}user}, \&
  {Hauschildt}}]{2007A&A...463..309S}
{Seifahrt}, A., {Neuh{\"a}user}, R., \& {Hauschildt}, P.~H. 2007, \aap, 463,
  309

\bibitem[{{Skemer} {et~al.}(2014){Skemer}, {Marley}, {Hinz}, {Morzinski},
  {Skrutskie}, {Leisenring}, {Close}, {Saumon}, {Bailey}, {Briguglio},
  {Defrere}, {Esposito}, {Follette}, {Hill}, {Males}, {Puglisi}, {Rodigas}, \&
  {Xompero}}]{2014ApJ...792...17S}
{Skemer}, A.~J., {Marley}, M.~S., {Hinz}, P.~M., {et~al.} 2014, \apj, 792, 17

\bibitem[{Skilling(2006)}]{skilling2006}
Skilling, J. 2006, Bayesian Anal., 1, 833

\bibitem[{{Snellen} {et~al.}(2014){Snellen}, {Brandl}, {de Kok}, {Brogi},
  {Birkby}, \& {Schwarz}}]{2014Natur.509...63S}
{Snellen}, I. A.~G., {Brandl}, B.~R., {de Kok}, R.~J., {et~al.} 2014, \nat,
  509, 63

\bibitem[{{Sparks} \& {Ford}(2002)}]{2002ApJ...578..543S}
{Sparks}, W.~B. \& {Ford}, H.~C. 2002, \apj, 578, 543

\bibitem[{{Stone} {et~al.}(2016){Stone}, {Eisner}, {Skemer}, {Morzinski},
  {Close}, {Males}, {Rodigas}, {Hinz}, \& {Puglisi}}]{2016ApJ...829...39S}
{Stone}, J.~M., {Eisner}, J., {Skemer}, A., {et~al.} 2016, \apj, 829, 39

\bibitem[{{Thatte} {et~al.}(2007){Thatte}, {Abuter}, {Tecza}, {Nielsen},
  {Clarke}, \& {Close}}]{2007MNRAS.378.1229T}
{Thatte}, N., {Abuter}, R., {Tecza}, M., {et~al.} 2007, \mnras, 378, 1229

\bibitem[{{Trotta}(2008)}]{2008ConPh..49...71T}
{Trotta}, R. 2008, Contemporary Physics, 49, 71

\bibitem[{{Uyama} {et~al.}(2020){Uyama}, {Currie}, {Hori}, {De Rosa}, {Mede},
  {Brandt}, {Kwon}, {Guyon}, {Lozi}, {Jovanovic}, {Martinache}, {Kudo},
  {Tamura}, {Kasdin}, {Groff}, {Chilcote}, {Hayashi}, {McElwain},
  {Asensio-Torres}, {Janson}, {Knapp}, \& {Serabyn}}]{2020AJ....159...40U}
{Uyama}, T., {Currie}, T., {Hori}, Y., {et~al.} 2020, \aj, 159, 40

\bibitem[{{Valletta} \& {Helled}(2018)}]{2018arXiv181110904V}
{Valletta}, C. \& {Helled}, R. 2018, arXiv e-prints, arXiv:1811.10904

\bibitem[{{van der Marel} {et~al.}(2019){van der Marel}, {Dong}, {di
  Francesco}, {Williams}, \& {Tobin}}]{2019ApJ...872..112V}
{van der Marel}, N., {Dong}, R., {di Francesco}, J., {Williams}, J.~P., \&
  {Tobin}, J. 2019, \apj, 872, 112

\bibitem[{{Viana Almeida} {et~al.}(2009){Viana Almeida}, {Santos}, {Melo},
  {Ammler-von Eiff}, {Torres}, {Quast}, {Gameiro}, \&
  {Sterzik}}]{2009A&A...501..965V}
{Viana Almeida}, P., {Santos}, N.~C., {Melo}, C., {et~al.} 2009, \aap, 501, 965

\bibitem[{{Wang} {et~al.}(2018){Wang}, {Graham}, {Dawson}, {Fabrycky}, {De
  Rosa}, {Pueyo}, {Konopacky}, {Macintosh}, {Marois}, {Chiang}, {Ammons},
  {Arriaga}, {Bailey}, {Barman}, {Bulger}, {Chilcote}, {Cotten}, {Doyon},
  {Duch{\^e}ne}, {Esposito}, {Fitzgerald}, {Follette}, {Gerard}, {Goodsell},
  {Greenbaum}, {Hibon}, {Hung}, {Ingraham}, {Kalas}, {Larkin}, {Maire},
  {Marchis}, {Marley}, {Metchev}, {Millar-Blanchaer}, {Nielsen}, {Oppenheimer},
  {Palmer}, {Patience}, {Perrin}, {Poyneer}, {Rajan}, {Rameau},
  {Rantakyr{\"o}}, {Ruffio}, {Savransky}, {Schneider}, {Sivaramakrishnan},
  {Song}, {Soummer}, {Thomas}, {Wallace}, {Ward-Duong}, {Wiktorowicz}, \&
  {Wolff}}]{2018AJ....156..192W}
{Wang}, J.~J., {Graham}, J.~R., {Dawson}, R., {et~al.} 2018, \aj, 156, 192

\end{thebibliography}

\end{document}